\documentclass[10pt]{elsarticle}
\usepackage{lineno,hyperref}
\usepackage{geometry}
\usepackage{setspace}
\usepackage{amsthm}
\usepackage{amsmath,bm}
\usepackage{anysize}
\usepackage{amsfonts}
\usepackage{amssymb}
\usepackage{algorithm}
\usepackage{algorithmic}
\usepackage{appendix}
\usepackage{bm}
\usepackage{color,colortbl,soul}
\usepackage{multirow}
\usepackage{graphics,graphicx}
\usepackage{caption}
\usepackage{epstopdf} 
\usepackage{eurosym}
\usepackage{float}
\usepackage{fancyhdr}
\usepackage{longtable}
\usepackage{mathrsfs}
\usepackage{enumerate}
\usepackage{flushend}
\usepackage[framemethod=tikz]{mdframed}
\usepackage{lipsum}
\usepackage{float}
\usepackage{verbatim}
\usepackage{natbib}
\usepackage{cases}
\usepackage{caption}
\usepackage{subcaption}
\usepackage{hyperref}
\usepackage{booktabs}
\usepackage{commath,amsmath}
\usepackage{multicol}
\usepackage{mathabx}
\usepackage{xcolor}
\usepackage{framed}
\usepackage{multicol}
\usepackage{nomencl}
\makenomenclature
\renewcommand*\nompreamble{\begin{multicols}{2}}
\renewcommand*\nompostamble{\end{multicols}}
\usepackage{tabu}

\journal{Engineering Fracture Mechanics}
\date{}
\makeatother
\setcitestyle{authoryear}
\usepackage{times}
\begin{document}
\begin{frontmatter}
\title{$\mathtt{E}^2$-TFA based multiscale analysis of failure in elasto-plastic composites}
\author[add1]{Harpreet Singh\corref{corrauth}}
\cortext[corrauth]{Corresponding author}
\ead{harpreet@iitgoa.ac.in}
\address[add1]{School of Mechanical Sciences, Indian Institute of Technology Goa, Farmagudi, Goa 403401 India}
\begin{abstract} 
\small \begin{spacing}{1.1}
This paper describes a novel homogenization methodology for analyzing the failure of elastoplastic composite materials based on elastic and eigen influence tensors-driven transformation field analysis ($\mathtt{E}^2$-TFA). The proposed technique considers the microscopic eigenstrain field accounting for intra-phase damage and inelastic strains. This results in realistic computations by alleviating the post-damage stiffness response, which is a drawback of TFA-based methods. We attain computational efficiency by identifying the preprocessing data solely from the elastic and eigen transformation functions and adopting a reduced order modelling technique with a piecewise constant eigenstrain field throughout the subdomains. The performance of the model is assessed by simulating the response for (a) the representative volume element (RVE) as a homogenized continuum and (b) the various composites under complex load histories with intricate macroscale morphologies. Furthermore, the nonlinear shear stress-strain response of a glass fiber composite is calculated and compared to experimentally measured fracture initiation parameters, failure plane orientation, and strain histories. Finally, we show that $\mathtt{E}^2$-TFA can accurately and efficiently capture damage and inelastic deformations in order to estimate the mechanical response of composite materials in a better way. \end{spacing} 
\end{abstract}

\begin{keyword} \small \begin{spacing}{1.1}
    Multiscale modeling, Homogenization, Transformation field analysis, Reduced order modeling, Damage, Plastic strain, Fiber-reinforced composites \end{spacing}
\end{keyword} 
\end{frontmatter}

\section{Introduction}

Tailored multiphase materials are vastly used in various industrial sectors for designing load-bearing components. The two-phase materials, such as fiber-reinforced composites, carry the significant advantages of high specific strength and specific stiffness over conventional mono-phase materials. \color{black}Several composite materials exhibit the typical ductile nature of the fracture, which consists of nucleation, growth and coalescence of small voids. \color{black}Besides, binder material demonstrates significant plastic deformation before the onset of interphase failure. Overall the mechanical behaviour of composite material is challenging to predict due to the complex interactions of microconstituents. Multiscale-based homogenization techniques offer the great advantage of incorporating these interaction details at a lower scale which is the scale of heterogeneities and including the effects of those at the macroscale, which is the scale of the naked eye and improves the accuracy of prediction capability of the numerical model. Numerous multiscale theories exist in the literature for obtaining the mechanical behaviour of these microstructure-dependent heterogeneous materials. Multiscale modelling efforts basically were started with the classical analytical theories originated by \citet{voigt1910lehrbuch}, \citet{reuss1930berucksichtigung}, \citet{hill1963elastic, hill1965continuum, hill1965self,hill1972constitutive}, \citet{budiansky1965elastic}, \citet{mori1973average}, \citet{willis1977bounds} and used for determining the linear properties of materials with reasonable accuracy. Microstructure-based approaches have gained tremendous popularity over the last three decades, where the Finite Element Method (FEM) deployed two-scale models for computational homogenization emerges as a promising tool for simulating the nonlinear response of the composites (see \citep{michel1999effective}, \citep{terada2001class}, \citep{miehe2003computational}, \citep{kouznetsova2004multi}). However, simulating non-linearity due to the plasticity of the constituents and fracture always poses difficulty to researchers regarding coupling the micro effects to the macro response. Several approaches have been developed for including the micro-fracture lead effects in a computational homogenization rooted framework. These studies could be categorised into two; 
\begin{enumerate}
\item 	First category employs the \textbf{fracture mechanics-based methodologies} where the crack is introduced as a discontinuity in the displacement field, and new surfaces are formed as cohesive interfaces. The response of the microstructure with embedded discontinuities is evaluated at one scale and processed to the larger scale by maintaining the equivalence of dissipated energies. (see  \citep{vsmilauer2011multiscale}, \citep{coenen2012multi}, \citep{coenen2012multia}, \citep{bosco2014multiscale}).
\item Second category uses the \textbf{continuum methodologies} where the stress-strain laws of micro-constituents are regulated to portray the effect of microvoids/cracks. Cracked surfaces are featured as material softening zones of localized strains  (see  \citep{fish1999computational}, \citep{fish2001multiscale}, \citep{oliver2004continuum}, \citep{oliver2008two}, \citep{oliver2014crack}, \citep{bogdanor2015multiscale}).
\end{enumerate}
On the other hand, separate studies were carried out to include the plastic deformations of the micro constituents at the macro scale (\citep{fish1997computational}, \citep{fish1998computational}, \citep{fish1999finite},  \citep{miehe2002strain}). However, due to the path dependency of inelastic strains and complex underlying inelastic strain fields, these numerical techniques demand many internal variables and, prohibitively, increase the computational cost in terms of enormous data storage and large solution time. To circumvent this issue, reduced-order techniques have been adopted as a viable solution for microscale simulations which are undoubtedly proven as computationally efficient without compromising the accuracy of the solution (\citep{abdulle2014reduced}, \citep{yuan2014reduced}, \citep{fritzen2015nonlinear}, \citep{oliver2017reduced}, \citep{caicedo2019high}). Although concurrent efforts are being established to incorporate the plasticity and damage effects in the multiscale framework, only a few are available in the literature, encompassing both in a single formulation.   

Meanwhile, \textbf{semi-analytical} techniques are attaining acceptability in the multiscale research community due to their ease of implementation and low computational burden. Among these methods, Transformation Field Analysis (TFA), proposed by \citet{dvorak1992transformation} and \citet{dvorak1994modeling}, captured the behaviour of the microstructural nonlinearities caused by the inelastic strain field. In TFA, the microscale domain is discretized into subdomains, and the plastic strain field is considered constant in each subdomain. Overall, the number of variables defining the plastic strain field at the microscale is reduced. The proposed technique leads to acceptable numerical cost. However, when TFA results were compared with FE, it showed the stiffer behaviour of the microdomain. \citet{chaboche2001towards} proposed another version of this, named piecewise uniform TFA (PWUTFA), and simulated the damage behaviour of two-phase composite material. Many other researchers used PWUTFA (\citet{sacco2009nonlinear}, \citet{marfia2018multiscale}, \citet{gopinath2018common}) to derive the behaviour of two-phase inelastic materials and found the reasonably good prediction capability. \citet{alaimo2019optimization} proposed a multi-objective optimization methodology for the selection of the number of sub-domains or clusters efficiently to reduce the inelastic strain field approximation error. To improve the accuracy, \citet{michel2003nonuniform} proposed non-uniform TFA (NTFA) by introducing a more extensive estimation of the inelastic strain field calculated as the linear combination of several modes which are extracted during preprocessing phase. Nonuniform TFA was adopted by \citet{fritzen2011nonuniform} for obtaining the response of materials with nonmorphological anisotropic heterogeneities. Furthermore, another technique, called piece-wise nonuniform TFA (PWNTFA), was proposed by \citet{sepe2013nonuniform} where history variables are calculated based on the linear combination of inelastic modes of deformation, which are obtained from analytical functions.\color{black} 

The TFA, which is also referred to as the eigenstrain field-based approach, is adopted by \cite{fish1997computational} where an asymptotic analysis-based mathematical homogenization framework for periodic microstructures \citep{bakhvalov1989homogenisation} \citep{papanicolau1978asymptotic} is formulated to include damage effects. \cite{fish1999computational} developed a closed-form solution for getting the macro-scale fields from the local solutions by proposing a theory based on the two-scale asymptotic expansion of damage. Similar to this approach \cite{oskay2007eigendeformation} devised another asymptotic expansion method with TFA for periodic heterogeneous material, which accounts for the interphase and intraphase failures using a concept of eigendeformation. \color{black}However, it has been observed that the lower-order approximation of the eigenstrain field in TFA raises the issue of spurious post-damage stiffness, causing inaccuracies in the predictions. \color{black}This problem can be alleviated by either reconstructing the eigenstrain field using recomputed influence tensors \citep{fish2013hybrid} and making a compatible or impotent eigenstrain field or using the enhanced stress-strain laws for the subdomains based on their shapes and sizes \citep{singh2020strain}. Essentially, the eigenstrain or eigendeformation-based formulations consist of two steps; 1). In the first step, also called the offline stage, influence tensors are calculated, which do not change with macroscopically applied loading and boundary conditions and 2). On-the-go macroscale calculations are performed in the second stage using those influence tensors. Many studies were carried out using the eigendeformation-based TFA approach for calculating the response of fractured multi-phase media (see \citep{crouch2010symmetric}, \citep{bogdanor2013uncertainty}, \citep{bogdanor2015multiscale}).

Furthermore, \cite{zhang2015eigenstrain} developed an enhanced eigenstrain-based reduced order homogenization method using two-scale asymptotic analysis and calculated the macroscopic response for polycrystalline materials. They used the influence tensors for capturing the local inelastic stress and strain fields in the microstructure and mentioned using this for intergrain damage in future. By introducing sparsity, the computational efficiency of the same formulation was improved by \cite{zhang2017sparse}, which is driven by an improved grain clustering scheme. Recently, \cite{oskay2020discrete} presented an eigendeformation-based reduced order homogenization technique for the failure analysis of composite material where they used a specific number of cohesive surfaces for the evolution of microscale failure.  Another multiscale discrete damage theory (MDDT) is developed by \cite{su2022modeling} for simulating the randomly oriented and progressively reoriented cracks. A cohesive surface approach is adopted for tracking the fracture phenomena in cross-ply laminates. Mesh size-objectivity is achieved by introducing a length scale parameter \citep{su2021mesh}. This model also simulates the behaviour under fatigue loading and captures multiple failure modes such as splitting, transverse matrix cracking and delamination. 

\cite{labat2023multiscale} used reduced order NTFA with tangent second order approximation (TSO-NTFA) for studying the macroscopic behaviour of UO$_2$ embedded in the polycrystalline ceramic. The heterogeneities are modelled as temperature and strain-rate-dependent elasto-viscoplastic material without considering any failure in the constituents. \cite{ju2022ntfa} formulated a model reduction approach using space-time decomposition with NTFA for microscopic inelastic strain fields. This reduced order homogenisation technique deduces the closed-form constitutive relations and gives error estimation resulting from modelling and discretisation. Reduced-order NUTFA-based multiscale model was exploited by \cite{addessi2023non} for analysing the masonry material, which is modelled with shell elements, subjected to out-of-plane and in-plane loading. Interface failure is characterised by cohesive-frictional zero-thickness contact surfaces.\color{black}

Clearly, there are two approaches predominantly existing in literature, i.e. one group is eigendeformation based, which predicts the damage in the heterogeneous media disregarding the plasticity in the material, and the other is eigenstrain based, which portrays the inelastic strain field from the plasticity of microconstituents in the absence of damage. 

Comprehensively, to date, numerous studies with different versions of TFA have been carried out. Still, a few key aspects of this approach need attention and are the focus of research:
\begin{enumerate}
	\item Approximation of inelastic and damage-equivalent eigenstrain field using the optimum number of history variables \citep{sparks2013identification} \citep{bogdanor2015multiscale}.
	\item Problem of spurious eigenstrain field and macroscale response caused by the unrealistic post damage stiffness \citep{fish2013hybrid}  \citep{singh2017reduced} \citep{singh2020strain}. 
\end{enumerate}
In this manuscript, the concerns mentioned above are addressed by proposing a new TFA approach for obtaining the mechanical response of elastoplastic composites in the presence of damage. In the present multiscale homogenization procedure, order reduction is achieved by subdividing the RVE domain into subdomains with a constant eigenstrain field similar to the PWUTFA approach. Innovatively, eigenstrain influence tensors and homogenized properties are calculated by using a single elastic influence tensor which simplifies the overall preprocessing (sometimes called offline) process. It is demonstrated by performing the RVE response check that this procedure for calculating influence tensors and homogenized properties also alleviates the problem of unrealistic post-damage stiffness.\color{black} 

Furthermore, to assess the efficacy of the proposed TFA-based homogenization technique, numerical results are presented and compared with the available experimental results. Meanwhile, checking the capability in case of complex loading histories, a multiscale simulation is performed for a two-phase composite material considering the elastoplastic material properties and damage. 

In Section \ref{sec:2}, multiscale philosophy in the context of the homogenization framework is introduced, and coupling between scales is explained. Section \ref{sec:3} focuses primarily on the microscale model and elastoplastic damage constitutive laws followed for phase materials. In Section \ref{sec:4}, TFA based reduced order model is illustrated, and the methodology for calculating various influence tensors is discussed in detail. The numerical procedure to implement the proposed method for preprocessing and solution stage is provided in Section \ref{sec:5}. Section \ref{sec:6} includes the numerical studies performed at the RVE and full scale to verify the proposed approach. Four verification studies are carried out; 1). RVE,  2). Open-hole laminated composite, 3). Double notched laminated composite, and 4). $10^\circ$ unidirectional fiber-reinforced composite. Finally, Section \ref{sec:7} reports the conclusions.      
\vspace{0.5 cm}              
\begin{table*}[!htbp]
\small
\begin{framed}
\begin{tabular}{p{1.15cm} p{5.5cm} p{0.2cm} p{1.15cm} p{5.5cm}}
\textbf{Nomenclature} &  & & &\\
\vspace{0.2mm}\\ 
{$\boldsymbol{\varepsilon}^o$} & {Strain at macroscale} &&
{$\boldsymbol{\sigma}^o$} & {Stress at macroscale} \\
{$\boldsymbol{\sigma}, \sigma_{ij}$} & {Stress at microscale}&&
{$\boldsymbol{\varepsilon}, \varepsilon_{ij}$} & {Strain at microscale}\\
{$\Omega$} & {Microscale domain} &&
{$\partial\Omega$} & {Microscale domain boundary} \\
{$\boldsymbol{y}$} & {Position vector of a point in micoscopic domain}&&
{$\boldsymbol{x}$} & {Position vector of a point in macoscopic domain} \\
{$\Gamma$} & {Macroscale domain}&&
{$\boldsymbol{n}$} & {A unit normal vector over a surface}\\
{$\Omega^1,\Omega^2,..$} & {Microscopic subdomains}&&
{$\boldsymbol{u}, u_i$}&{Displacement field}\\
{$\tilde{\boldsymbol{u}}$}&{Fluctuating displacement field}&&
{$\bar{\boldsymbol{u}}$}&{Average displacement field}\\
{$\boldsymbol{\tilde{\varepsilon}}, \tilde{\varepsilon}_{ij}$}&{Effective strain}&&
{$\tilde{\boldsymbol{\sigma}}, \tilde{\sigma}_{ij}$}&{Effective stress}\\
{$\bar{\boldsymbol{\sigma}}$}&{Average stress}&&
{$\bar{\boldsymbol{\varepsilon}}$}&{Average strain}\\
{$\boldsymbol{\varepsilon}^e$}&{Elastic strain}&&
{$\boldsymbol{\varepsilon}^p$}&{Plastic strain}\\
{$\mathbb{L}$}&{Elastic contitutive tensor}&&
{$\mathbb{L}^{ep}$}&{Elastoplatic constitutive tensor}\\
{$\mathbb{D}$}&{Damage tensor}&&
{$\mathbb{I}$}&{Identity tensor}\\
{$\overline{\mathbb{M}}$}&{Homogenized eigen constitutive tensor}&&
{$\overline{\mathbb{L}}$}&{Homogenized elastic constitutive tensor}\\
{$\overline{\mathbb{G}}$}&{Stress concentration tensor}&&
{$\mathbb{T}$}&{Transformation tensor which relates the eigenstrain with physical strain}\\
{$\mathbb{E}$}&{Elastic influence tensor}&&
{$\mathbb{S}$}&{Eigenstrain influence tensor}\\
{$\overline{\mathbb{E}}$}&{Weighted average elastic influence tensor}&&
{$\overline{\mathbb{S}}$}&{Eigen influence tensor or weighted average eigenstrain tensor}\\
{$\omega$}&{Damage variable}&&
{$[M]$}&{Damage effect matrix}\\
{$\Psi$}&{Helmholtz free energy potential function}&&
{$T$}&{Temperature}\\
{$r$}&{Isotropic hardening variable}&&
{$\boldsymbol{\alpha}$}&{Kinematic hardening variable}\\
{$\rho$}&{Density}&&
{$s$}&{Entropy}\\
{$G$}&{Shear modulus}&&
{$K$}&{Bulk modulus}\\
{$\delta_{ij}$}&{Kronecker delta}&&
{$\boldsymbol{q}$}&{Heat flux}\\
{$R$}&{A variable associated to state variable $r$}&&
{$Y$}&{A variable associated to state variable $\omega$}\\
{$\mathfrak{F}$}&{Dissipation potential function}&&
{$\mathfrak{F}^P$}&{Dissipation potential function accounting plastic deformation}\\
{$\mathfrak{F}^D$}&{Dissipation potential function accounting damage}&&
{$\dot{\zeta}^P$}&{Plastic multiplier}\\
{$\dot{\zeta}^D$}&{Damage multiplier}&&
{$\boldsymbol{\sigma}^{DEV}$}&{Deviatoric stress}\\
{${\sigma}_{EQ}$}&{Equivalent stress}&&
{$\kappa$}&{Maximum principal strain}\\
{$\kappa_{D}$}&{Damage initiation strain}&&
{$\kappa_{F}$}&{Strain at complete failure}\\
{$\hat{\epsilon}_1, \hat{\epsilon}_2, \hat{\epsilon}_3$}&{Principal strain values}&&
{$\boldsymbol{\mathcal{H}}$}&{Heaviside step function}\\
{$\boldsymbol{\mu}$}&{Eigenstrain}&&
{$\bar{\boldsymbol{\mu}}$}&{Weighted average eigenstrain}\\
{$\boldsymbol{\lambda}$}&{Eigenstress}&&
{$v_f$}&{Volume fraction}\\
{$N$}&{Shape function}&&
{$\phi$}&{Weighing function}\\
{$\boldsymbol{t}^o$}&{Equivalent traction vector corresponding to $\boldsymbol{\sigma}^o$}&&
{$\boldsymbol{R}$}&{Residual vector in Newton-Raphson method}\\
{$\boldsymbol{U}$}&{Unknown vector in Newton-Raphson method}\\

\end{tabular}
\end{framed}
\normalfont
\end{table*}
\subsection*{Notation:}      
The scalar variables are written in Greek, or Latin lowercase letters, i.e. $\alpha$, $a\ldots$ and the vectors are donated as boldface lowercase letters, i.e. $\boldsymbol{u}$, $\boldsymbol{v}\ldots$. The dot product of two vectors is $ \boldsymbol{u}$ $\cdot$ $\boldsymbol{v} = u_iv_i$, and the Einstein summation convention is adopted for repeated indices. Second-order tensors are represented either in boldface Greek letters in lowercase or in Latin capital letters, i.e. $\boldsymbol{\sigma}$, $\bf{U}$. Double contraction or scalar product of two second-order tensors is $ \bold{U} : \bold{V} = U_{ij}V_{ij}$ or $\boldsymbol{\sigma} : \boldsymbol{\varepsilon}=\sigma_{ij}\varepsilon_{ij} $. Fourth-order tensors are shown as Latin blackboard capital letters in bold form, i.e. $\mathbb{U}$. Double contraction of fourth-order tensor with second-order tensor is $ \mathbb{U} : \bold{V} = U_{ijkl}V_{kl}$. The symbol $\norm{ \bullet }_2$ represents the $\mathcal{L}_2$ norm, which is defined as $\norm{ \boldsymbol{u} }_2=\sqrt{\sum_{i=1}^{3}u_i^2}$. Superscripted quantity in parentheses, e.g.  $\bold{U}^{(n+1)}$, corresponds to the $n+1$ time/load step number. Subscripted quantity is a square bracket, e.g.  $\bold{U}_{[m+1]}$ refers to $m+1$ iteration number. Superscripted letters $i$ and $j$ refer to the specific domain number.

\section{Micro-Macro Coupling}\label{sec:2}
\begin{figure}[H]
\centering
\includegraphics[width=0.75\textwidth]{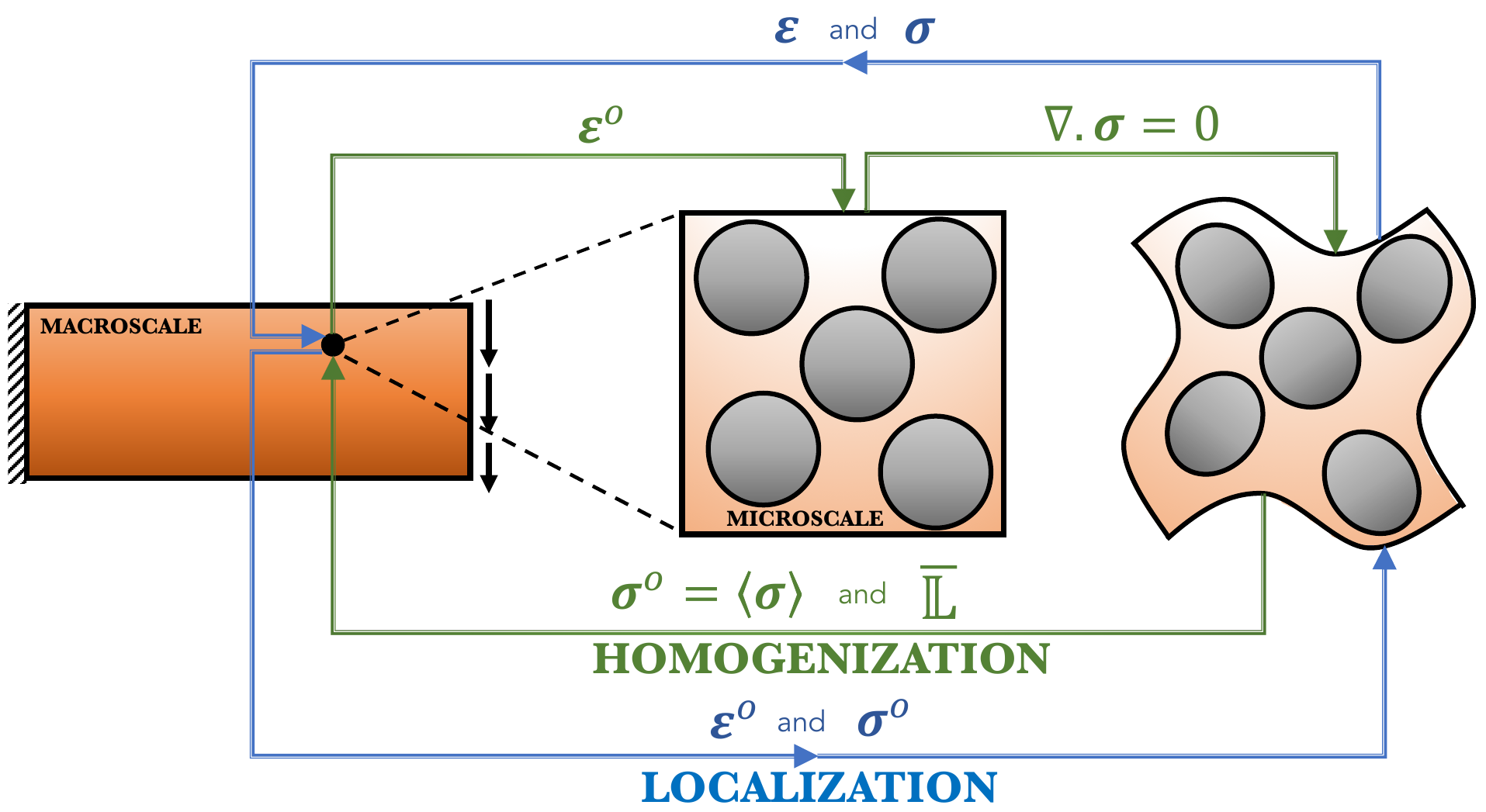}
\caption{Macroscale and microscale coupling process flow in two settings 1). Homogenization, where macroscale information in terms of macro-strain, $\boldsymbol{\varepsilon}^o$ is transferred to microscale, which results in deformed configuration and provides homogenized stress, $\boldsymbol{\sigma}^o$ at the macroscale (shown in Green), and 2). Localization, where pre-obtained homogenized data is supplied at the microscale, and localized stress and strain distribution are evaluated from microscale analyses (shown in Blue).}
\label{fig:1}
\end{figure}

\begin{figure}[H]
\centering
\includegraphics[width=0.65\textwidth]{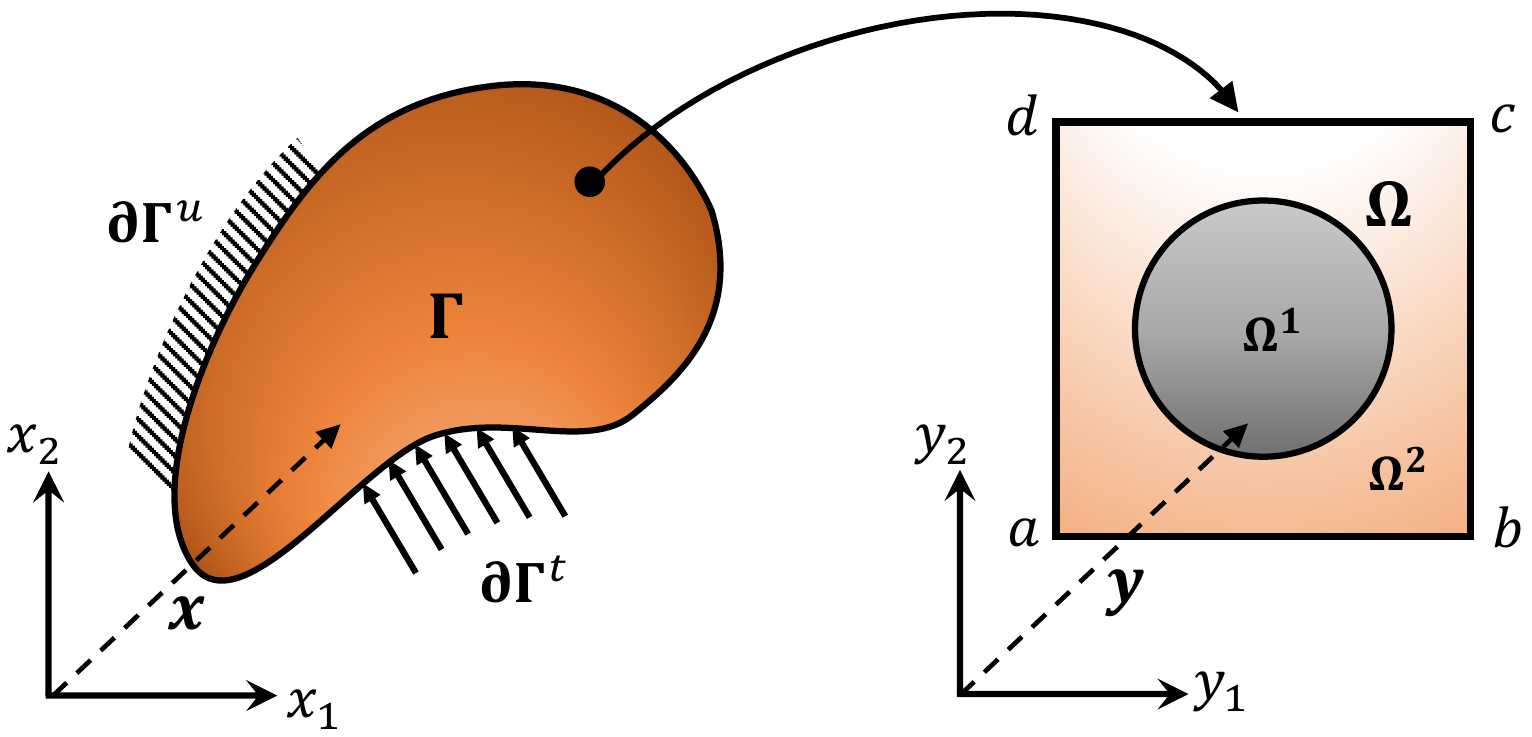}
\caption{Macroscale boundary value problem where $\boldsymbol{x}$ represents a position in the macroscale domain, $\Gamma$. $\partial\Gamma^u$ and $\partial\Gamma^t$ show the Dirichlet boundary and the Neumann boundary, respectively. For any point in the macro-domain, there exists a representative volume element with domain, $\Omega$ (marked $abcd$) with position representation as $\boldsymbol{y}$. $\Omega^1$ and $\Omega^2$ depict the heterogeneous nature of the micro-domain.}
\label{fig:2}
\end{figure}

Multiscale techniques consist of two-way approaches; 1). Homogenization; here, at each point of the macroscale continuum domain, macro-strain $\boldsymbol{\varepsilon}^o$ is known, and macro-stress $\boldsymbol{\sigma}^o$ is calculated by solving a boundary value problem at the micro-scale for the micro-structural stress field. 2). Localization; here, at each point of the macroscale domain, the stress and strain micro-structural fields associated with a given macrostructural equilibrium state are calculated. This is the reverse of the homogenization approach as shown in Fig. \ref{fig:1}. 
A Representative Volume Element (RVE) consisting of the domain $\Omega$ and boundary $\partial\Omega$ is considered corresponding to each macroscale material point, denoted with position vector $\boldsymbol{x}$ of macroscale domain $\Gamma$ which defines the underlying microstructure of the material (refer Fig. \ref{fig:2}). By using averaging theorem, the coupling between macroscopic and microscopic quantities can be established, which gives macroscopic field quantity in terms of volume averaged microscale field variable. When this is applied to the stress field, it can be expressed as
\begin{equation}
	\boldsymbol{\sigma}^o=\langle \boldsymbol{\sigma} \rangle=\dfrac{1}{\vert{\Omega}\vert}\int_\Omega \boldsymbol{\sigma}(\boldsymbol{y})\, d\boldsymbol{y}  \label{eq:1}
\end{equation}
where $\boldsymbol{\sigma}^o$ is the macrostructural stress tensor for a position $\boldsymbol{x}$ in $\Gamma$ and $\boldsymbol{\sigma}(\boldsymbol{y})$ is the microstructural stress field in the RVE when subjected to surface tractions as $\boldsymbol{\sigma}^o \boldsymbol{n}\Big\vert_{\partial \Omega}$. The outward normal to the boundary is denoted by $\boldsymbol{n}\Big\vert_{\partial \Omega}$. 
Similarly, the average strain field is given as 
\begin{equation}
	\boldsymbol{\varepsilon}^o=\langle \boldsymbol{{\varepsilon}} \rangle=\dfrac{1}{\vert{\Omega}\vert}\int_\Omega \boldsymbol{{\varepsilon}}(\boldsymbol{y})\, d\boldsymbol{y} \label{eq:2}
\end{equation}
where $\boldsymbol{\varepsilon}^o$ is the macrostructural strain tensor for a position $\boldsymbol{x}$ in $\Gamma$ and $\boldsymbol{\varepsilon}(\boldsymbol{y})$ is the microstructural strain field in the RVE when subjected to boundary displacements as $\boldsymbol{\varepsilon}^o\boldsymbol{y}\Big\vert_{\partial \Omega}$. The position vector for the boundary is denoted by $\boldsymbol{y}\Big\vert_{\partial \Omega}$.
\nomenclature{$\boldsymbol{\varepsilon}^o$}{Strain at macroscale} 
\nomenclature{$\boldsymbol{\sigma}^o$}{Stress at macroscale} 
\nomenclature{$\boldsymbol{\sigma}, \sigma_{ij}$}{Stress at microscale}
\nomenclature{$\boldsymbol{\varepsilon}, \varepsilon_{ij}$}{Strain at microscale}
\nomenclature{$\Omega$}{Microscale domain} 
\nomenclature{$\partial\Omega$}{Microscale domain boundary} 
\nomenclature{$\boldsymbol{y}$}{Position vector of a point in micoscopic domain}
\nomenclature{$\boldsymbol{x}$}{Position vector of a point in macoscopic domain} 
\nomenclature{$\Gamma$}{Macroscale domain}
\nomenclature{$\boldsymbol{n}$}{A unit normal vector over a surface}
\nomenclature{$\Omega^1,\Omega^2,...$}{Microscopic subdomains}

\section{Microscale Model}\label{sec:3}
At Microscale, which is also the scale of heterogeneities, the two phases, i.e., fiber\textsuperscript{1} and matrix\textsuperscript{2} are considered, and the random distribution is replaced with the regular arrangement of the heterogeneities for the sake of simplicity. Instead of analyzing the full domain with microstructural details, a single  RVE is considered. A two-phase RVE may have either multiple fibers or a single fiber embedded in the matrix. RVE, with a single fiber, represented as domain $\Omega$ containing fiber phase as $\Omega^1$ and matrix phase as $\Omega^2$ subdomains (see Fig. {\ref{fig:2}), also represents a unit cell for a periodic arrangement of heterogeneities which generates the full domain by repeating this unit cell. Material continuity is maintained between $\Omega^1$ and $\Omega^2$. The displacement field for this microscale domain is denoted by $\boldsymbol{u}(\boldsymbol{y})$ with components as \{$u_1,$ $u_2,$ $u_3$\} in $\mathbb{R}^3$-Eucleidian space. Similarly, the strain field at the microscale is denoted by a second-order symmetric tensor $\boldsymbol{\varepsilon}(\boldsymbol{y})$, which is often written as $[\varepsilon_{11},$ $\varepsilon_{22},$ $\varepsilon_{33},$ $\gamma_{12},$ $\gamma_{23},$ $\gamma_{13}]^\text{T}$ in the Voigt form. Stress field $\boldsymbol{\sigma}(\boldsymbol{y})$ can be calculated by solving the equilibrium equation for a RVE;\color{black}
\begin{equation}
	\text{div}\hspace{1mm}{ \boldsymbol{\sigma}(\boldsymbol{y})} = 0 \label{eq:3}
\end{equation}\color{black}
when subjected to boundary conditions as per the following
\begin{equation}
	\boldsymbol{u}(\boldsymbol{y})\Big\vert_{\partial \Omega} = \tilde{\boldsymbol{u}}(\boldsymbol{y})\Big\vert_{\partial \Omega} + \bar{\boldsymbol{u}}(\boldsymbol{y})\Big\vert_{\partial \Omega}  \label{eq:4}
\end{equation}
Here the displacement field mentioned earlier is decomposed into two parts. One part $\bar{\boldsymbol{u}}(\boldsymbol{y})$ accounts average displacement field, whereas $\tilde{\boldsymbol{u}}(\boldsymbol{y})$ accounting for the fluctuations due to the heterogeneous structure of the domain.
The average displacement field $\bar{\boldsymbol{u}}(\boldsymbol{y})$ is obtained from the macroscale deformation field. To enforce the periodicity of this microscale volume element, $\tilde{\boldsymbol{u}}(\boldsymbol{y})$ is prescribed to follow the following relations:
\begin{eqnarray*}
	\tilde{\boldsymbol{u}}(\boldsymbol{y})\bigg\vert_{\boldsymbol{y}\in\partial\Omega^{ab}} &=& \tilde{\boldsymbol{u}}(\boldsymbol{y})\bigg\vert_{\boldsymbol{y}\in\partial\Omega^{dc}}\\
	\tilde{\boldsymbol{u}}(\boldsymbol{y})\bigg\vert_{\boldsymbol{y}\in\partial\Omega^{bc}} &=& \tilde{\boldsymbol{u}}(\boldsymbol{y})\bigg\vert_{\boldsymbol{y}\in\partial\Omega^{ad}}
\end{eqnarray*}
\subsection{Constitutive Behavior of Matrix and Inclusion}
Each phase's constitutive response is characterized by nonlinear functions. Plastic deformations and micro-cracking are considered as the source of the nonlinear constitutive response. Continuum Damage Mechanics (CDM) based framework is implemented for accounting for the softening of material caused by micro cracks. As per the strain equivalence principle, elastic strain ($\boldsymbol{\varepsilon}^e$) and plastic strain ($\boldsymbol{\varepsilon}^p$) in damaged and effective configurations are equal \citep{krajcinovic1987continuum}; and decomposition of the total strain gives
\begin{equation}
	\boldsymbol{\tilde{\varepsilon}}=\boldsymbol{\varepsilon}=\boldsymbol{\tilde{\varepsilon}}^e+\boldsymbol{\tilde{\varepsilon}}^p=\boldsymbol{\varepsilon}^e+\boldsymbol{\varepsilon}^p \label{eq:5}
\end{equation}
Superscript ($\boldsymbol{\tilde{\bullet}}$) represents the effective configuration. 
The effective and nominal stress is given as 
\begin{eqnarray}
	\color{black}\boldsymbol{\tilde{\sigma}}=\mathbb{L}:{\boldsymbol{\varepsilon}^e}    \qquad    & &\textsf{Equivalent Undamaged State}\\
	\color{black}\boldsymbol{\sigma}=\mathbb{L}(\mathbb{D}):{\boldsymbol{\varepsilon}^e}               \qquad    & &\textsf{Damaged State} \label{constitutive}
\end{eqnarray}
where $\mathbb{L}$ is a fourth-order isotropic elasticity tensor and $\mathbb{D}$ is a fourth-order damage tensor. $\mathbb{L}$ can also be represented as 
\begin{equation}
	\mathbb{L}=2G\delta_{ik}\delta_{jl}+\left(K-\dfrac{1}{3}\right)\delta_{ij}\delta_{kl} \label{eq:8}
\end{equation} 
$G$ and $K$ are shear and bulk modulus, respectively and $\delta_{ij}=1 $ if $i=j$ otherwise $\delta_{ij}=0 $. Effective and nominal stress states are expressed as 
\begin{equation}
	\boldsymbol{{\sigma}}=\left[\mathbb{L}(\mathbb{D}):\mathbb{L}^{-1}\right]:\boldsymbol{\tilde{\sigma}} \label{eq:9}
\end{equation}  
Transformation of undamaged constitutive tensor, $\mathbb{L}$ to damaged constitutive tensor $\mathbb{L}(\mathbb{D})$ is given as 
\begin{equation}
	\mathbb{L}(\mathbb{D})=(\mathbb{I}-\mathbb{D}):\mathbb{L} \label{eq:10}
\end{equation}
Here, $\mathbb{I}$ is the fourth-order unit tensor. Using Eq. (\ref{eq:10}), Eq. (\ref{eq:9}) is expressed as 
\begin{equation}
	\boldsymbol{{\sigma}}=\mathbb{M}^{-1}:\boldsymbol{\tilde{\sigma}} \label{eq:11}
\end{equation}
where $\mathbb{M}=(\mathbb{I}-\mathbb{D})^{-1}$ is damage effect tensor. Considering the isotropic damage state in the matrix and inclusion $\mathbb{M}$ can be denoted in terms of a single damage variable, $\omega$;
\begin{equation}
	\mathbb{M}=(1-\omega)^{-1}\mathbb{I} \label{eq:12}
\end{equation}
By representing fourth order tensor $\mathbb{M}$ as symmetric tensor and in Voigt representation as $6\times 6$ matrix $[M]$;
\begin{equation}
	[M]=\begin{bmatrix}
\frac{1}{1-\omega} & 0 & 0 & 0 & 0 & 0\\
0 & \frac{1}{1-\omega} & 0 & 0 & 0 & 0\\
0 & 0 & \frac{1}{1-\omega} & 0 & 0 & 0\\
0 & 0 & 0 & \frac{1}{1-\omega} & 0 & 0\\
0 & 0 & 0 & 0 & \frac{1}{1-\omega} & 0\\
0 & 0 & 0 & 0 & 0 & \frac{1}{1-\omega}
\end{bmatrix} \label{eq:13}
\end{equation}
Incremental form of Eq. (\ref{constitutive}) is 
\begin{equation}
	\boldsymbol{\dot{\sigma}}={\mathbb{L}}(\mathbb{D}):\boldsymbol{\dot{\varepsilon}}^e+\dot{\mathbb{L}}(\mathbb{D}):\boldsymbol{\varepsilon}^e \label{eq:14}
\end{equation} 
or
\begin{equation}
	\boldsymbol{\dot{\sigma}}=[(\mathbb{I}-\mathbb{D}):\mathbb{L}]:\boldsymbol{\dot{\varepsilon}}^e-\dot{\omega}\mathbb{I}:\boldsymbol{\varepsilon}^e \label{eq:15}
\end{equation} 
Eq. (\ref{eq:15}) shows that stress increment depends on the evolution of two state variables (a). plastic strain, $\boldsymbol{\varepsilon}^p$ or elastic strain, $\boldsymbol{\varepsilon}^e$, and (b). damage, $\omega$. Furthermore, the evolution of these state variables can be formulated using a thermodynamic framework which defines Helmholtz free energy potential function as 
\begin{equation}
	\Psi=\Psi(\boldsymbol{\varepsilon},\hspace{1mm} T, \hspace{1mm} r, \hspace{1mm} \boldsymbol{\alpha}, \hspace{1mm} \omega) \label{eq:16}
\end{equation}  
where $T$ denotes temperature. $r$ and $\boldsymbol{\alpha}$ are internal variables referring to isotropic hardening variable and kinematic hardening variable, respectively. In addition to this, for the dissipative process of a continuum domain, Clausius-Duhem inequality imposes the following restriction:
\begin{equation}
	\left(\boldsymbol{\sigma}-\rho\dfrac{\partial\Psi}{\partial\boldsymbol{\varepsilon}^e}\right):\boldsymbol{\dot{\varepsilon}}^e-\rho\left(s+\dfrac{\partial\Psi}{\partial T}\right):\dot{T}+\boldsymbol{\sigma}:\boldsymbol{\dot{\varepsilon}}^p-\rho\dfrac{\partial\Psi}{\partial r}\dot{r}-\rho\dfrac{\partial\Psi}{\partial \boldsymbol{\alpha}}\boldsymbol{\dot{\alpha}}-\rho\dfrac{\partial\Psi}{\partial \omega}\dot{\omega}-\dfrac{\text{grad} T}{T}\cdot\boldsymbol{q}\geq 0 \label{clausius}
\end{equation} 
Here, $\rho$ is the density, and $s$ denotes the entropy of the material. $\boldsymbol{q}$ represents the heat flux. Using the state laws $\boldsymbol{\sigma}=\rho\dfrac{\partial\Psi}{\partial\boldsymbol{\varepsilon}^e}$ and $s=-\dfrac{\partial\Psi}{\partial T}$ and by ignoring the kinematic hardening process, Eq. (\ref{clausius}) takes the following form under isothermal conditions:
\begin{equation}
	\boldsymbol{\sigma}:\boldsymbol{\dot{\varepsilon}}^p-\rho\dfrac{\partial\Psi}{\partial r}\dot{r}-\rho\dfrac{\partial\Psi}{\partial \omega}\dot{\omega}\geq 0 \label{clausius2}
\end{equation} 
Concisely, Eq. (\ref{clausius2}) is written as 
\begin{equation}
	\Phi=\boldsymbol{\sigma}:\boldsymbol{\dot{\varepsilon}}^p-R\dot{r}+Y\dot{\omega}\geq 0 \label{clausius3}
\end{equation}
where $R=\rho\frac{\partial\Psi}{\partial r}$ and $Y=-\rho\frac{\partial\Psi}{\partial \omega}$ are used in writing Eq. (\ref{clausius3}). 

For damaged material, Helmholtz free energy depends on the state of strain, damage and plastic strain. \color{black}The following form of Helmholtz free energy function is considered \citep{lemaitre2012course}\color{black}: 
\begin{equation}
	\rho\Psi=\dfrac{1}{2}\boldsymbol{\varepsilon}^e:\mathbb{L}(\mathbb{D}):\boldsymbol{\varepsilon}^e+\dfrac{1}{2}R_{\infty}r^2 \label{eq:20}
\end{equation}   
$R_{\infty}$ defines another constant for isotropic hardening. Based on this free energy function, the constitutive equations are formulated as below:
\begin{eqnarray}
	\boldsymbol{\sigma}&=\rho\dfrac{\partial\Psi}{\partial\boldsymbol{\varepsilon}^e}&=\mathbb{L}(\mathbb{D}):{\boldsymbol{\varepsilon}}^e   \\
	R&=\rho\dfrac{\partial\Psi}{\partial r}&=R_{\infty}r \label{eq:22}\\
	Y&=-\rho\dfrac{\partial\Psi}{\partial\omega}&=\dfrac{1}{2}{\boldsymbol{\varepsilon}}^e:\mathbb{L}:{\boldsymbol{\varepsilon}}^e				\label{constitutive3}
\end{eqnarray} 
A dissipation potential function, $\mathfrak{F}$, accounting for the growth of three dissipation mechanisms a). damage, b). plastic deformation, and c). isotropic hardening is considered as 
\begin{equation}
	\mathfrak{F}(\boldsymbol{\sigma},\hspace{1mm} R, \hspace{1mm}Y)=\mathfrak{F}^P(\boldsymbol{\sigma}, \hspace{1mm}R, \hspace{1mm}\omega)+\mathfrak{F}^D(Y,\hspace{1mm}\omega) \label{eq:24}
\end{equation}  
The evolution equations for plastic strain, isotropic hardening and damage are given as 
\begin{eqnarray}
		\dot{\boldsymbol{\varepsilon}}^p&=&\dot{\zeta}^P\dfrac{\partial \mathfrak{F}}{\partial\boldsymbol{\sigma}} \label{eq:25} \\
	\dot{r}&=&-\dot{\zeta}^P\dfrac{\partial \mathfrak{F}}{\partial R} \label{eq:26}\\
	\dot{\omega}&=&\dot{\zeta}^D\dfrac{\partial \mathfrak{F}}{\partial Y}			\label{constitutive4}
\end{eqnarray}
where $\dot{\zeta}^P$ and $\dot{\zeta}^D$ are plastic and damage multiplier respectively. Growth of the dissipation potential surface (Eq. (\ref{eq:24})) starts when the following Kuhn-Tucker conditions are satisfied:
\begin{eqnarray}
	\dot{\zeta}^P\geq0,\qquad \mathfrak{F}\leq0, \qquad \dot{\zeta}^P\mathfrak{F}=0 \qquad & &\textsf{Plastic Strain}\\
	\dot{\zeta}^D\geq0,\qquad \mathfrak{F}\leq0, \qquad \dot{\zeta}^D\mathfrak{F}=0 \qquad & &\textsf{Damage}
\end{eqnarray} 
Writing the following form of plastic dissipation surface in the effective stress space as 
\begin{equation}
	\mathfrak{F}^P(\boldsymbol{\sigma},\hspace{1mm} R, \hspace{1mm}\omega) = \tilde{\sigma}_{EQ}-R-\sigma_Y \label{eq:30}
\end{equation}
where $\tilde{\sigma}_{EQ} = \left(\dfrac{3}{2}\boldsymbol{\tilde{\sigma}}^{DEV}:\boldsymbol{\tilde{\sigma}}^{DEV}\right)^{\frac{1}{2}}$ and $\boldsymbol{\tilde{\sigma}}^{DEV}$ denotes the deviatoric component of $\boldsymbol{\tilde{\sigma}}$. Using Eq. (\ref{eq:25}) and (\ref{eq:30}), plastic strain increment is calculated as 
\begin{equation}
	\dot{\boldsymbol{\varepsilon}}^p=\dot{\zeta}^P\dfrac{\partial \mathfrak{F}^P}{\partial\boldsymbol{\sigma}}=\dot{\zeta}^P\dfrac{\partial \tilde{\sigma}_{EQ}}{\partial\boldsymbol{\sigma}}=\dot{\zeta}^P\dfrac{\partial}{\partial \boldsymbol{\sigma}}\left(\dfrac{3}{2}(\mathbb{M}:\boldsymbol{\sigma}^{DEV}):(\mathbb{M}:\boldsymbol{\sigma}^{DEV})\right)^{1/2} \label{eq:31}
\end{equation} 
Finally, from Eq. (\ref{eq:11}) and Eq. (\ref{eq:12}), we can rewrite Eq. (\ref{eq:31}) as 
\begin{equation}
	\dot{\boldsymbol{\varepsilon}}^p=\dfrac{3}{2}\dfrac{\boldsymbol{\tilde{\sigma}}^{DEV}}{\tilde{\sigma}_{EQ}}\left(\dfrac{\dot{\zeta}^P}{1-\omega}\right) \label{eq:32}
\end{equation}
which gives the equivalent plastic strain, $\dot{{\varepsilon}}^p_{EQ}$
\begin{equation}
\dot{{\varepsilon}}^p_{EQ}=\left(\dfrac{2}{3}\dot{\boldsymbol{\varepsilon}}^p:\dot{\boldsymbol{\varepsilon}}^p\right)^{1/2}=\dfrac{\dot{\zeta}^P}{1-\omega} \label{eq:33}
\end{equation}
Similarly, the evolution of the isotropic hardening variable is calculated as
\begin{equation}
\dot{r}=-\dot{\zeta}^P\dfrac{\partial \mathfrak{F}}{\partial R}=\dot{\zeta}^P \label{eq:34}
\end{equation}
Considering the following form of damage dissipation surface in principal strain space as 
\begin{equation}
	\mathfrak{F}^D(Y,\hspace{1mm} \omega) = \dfrac{Y}{(1-\omega)}\left(\dfrac{\kappa_F}{\kappa_F-\kappa_D}\right)\dfrac{1}{\kappa_D} \label{eq:35}
\end{equation}
which gives the damage evolution as 
\begin{equation}
	\dot{\omega}=\dot{\zeta}^D\dfrac{\partial \mathfrak{F}^D}{\partial Y}	\label{eq:36}
\end{equation}
where 
\begin{equation}
	\dot{\zeta}^D=(1-\omega)\dot{\kappa} \boldsymbol{\mathcal{H}} (\kappa-\kappa_D)	 \label{eq:37}
\end{equation}
$\boldsymbol{\mathcal{H}}$ is Heaviside step function and $\kappa$ is maximum value of the principal strain ($\text{max}\{\hat{\epsilon}_1, \hat{\epsilon}_2, \hat{\epsilon}_3\}$). $\kappa_D$ is the threshold value of strain for initiation of damage, and $\kappa_F$ corresponds to strain for the state of complete failure/damage. 
\nomenclature{$\boldsymbol{u}, u_i$}{Displacement field} 
\nomenclature{$\tilde{\boldsymbol{u}}$}{Fluctuating displacement field} 
\nomenclature{$\bar{\boldsymbol{u}}$}{Average displacement field}
\nomenclature{$\boldsymbol{\tilde{\varepsilon}}, \tilde{\varepsilon}_{ij}$}{Effective strain}
\nomenclature{$\tilde{\boldsymbol{\sigma}}, \tilde{\sigma}_{ij}$}{Effective stress} 
\nomenclature{$\bar{\boldsymbol{\sigma}}$}{Average stress} 
\nomenclature{$\bar{\boldsymbol{\varepsilon}}$}{Average strain}
\nomenclature{$\boldsymbol{\varepsilon}^e$}{Elastic strain}
\nomenclature{$\boldsymbol{\varepsilon}^p$}{Plastic strain}
\nomenclature{$\mathbb{L}$}{Elastic contitutive tensor} 
\nomenclature{$\mathbb{L}^{ep}$}{Elastoplatic constitutive tensor} 
\nomenclature{$\mathbb{D}$}{Damage tensor} 
\nomenclature{$\omega$}{Damage variable}
\nomenclature{$[M]$}{Damage effect matrix} 
\nomenclature{$\mathbb{I}$}{Identity tensor}
\nomenclature{$\Psi$}{Helmholtz free energy potential function}
\nomenclature{$T$}{Temperature}
\nomenclature{$r$}{Isotropic hardening variable}
\nomenclature{$\boldsymbol{\alpha}$}{Kinematic hardening variable}
\nomenclature{$\rho$}{Density}
\nomenclature{$s$}{Entropy}
\nomenclature{$G$}{Shear modulus}
\nomenclature{$K$}{Bulk modulus}
\nomenclature{$\delta_{ij}$}{Kronecker delta}
\nomenclature{$\boldsymbol{q}$}{Heat flux}
\nomenclature{$R$}{A variable associated to state variable $r$}
\nomenclature{$Y$}{A variable associated to state variable $\omega$}
\nomenclature{$\mathfrak{F}$}{Dissipation potential function}
\nomenclature{$\mathfrak{F}^P$}{Dissipation potential function accounting plastic deformation}
\nomenclature{$\mathfrak{F}^D$}{Dissipation potential function accounting damage}
\nomenclature{$\dot{\zeta}^P$}{Plastic multiplier}
\nomenclature{$\dot{\zeta}^D$}{Damage multiplier}
\nomenclature{$\boldsymbol{\sigma}^{DEV}$}{Deviatoric stress} 
\nomenclature{${\sigma}_{EQ}$}{Equivalent stress} 
\nomenclature{$\kappa$}{Maximum principal strain} 
\nomenclature{$\kappa_{D}$}{Damage initiation strain} 
\nomenclature{$\kappa_{F}$}{Strain corresponding to complete failure} 
\nomenclature{$\hat{\epsilon}_1, \hat{\epsilon}_2, \hat{\epsilon}_3$}{Principal strain values} 
\nomenclature{$\boldsymbol{\mathcal{H}}$}{Heaviside step function} 

\section{$\mathtt{E}^2$-Transformation Field Analysis based Homogenization}\label{sec:4}
The micromechanical model described in Section \ref{sec:2} can be solved by the Finite Element (FE) technique, where the micromechanical domain is discretized into elements, and the solution is obtained at integration points. Certainly, the accuracy of the solution depends on the refinement of the discretization or the number of integration points in the domain, which directly leads to the requirement of computational power. Reduction of order can be used to reduce the computational cost. There are various Reduced Order Modeling (ROM) methods available in the literature. Transformation Field Analysis (TFA) by \citet{dvorak1992transformation} has been adopted vastly for multiscale analyses, which has been used in various other forms such as NTFA (Nonuniform TFA) by \citet{chaboche2001towards}, Reduced Order NTFA by \citet{fritzen2013reduced}, PWUTFA (Piecewise Uniform TFA) by \citet{sacco2009nonlinear}, \citet{marfia2018multiscale} and \citet{gopinath2018common}, PWUHS (Piecewise Uniform Hashin-Shtrikman Homogenization) by \citet{castrogiovanni2021tfa}, etc. Transformation field analysis uses the concept of transformation strain or eigenstrain, a stress-free strain. Eigenstrain can be categorized either as "physical", such as residual strain, thermal strain, plastic strain, etc., or "fictitious/equivalent" such as damage-equivalent strain.

 Eigenstrain denoted by $\boldsymbol{\mu}(\boldsymbol{y})$ is related to corresponding eigenstress field $\boldsymbol{\lambda}(\boldsymbol{y})$. The constitutive relation in the presence of this transformation field can be written as,
\begin{eqnarray}
	\boldsymbol{\sigma}(\boldsymbol{y})&=&\mathbb{L}:\boldsymbol{\varepsilon}(\boldsymbol{y})+\boldsymbol{\lambda}(\boldsymbol{y}) \label{eq38}\\
	\boldsymbol{\lambda}(\boldsymbol{y}) &=&-\mathbb{L}:\boldsymbol{\mu} (\boldsymbol{y})\\
	\boldsymbol{\sigma}(\boldsymbol{y})&=&\mathbb{L}:\left(\boldsymbol{\varepsilon}(\boldsymbol{y})-\boldsymbol{\mu}(\boldsymbol{y})\right) \label{eq:40}
\end{eqnarray}  
\begin{figure}[H]
\centering
\includegraphics[width=0.75\textwidth]{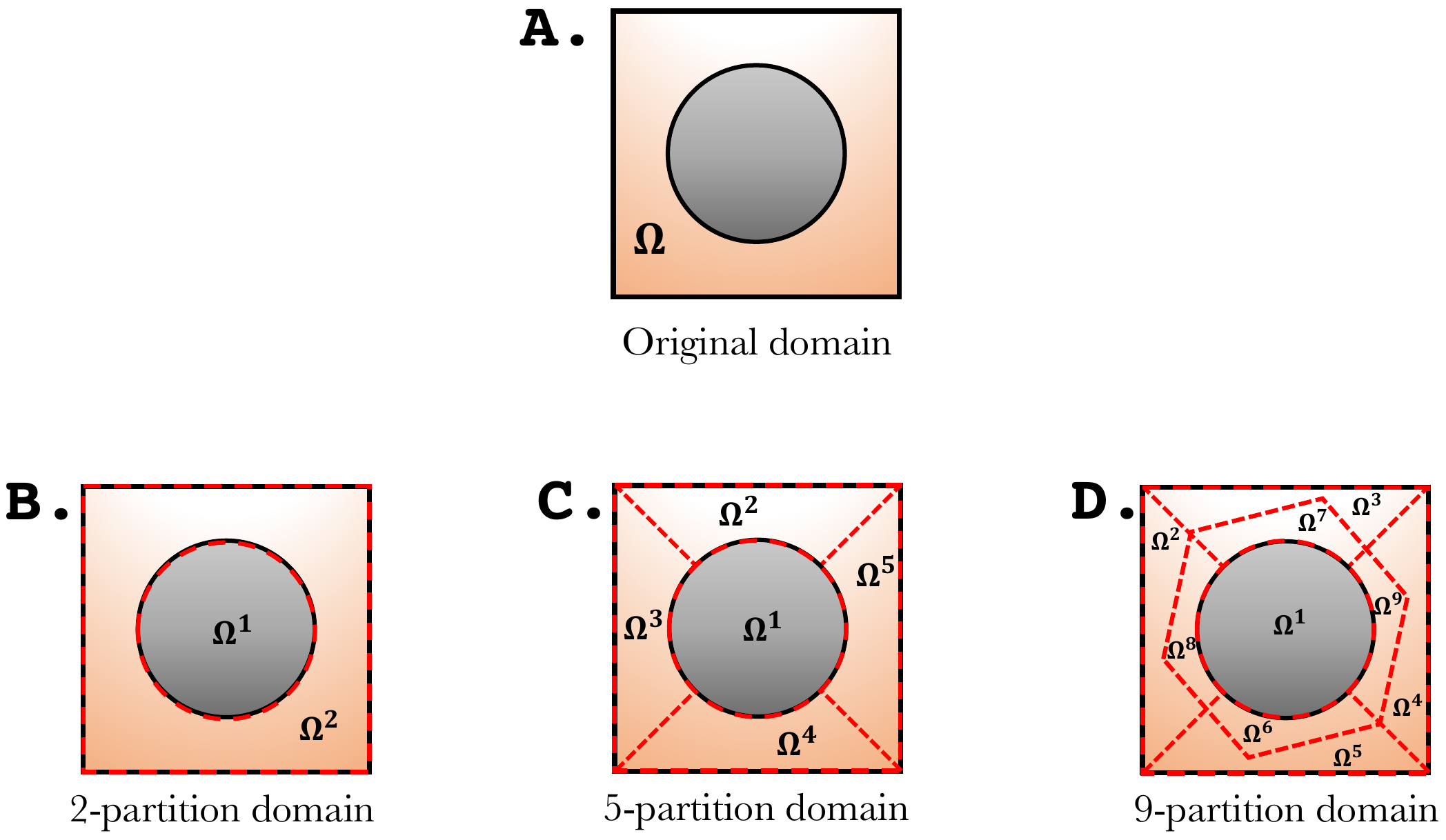}
\caption{Order reduction approach by performing the partitioning and creating the subdomains from $\bf{A}$. Original domain, variety of partitioned domains $\bf{B}$. 2-partition domain, $\bf{C}$. 5-partition domain and $\bf{D}$. 9-partition domain. In every variation, the phase boundaries are regarded merely to avoid two different phases being part of a single subdomain.}
\label{fig:3}
\end{figure}
Further, for \color{black}reduction of the order of the model, \color{black}the given domain is partitioned into $M$ number of subdomains; 
\begin{equation}
	\Omega=\bigcup_{i=1}^{M}\Omega^{i} \label{eq:41}
\end{equation}
with each subdomain referring to a volume fraction $v_f^i$. 
\subsection{Influence Tensors}
The macroscale strain at a point, $\boldsymbol{\varepsilon}^o$ can be transferred to RVE, which represents the material state, by prescribing the displacement as $\boldsymbol{u}(\boldsymbol{y})=\boldsymbol{\varepsilon}^o \boldsymbol{y}$ at the surface boundary. This applied displacement generates a strain field, $\boldsymbol{\varepsilon}(\boldsymbol{y})$ in the RVE, which can be expressed as  
\begin{equation}
	\boldsymbol{\varepsilon}(\boldsymbol{y})=\sum_{i=1}^M N^{i}(\boldsymbol{y})\bar{\boldsymbol{\varepsilon}}^{i} \label{eq:reduced_order}
\end{equation}
The shape function, $N^{i}(\boldsymbol{y})$ satisfies the partition of unity property i.e. $\sum_{i=1}^M N^{i}(\boldsymbol{y})=1$. Overall, the order of the RVE system reduces to $M$. The strain value for each partition $\bar{\boldsymbol{\varepsilon}}^i$ is expressed in terms of weighing function, $\phi^i (\boldsymbol{y})$ as
\begin{equation}
	\bar{\boldsymbol{\varepsilon}}^i=\int_{\Omega} \phi^i (\boldsymbol{y})\boldsymbol{\varepsilon} (\boldsymbol{y})\,d\boldsymbol{y} \label{eq:nonlocal}
\end{equation}
$\phi^i (\boldsymbol{y})$ ensures the nonlocal nature of the strain distribution and $\phi^i (\boldsymbol{y})$ should be positive ($\ge$0) with normality as $\displaystyle\int_{\Omega} \phi^i (\boldsymbol{y})d\boldsymbol{y}=1$. Eq. (\ref{eq:reduced_order}) and (\ref{eq:nonlocal}) leads to condition of orthonormality of $\phi^i (\boldsymbol{y})$ with $N^{j}(\boldsymbol{y})$ as 
\begin{equation}
	\int_{\Omega} \phi^i (\boldsymbol{y})N^{j}(\boldsymbol{y})d\boldsymbol{y}=\delta_{ij} \label{eq:44}
\end{equation}
Here $\delta_{ij}$ is the Kronecker delta with maximum indices values equal to the number of partitions. 
\color{black}
The choice of $N^{j}(\boldsymbol{y})$ and $\phi^i (\boldsymbol{y})$ controls both the computational complexity and accuracy. In the proposed formulation, the following form of shape function and weighing function is taken:
\begin{align*}
N^{i}(\boldsymbol{y})=\begin{cases}
    1, & \text{if $\boldsymbol{y}\in \Omega^i$}\\
    0, & \text{if $\boldsymbol{y}\notin \Omega^i$}
  \end{cases}           & \hspace{14mm} \phi^i (\boldsymbol{y})=\dfrac{1}{\vert\Omega^i\vert}N^{i}(\boldsymbol{y})         
\end{align*}
Using this form of $N^{i}(\boldsymbol{y})$ and $\phi^i(\boldsymbol{y})$ leads to the piecewise distribution of strain field for the partitions in the RVE, which is the same as proposed by \citet{sacco2009nonlinear} and $\bar{\boldsymbol{\varepsilon}}^i$ basically yields the volume average strain value corresponding to a partition, $\Omega^i$. 

In TFA, a fourth-order tensor, $\mathbb{E}(\boldsymbol{y})$, which is called an elastic influence tensor, is defined to relate macroscopic strain, $\boldsymbol{\varepsilon}^o$ with microscale distribution of the strain, $\boldsymbol{\varepsilon}(\boldsymbol{y})$ as
\begin{equation}
	\boldsymbol{\varepsilon}(\boldsymbol{y})=\mathbb{E}(\boldsymbol{y}):\boldsymbol{\varepsilon}^o 		\label{eq:45}
\end{equation}
$\boldsymbol{y}$ is the position vector for any point in the RVE domain. Multiplying both the sides by $\phi^i(\boldsymbol{y})$ and integrating it over domain $\Omega$ gives
\begin{equation}
	\int_{\Omega} \phi^i (\boldsymbol{y})\boldsymbol{\varepsilon}(\boldsymbol{y})\,d\boldsymbol{y}=\left(\int_{\Omega} \phi^i(\boldsymbol{y})  \mathbb{E}(\boldsymbol{y})\, d\boldsymbol{y} \right):\boldsymbol{\varepsilon}^o \label{eq:st_dist}
\end{equation}
using Eq. (\ref{eq:nonlocal}), Eq. (\ref{eq:st_dist}) is expressed as 
\begin{equation}
	{\bar{\boldsymbol{\varepsilon}}^i=\overline{\mathbb{E}}^i:\boldsymbol{\varepsilon}^o}  \label{eq:47}
\end{equation}
where 
\begin{equation}
	\overline{\mathbb{E}}^i=\int_{\Omega} \phi^i(\boldsymbol{y}) \mathbb{E}(\boldsymbol{y})\,d\boldsymbol{y}   \label{eq:48}
\end{equation}
But in the presence of eigenstrains, the Eq. (\ref{eq:45}) takes the form as  
\begin{equation}
	\boldsymbol{\varepsilon}(\boldsymbol{y})=\mathbb{E}(\boldsymbol{y}):\boldsymbol{\varepsilon}^o+\int_{\Omega}\mathbb{S}(\boldsymbol{y},\tilde{\boldsymbol{y}}):\boldsymbol{\mu}^o(\tilde{\boldsymbol{y}})\, d\tilde{\boldsymbol{y}}   \label{eq:49}
\end{equation}
$\mathbb{S}(\boldsymbol{y},\tilde{\boldsymbol{y}})$ is eigenstrain influence tensor function that relates the mechanical strain at a position $\boldsymbol{y}$ caused due to the eigenstrain $\boldsymbol{\mu}^o$ at another position $\tilde{\boldsymbol{y}}$. Furthermore, applying the order reduction technique for eigenstrain gives:
\begin{equation}
	\boldsymbol{\mu}^o(\tilde{\boldsymbol{y}})=\sum_{i=1}^M N^{i}(\tilde{\boldsymbol{y}})\bar{\boldsymbol{\mu}}^{i}      \label{eq:50}
\end{equation}
Similar to Eq. (\ref{eq:nonlocal}), eigenstrain for each partition $\bar{\boldsymbol{\mu}}^i$ is expressed in terms of weighing function, $\phi^i (\boldsymbol{y})$ as
\begin{equation}
	\bar{\boldsymbol{\mu}}^i=\int_{\Omega} \phi^i (\boldsymbol{y})\boldsymbol{\mu} ^o(\boldsymbol{y})\,d\boldsymbol{y}    \label{eq:51}
\end{equation}
Finally, the Eq. (\ref{eq:49}) is expressed as 
\begin{equation}
	\boldsymbol{\varepsilon}(\boldsymbol{y})=\mathbb{E}(\boldsymbol{y}):\boldsymbol{\varepsilon}^o+\sum_{j=1}^M\left(\int_{\Omega} N^{j}(\tilde{\boldsymbol{y}})\mathbb{S}(\boldsymbol{y},\tilde{\boldsymbol{y}}) \, d\tilde{\boldsymbol{y}}\right):\bar{\boldsymbol{\mu}}^{j}  \label{eq:52}
\end{equation}
replacing $\displaystyle\int_\Omega N^{j}(\tilde{\boldsymbol{y}})\mathbb{S}(\boldsymbol{y},\tilde{\boldsymbol{y}}) \, d\tilde{\boldsymbol{y}}= \mathbb{S}^j(\boldsymbol{y})$ and multiplying both the sides by $\phi^i (\boldsymbol{y})$  and integrating it over sub-domain $\Omega^i$ gives
\begin{equation}
	\int_{\Omega} \phi^i (\boldsymbol{y})  \boldsymbol{\varepsilon} (\boldsymbol{y})\,d\boldsymbol{y}=\left(\int_{\Omega} \phi^i (\boldsymbol{y})  \mathbb{E}(\boldsymbol{y})\, d\boldsymbol{y} \right):\boldsymbol{\varepsilon}^o+ \sum_{j=1}^M\left(\int_{\Omega}\phi^i (\boldsymbol{y})\mathbb{S}^j(\boldsymbol{y})\, d\boldsymbol{y} \right):\bar{\boldsymbol{\mu}}^{j}   \label{eq:53}
\end{equation}
Defining eigen influence tensor, $\overline{\mathbb{S}}^{ij}$ as 
\begin{equation}
	\overline{\mathbb{S}}^{ij}=\int_{\Omega}\phi^i (\boldsymbol{y})\mathbb{S}^j(\boldsymbol{y})\, d\boldsymbol{y}   \label{eq:54}
\end{equation}
Using Eq. (\ref{eq:nonlocal}), (\ref{eq:48}) and (\ref{eq:54}), Eq. (\ref{eq:53}) becomes

\begin{equation}
	{\bar{\boldsymbol{\varepsilon}}^i=\overline{\mathbb{E}}^i:\boldsymbol{\varepsilon}^o+ \sum_{j=1}^M\overline{\mathbb{S}}^{ij}:\bar{\boldsymbol{\mu}}^{j}} \label{eq:55}
\end{equation}
\color{black}
\subsection{Microscale Constitutive Response}
Further, the constitutive relation, in the presence of eigenstrain, is
\begin{equation}
	\sigma(\boldsymbol{y})=\mathbb{L}(\boldsymbol{y}):\left(\boldsymbol{\varepsilon}(\boldsymbol{y})-\boldsymbol{\mu}(\boldsymbol{y})\right)   \label{eq:56}
\end{equation}
For calculating the stress in a subvolume $i$, multiply both the sides with $ \phi^i (\boldsymbol{y})$ and take the integration over the subvolume;
\begin{equation}
		\int_{\Omega}  \phi^i (\boldsymbol{y}) \sigma(\boldsymbol{y})\,d\boldsymbol{y}=\int_{\Omega}  \phi^i (\boldsymbol{y})  \mathbb{L}(\boldsymbol{y}):\boldsymbol{\varepsilon}(\boldsymbol{y})\, d\boldsymbol{y} - \int_{\Omega} \phi^i (\boldsymbol{y})\mathbb{L}(\boldsymbol{y}):\boldsymbol{\mu}(\boldsymbol{y})\, d\boldsymbol{y} \label{eq:57}
\end{equation}
A special partitioning scheme is adopted where the phase boundaries always coincide with the subvolume boundaries, and the existence of two different phases in a single subvolume is prohibited. Considering that
\begin{equation}
	\mathbb{L}(\boldsymbol{y})=\mathbb{L}_\alpha^i \label{eq:58}
\end{equation}
where $\alpha$ is a number assigned to a phase material. Using Eq. (\ref{eq:nonlocal}), (\ref{eq:51}) and (\ref{eq:58}), Eq. (\ref{eq:57}) can be written as 
\begin{equation}
	\bar{\boldsymbol{\sigma}}^i=\mathbb{L}_{\alpha}^i:\left(\bar{\boldsymbol{\varepsilon}}^i-\bar{\boldsymbol{\mu}}^{i}\right)      \label{eq:59}
\end{equation}
where $\bar{\boldsymbol{\sigma}}^i=\displaystyle\int_{\Omega} \phi^i (\boldsymbol{y})\boldsymbol{\sigma} (\boldsymbol{y})\,d\boldsymbol{y}$. Finally, using Eq. (\ref{eq:55}), stress in any partition $i$ is 
\begin{equation}
	\bar{\boldsymbol{\sigma}}^i=\mathbb{L}_{\alpha}^i:\left(\overline{\mathbb{E}}^i:\boldsymbol{\varepsilon}^o+ \sum_{j=1}^M\overline{\mathbb{S}}^{ij}:\bar{\boldsymbol{\mu}}^{j}-\bar{\boldsymbol{\mu}}^{i}\right)     \label{eq:60}
\end{equation}
\begin{equation}
	{\bar{\boldsymbol{\sigma}}^i=\mathbb{L}_{\alpha}^i:\left[\overline{\mathbb{E}}^i:\boldsymbol{\varepsilon}^o+ \sum_{j=1}^M\left[\left(\overline{\mathbb{S}}^{ij}-\mathbb{I}\delta_{ij}\right):\bar{\boldsymbol{\mu}}^{j}\right]\right]}    \label{eq:61}
\end{equation}
\subsection{Macroscale Constitutive Response} \label{Macro_Constitutive}
As per the average stress theorem, the average stress in the RVE is equal to a constant stress tensor, $\boldsymbol{\sigma}^o$ with equivalent tractions, $\boldsymbol{t}^o\big\vert_{\partial\Omega}=\boldsymbol{\sigma}^o\boldsymbol{n}\big\vert_{\partial\Omega}$ prescribed over the domain boundary. Average stress in the RVE is 
\begin{equation}
	\bar{\boldsymbol{\sigma}}=\dfrac{1}{\vert \Omega\vert}\int_\Omega\mathbb{L}(\boldsymbol{y}):\left(\boldsymbol{\varepsilon}(\boldsymbol{y})-\boldsymbol{\mu}(\boldsymbol{y})\right)\,d\boldsymbol{y}     \label{eq:62}
\end{equation}
substituting $\boldsymbol{\varepsilon}(\boldsymbol{y})$ from Eq. (\ref{eq:52}), Eq. (\ref{eq:62}) becomes:
\begin{equation}
	\bar{\boldsymbol{\sigma}}=\dfrac{1}{\vert \Omega\vert}\int_\Omega \mathbb{L}(\boldsymbol{y}):\left(\mathbb{E}(\boldsymbol{y}):\boldsymbol{\varepsilon}^o+\sum_{i=1}^M\left(\int_\Omega N^{i}(\tilde{\boldsymbol{y}})\mathbb{S}(\boldsymbol{y},\tilde{\boldsymbol{y}}) \, d\tilde{\boldsymbol{y}}\right):\bar{\boldsymbol{\mu}}^{i}-\boldsymbol{\mu}(\boldsymbol{y})\right)\,d\boldsymbol{y}    \label{eq:63}
\end{equation}
Furthermore, using Eq. (\ref{eq:50}), Eq. (\ref{eq:63}) is expressed as 
\begin{equation}
	\bar{\boldsymbol{\sigma}}=\dfrac{1}{\vert \Omega\vert}\int_\Omega \mathbb{L}(\boldsymbol{y}):\left(\mathbb{E}(\boldsymbol{y}):\boldsymbol{\varepsilon}^o+\sum_{i=1}^M\left(\int_\Omega N^{i}(\tilde{\boldsymbol{y}})\mathbb{S}(\boldsymbol{y},\tilde{\boldsymbol{y}}) \, d\tilde{\boldsymbol{y}}\right):\bar{\boldsymbol{\mu}}^{i}-\sum_{i=1}^M\delta(\boldsymbol{y}-\tilde{\boldsymbol{y}}) N^i(\tilde{\boldsymbol{y}})\bar{\boldsymbol{\mu}}^{i}\right)\,d\boldsymbol{y} \label{eq:64}
\end{equation} 
replacing $N^i(\tilde{\boldsymbol{y}})\delta(\boldsymbol{y}-\tilde{\boldsymbol{y}}) = \mathbb{I}^i(\boldsymbol{y})$, which is basically $\mathbb{I}^{i}(\boldsymbol{y})=\begin{cases}
    \mathbb{I}, & \text{if $\boldsymbol{y}\in \Omega^i$}\\
    0, & \text{if $\boldsymbol{y}\notin \Omega^i$}
  \end{cases}$. Finally it results: 
\begin{equation}
	\bar{\boldsymbol{\sigma}}=\left(\dfrac{1}{\vert \Omega\vert}\int_\Omega \mathbb{L}(\boldsymbol{y}):\mathbb{E}(\boldsymbol{y})\,d\boldsymbol{y}\right):\boldsymbol{\varepsilon}^o+\sum_{i=1}^M\left(\dfrac{1}{\vert \Omega\vert}\int_\Omega \mathbb{L}(\boldsymbol{y}):(\mathbb{S}^i(\boldsymbol{y}) -\mathbb{I}^i(\boldsymbol{y}))\,d\boldsymbol{y}\right):\bar{\boldsymbol{\mu}}^{i} \label{eq:65}
\end{equation} 
which eventually becomes: 
\begin{equation}
	{\bar{\boldsymbol{\sigma}}=\overline{\mathbb{L}}:\boldsymbol{\varepsilon}^o+\sum_{i=1}^M\overline{\mathbb{M}}^i:\bar{\boldsymbol{\mu}}^{i}} \label{eq:66}
\end{equation} 
where $\overline{\mathbb{L}}=\left(\dfrac{1}{\vert \Omega \vert}\displaystyle\int_\Omega \mathbb{L}(\boldsymbol{y}):\mathbb{E}(\boldsymbol{y})\,d\boldsymbol{y}\right)$ and $\overline{\mathbb{M}}^i=\left(\dfrac{1}{\vert {\Omega}\vert}\displaystyle\int_\Omega \mathbb{L}(\boldsymbol{y}):(\mathbb{S}^i(\boldsymbol{y}) -\mathbb{I}^i(\boldsymbol{y}))\,d\boldsymbol{y}\right)$.
\subsection{Relation between Influence Tensors}
\subsubsection{Calculation of $\overline{\mathbb{M}}^i$}
Rewriting Eq. (\ref{eq:40}) for macroscale domain as 
\begin{equation}
	\bar{\boldsymbol{\sigma}}=\overline{\mathbb{L}}:\boldsymbol{\varepsilon}^o-\overline{\mathbb{L}}:\bar{\boldsymbol{\mu}} \label{eq:67}
\end{equation}
Comparing Eq. (\ref{eq:66}) and Eq. (\ref{eq:67}) gives
\begin{equation}
 \sum_{i=1}^M\overline{\mathbb{M}}^i:\bar{\boldsymbol{\mu}}^{i} = -\overline{\mathbb{L}}:\bar{\boldsymbol{\mu}}     \label{eq:68}
\end{equation}
Using Levin's formula \citep{levin1967} for piecewise distribution of eigenstrains, $\bar{\boldsymbol{\mu}}$ can be written as
\begin{equation}
 \bar{\boldsymbol{\mu}} = \sum_{i=1}^M -v_f^i{\overline{\mathbb{G}}^i}^{\text{T}}:\bar{\boldsymbol{\mu}}^{i}     \label{eq:69}
\end{equation}
where $v_f^i$ denotes the volume fraction of $i$ partition and $\overline{\mathbb{G}}^i$ defines the relation between $\bar{\boldsymbol{\sigma}}^i$ and $\boldsymbol{\sigma}^o$ as 
\begin{equation}
\bar{\boldsymbol{\sigma}}^i=\overline{\mathbb{G}}^i:	\boldsymbol{\sigma}^o \label{eq:70}
\end{equation}
Using Eq. (\ref{eq:47}), ${\overline{\mathbb{G}}^i}^{\text{T}}$ can be deduced as 

\begin{equation}
{\overline{\mathbb{G}}^i}^{\text{T}}=\overline{\mathbb{L}}^{-1}:{\overline{\mathbb{E}}^i}^{\text{T}}:\mathbb{L}_{\alpha}^i	  \label{eq:71}
\end{equation} 
\color{black}
Replacing ${\overline{\mathbb{G}}^i}^{\text{T}}$ from Eq. (\ref{eq:71}) in Eq. ({\ref{eq:69}}) and using Eq. (\ref{eq:68}), finally $\overline{\mathbb{M}}^i$ can be written as 
\begin{equation}
	{\overline{\mathbb{M}}^i=-v_f^i \mathbb{L}_{\alpha}^i:{\overline{\mathbb{E}}^i}}   \label{eq:72a}
\end{equation} 

\subsubsection{Calculation of $\overline{\mathbb{S}}^{ij}$}
As described in section \ref{Macro_Constitutive}, $\overline{\mathbb{M}}^i$ is obtained as:  
\begin{equation}
	\overline{\mathbb{M}}^i=\dfrac{1}{\vert {\Omega}\vert}\int_\Omega \mathbb{L}(\boldsymbol{y}):(\mathbb{S}^i(\boldsymbol{y}) -\mathbb{I}^i(\boldsymbol{y}))\,d\boldsymbol{y}  \label{eq:72}
\end{equation} 
Substituting Eq. (\ref{eq:72a}) and Eq. (\ref{eq:66}) in Eq. (\ref{eq:72}) gives
\begin{equation}
	-v_f^i \int_{\Omega} \phi^i(\boldsymbol{y})\mathbb{L}(\boldsymbol{y}):\mathbb{E}(\boldsymbol{y})\,d\boldsymbol{y}=\dfrac{1}{\vert \Omega\vert}\int_\Omega \mathbb{L}(\boldsymbol{y}):(\mathbb{S}^i(\boldsymbol{y}) -\mathbb{I}^i(\boldsymbol{y}))\,d\boldsymbol{y} \label{eq:73}
\end{equation}
which can be expressed as 
\begin{equation}
	\int_\Omega \mathbb{L}(\boldsymbol{y}):(\mathbb{S}^i(\boldsymbol{y}) -\mathbb{I}^i(\boldsymbol{y})+v_f^i{\vert {\Omega}\vert}\phi^i(\boldsymbol{y}) \mathbb{E}(\boldsymbol{y}))\,d\boldsymbol{y} =0  \label{eq:74}
\end{equation}
The equation holds when it satisfies:
\begin{equation}
	\mathbb{S}^i(\boldsymbol{y}) -\mathbb{I}^i(\boldsymbol{y})+{\vert {\Omega}^i\vert}\phi^i(\boldsymbol{y})\mathbb{E}(\boldsymbol{y}) =0   \label{eq:75}
\end{equation}
where ${\vert {\Omega}^i\vert}=v_f^i {\vert {\Omega}\vert}$ is the volume of $i^{th}$ sub-domain. Next, multiplying Eq. (\ref{eq:75}) by $\phi^j(\boldsymbol{y})$ and integrating it over domain $\Omega$ gives
\begin{equation}
\overline{\mathbb{S}}^{ij} = \int_{\Omega}\phi^j(\boldsymbol{y}) \mathbb{I}^i(\boldsymbol{y})\,d\boldsymbol{y}-\int_{\Omega} N^{i}(\boldsymbol{y})\phi^j(\boldsymbol{y})\mathbb{E}(\boldsymbol{y})\,d\boldsymbol{y}   \label{eq:76}
\end{equation}
The term $\int_{\Omega} N^{i}(\boldsymbol{y})\phi^j(\boldsymbol{y})\mathbb{E}(\boldsymbol{y})\,d\boldsymbol{y}$ can be approximated as $v_f^i \overline{\mathbb{E}}^j$ which, finally, gives Eq. (\ref{eq:76}): 
\begin{equation}
	{\overline{\mathbb{S}}^{ij} = \int_{\Omega}\phi^j(\boldsymbol{y}) \mathbb{I}^i(\boldsymbol{y})\,d\boldsymbol{y}-v_f^i \overline{\mathbb{E}}^j}   \label{eq:76a}
\end{equation}
\subsection{Discussion on $\mathtt{E}^2$-TFA Approach}
The two essential functions elastic influence tensor, $\mathbb{E}(\boldsymbol{y})$, and eigen influence tensor, $\mathbb{S}^i(\boldsymbol{y})$, are utilised to determine the other preprocessing and homogenization tensors. Eq. (\ref{eq:48}) demonstrates that using the reduction of order technique, it is possible to derive the averaged elastic influence tensor, $\overline{\mathbb{E}}^i$, for a subdomain using $\mathbb{E}(\boldsymbol{y})$. Similarly, by using Eq. (\ref{eq:76a}), averaged elastic influence tensor $\overline{\mathbb{E}}^i$ may be used to get averaged eigen influence tensor $\overline{\mathbb{S}}^{ij}$. According to the description of the localization method in Section 2, $\overline{\mathbb{E}}^j$ and $\overline{\mathbb{S}}^{ij}$ determine the microscale elastic and eigenstrain from the macroscale strain. 
 
As outlined in Eq. (\ref{eq:66}), from given phase material data, $\mathbb{L}(\boldsymbol{y})$ and $\mathbb{E}(\boldsymbol{y})$, homogenized constitutive elastic function, $\overline{\mathbb{L}}$, can be calculated. Additionally, the homogenized constitutive eigenfunction, $\overline{\mathbb{M}}^i$, for a subdomain is represented by Eq. (\ref{eq:72a}) in terms of the homogenized elastic constitutive function, $\overline{\mathbb{L}}$ and the corresponding volume fraction $v_f^i$. \color{black} It can be checked that Eq. (\ref{eq:72a}) and Eq. (\ref{eq:76a}) satisfy strain averaging theorem (Eq. \ref{eq:2}) in two cases: 1) when there is no eigenstrain present in the domain/sub-domains, i.e. $\bar{\boldsymbol{\mu}}^{i} =0$, additionally, 2) when the partition/sub-domain becomes fully damaged, and the eigenstrain is the same as the applied strain, i.e. $\bar{\boldsymbol{\mu}}^{i} = \bar{\boldsymbol{\varepsilon}}^{i} $. It is anticipated that the approximation of $\int_{\Omega} N^{i}(\boldsymbol{y})\phi^j(\boldsymbol{y})\mathbb{E}(\boldsymbol{y})\,d\boldsymbol{y} = v_f^j \overline{\mathbb{E}}^i$ may lead to some inaccuracies when the partition is in partially damaged state. For a partially damaged state of a subvolume, this can be sought by using the eigenstrain as $\bar{\boldsymbol{\mu}}^{i} = \mathbb{T}: \bar{\boldsymbol{\varepsilon}}^{i} $, where $\mathbb{T}$ is a transformation tensor which relates the eigenstrain with physical strain. In the present formulation, this transformation tensor, $\mathbb{T}$ is considered as identity tensor, $\mathbb{I}$ for brevity.  

\color{black}
\nomenclature{$\boldsymbol{\mu}$}{Eigenstrain} 
\nomenclature{$\bar{\boldsymbol{\mu}}$}{Weighted average eigenstrain} 
\nomenclature{$\boldsymbol{\lambda}$}{Eigenstress} 
\nomenclature{$v_f$}{Volume fraction}
\nomenclature{$N$}{Shape function}
\nomenclature{$\phi$}{Weighing function}
\nomenclature{$\mathbb{E}$}{Elastic influence tensor} 
\nomenclature{$\overline{\mathbb{E}}$}{Weighted average elastic influence tensor}
\nomenclature{$\mathbb{S}$}{Eigenstrain influence tensor}
\nomenclature{$\overline{\mathbb{S}}$}{Eigen influence tensor or weighted average eigenstrain tensor}
\nomenclature{$\boldsymbol{t}^o$}{Equivalent traction vector corresponding to $\boldsymbol{\sigma}^o$} 
\nomenclature{$\overline{\mathbb{M}}$}{Homogenized eigen constitutive tensor} 
\nomenclature{$\overline{\mathbb{L}}$}{Homogenized elastic constitutive tensor} 
\nomenclature{$\overline{\mathbb{G}}$}{Stress concentration tensor} 
\nomenclature{$\mathbb{T}$}{Transformation tensor which relates the eigenstrain with physical strain} 
\nomenclature{$\boldsymbol{R}$}{Residual vector in Newton-Raphson method} 
\nomenclature{$\boldsymbol{U}$}{Unknown vector in Newton-Raphson method} 

\section{Numerical Procedure}\label{Sec_NumericalP}\label{sec:5}
The numerical procedure for this formulation consists of two main steps. The first step is called the preprocessing stage, where the influence and other concentration tensors are calculated along with the homogenised property tensors. In the second step, macroscale stress evaluation is performed by solving the nonlinear equations using the Newton-Raphson algorithm. 
\subsection{Preprocessing Stage: Microscale Calculations}
Preprocessing stage starts with the characterisation of RVE by defining its morphology corresponding to a fixed volume fraction of heterogeneities. The next step is the discretisation of the RVE domain, which is a two-step process. The first step is the a-priori selection of the number of partitions and designation of these sub-domains with an identifier. The RVE subdomains are discretised into elements for FE analyses, and material properties, in terms of $\mathbb{L}(\boldsymbol{y})$, are assigned in the next step. A single-fiber RVE with \color{black}2-partition \color{black} domain has been utilised throughout this manuscript. Furthermore, elastic influence tensor $\mathbb{E}(\boldsymbol{y})$ or $\overline{\mathbb{E}}^i$ are calculated by performing FE analyses corresponding to six sets of macroscale strain components as per Eq. (\ref{eq:45}). Eigenstrain influence tensors, $\overline{\mathbb{S}}^{ij}$ are computed using the elastic influence tensors as given in Eq. (\ref{eq:76}). The homogenised properties of RVE, first, 1). elastic constitutive tensor $\overline{\mathbb{L}}$ is evaluated using $\mathbb{E}(\boldsymbol{y})$ and microconstituents's material properties $\mathbb{L}(\boldsymbol{y})$ and then, 2). eigen constitutive tensor, $ \overline{\mathbb{M}}^i$ from the given volume fraction and calculated constitutive homogenised tensor, $ \overline{\mathbb{L}}$ as per Eq. (\ref{eq:71}). BOX-1 illustrates the procedure followed in the preprocessing stage for solving the problem at the microscale.

\subsection{Solution Stage: Macroscale Calculations}
Eigenstrain caused by plasticity and damage is a history-dependent state variable and requires incremental forms of transformation field and constitutive equations. Now,
Rewriting Eq. (\ref{eq:55}) in incremental form as
\begin{equation}
	\dot{\boldsymbol{\bar{\varepsilon}}^i}-\overline{\mathbb{E}}^i:\dot{\boldsymbol{{\varepsilon}}}^o- \sum_{j=1}^M\overline{\mathbb{S}}^{ij}:\dot{\boldsymbol{\bar{\mu}}^{j}}=0 \label{eq:77}
\end{equation}
\begin{center}
\includegraphics[width=0.85\textwidth]{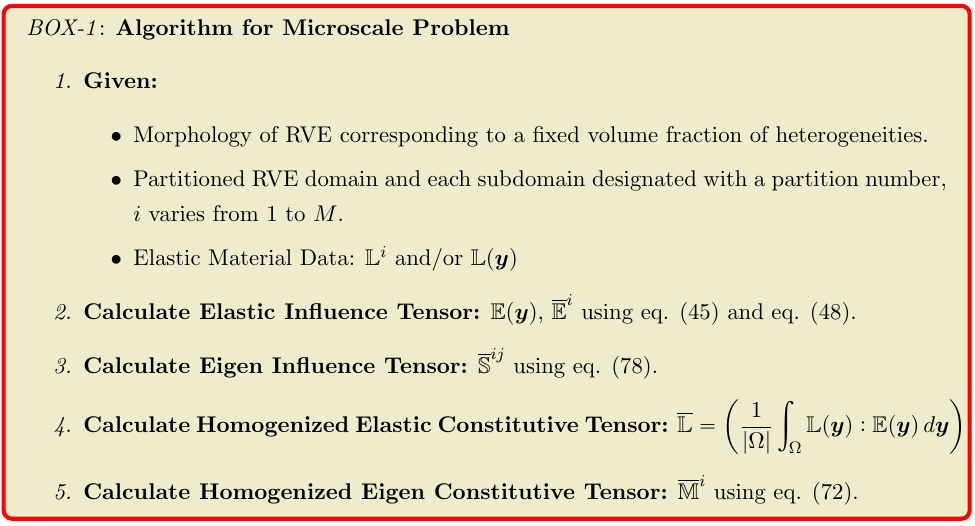}
\end{center}
To solve Eq. (\ref{eq:77}) for $\dot{\boldsymbol{\bar{\varepsilon}}^i}$ and $\dot{\boldsymbol{\bar{\mu}}^j}$, Newton-Raphson method is adopted. Using Eq. (\ref{eq:77}), defining the residual as
\begin{equation}
	\mathbf{R}^i=\dot{\boldsymbol{\bar{\varepsilon}}^i}-\overline{\mathbb{E}}^i:\dot{\boldsymbol{{\varepsilon}}}^o- \sum_{j=1}^M\overline{\mathbb{S}}^{ij}:\dot{\boldsymbol{\bar{\mu}}^{j}} \label{eq:78}
\end{equation}
which implies the unknown $\dot{\boldsymbol{\bar{\varepsilon}}^i}$, can be named as following:
\begin{equation}
	\mathbf{U}^i=\dot{\boldsymbol{\bar{\varepsilon}}^i} \label{eq:79}
\end{equation} 
For solving the RVE problem of order, $M$, the residual vector, $\mathbf{R}^i$ takes the form as  
\begin{equation}
	[\mathbf{R}]=\Big\{\mathbf{R}^1 \quad \mathbf{R}^2 \quad \mathbf{R}^3 \cdots  \mathbf{R}^M\Big\}  \label{eq:80}
\end{equation} 
and identically unknown vector, $\mathbf{U}^i$ is written in a complementary way:
\begin{equation}
	[\mathbf{U}]=\Big\{\mathbf{U}^1 \quad \mathbf{U}^2 \quad \mathbf{U}^3 \cdots  \mathbf{U}^M\Big\} \label{eq:81}
\end{equation}
In Eq. (\ref{eq:79}) and Eq. (\ref{eq:80}), $[\bullet ]$ denotes the matrix notation. Newton-Raphson procedure to solve this nonlinear system of equations for $[\mathbf{R}]=0$ in an iterative manner gives:
\begin{equation}
	[\mathbf{R}]^{[p]}+\left[\dfrac{\partial\mathbf{R}}{\partial\mathbf{U}}\right]\Bigg\vert_{[p]}:[\delta \mathbf{U}]=0  \label{eq:82}
\end{equation}
which is iterated for $p$ times for obtaining the solution. Tangent matrix, $\left[\dfrac{\partial\mathbf{R}}{\partial\mathbf{U}}\right]$ is expanded as 
\begin{equation}
\left[\dfrac{\partial\mathbf{R}}{\partial\mathbf{U}}\right]=
\left[\renewcommand\arraystretch{2.0}
\begin{matrix}
\dfrac{\partial\mathbf{R}^1}{\partial\mathbf{U}^1} & \dfrac{\partial\mathbf{R}^1}{\partial\mathbf{U}^2} & \cdots\\
\vdots & \ddots & \dfrac{\partial\mathbf{R}^{M-1}}{\partial\mathbf{U}^M} \\
\cdots & \cdots & \dfrac{\partial\mathbf{R}^M}{\partial\mathbf{U}^M}
\end{matrix}
\right]  \label{eq:83}
\end{equation}
Each component of Eq. (\ref{eq:83}) is deduced from:
\begin{equation}
	\dfrac{\partial\mathbf{R}^i}{\partial\mathbf{U}^j}=\mathbb{I}\delta_{ij}-\overline{\mathbb{E}}^i:\left(\dfrac{\partial\boldsymbol{\dot{\varepsilon}}^o}{\partial\dot{\boldsymbol{\bar{\varepsilon}}^j}}\right)- \sum_{k=1}^M\overline{\mathbb{S}}^{ik}:\left(\dfrac{\partial\dot{\boldsymbol{\bar{\mu}}^k}}{\partial\dot{\boldsymbol{\bar{\varepsilon}}^j}}\right) \label{eq:84}
\end{equation}
Using Eq. (\ref{eq:84}), Eq. (\ref{eq:83}) becomes:
\begin{equation}
\left[\dfrac{\partial\mathbf{R}}{\partial\mathbf{U}}\right]=
\renewcommand\arraystretch{1.75}
\begin{bmatrix}
\mathbb{I}-\overline{\mathbb{E}}^1:\left(\dfrac{\partial\boldsymbol{\dot{\varepsilon}}^o}{\partial\dot{\boldsymbol{\bar{\varepsilon}}^1}}\right)-\displaystyle\sum_{k=1}^M\overline{\mathbb{S}}^{1k}:\left(\dfrac{\partial\dot{\boldsymbol{\bar{\mu}}^k}}{\partial\dot{\boldsymbol{\bar{\varepsilon}}^1}}\right) & \cdots & \cdots & -\overline{\mathbb{E}}^1:\left(\dfrac{\partial\boldsymbol{\dot{\varepsilon}}^o}{\partial\dot{\boldsymbol{\bar{\varepsilon}}^M}}\right)-\displaystyle\sum_{k=1}^M\overline{\mathbb{S}}^{1k}:\left(\dfrac{\partial\dot{\boldsymbol{\bar{\mu}}^k}}{\partial\dot{\boldsymbol{\bar{\varepsilon}}^M}}\right) \\
-\overline{\mathbb{E}}^2:\left(\dfrac{\partial\boldsymbol{\dot{\varepsilon}}^o}{\partial\dot{\boldsymbol{\bar{\varepsilon}}^1}}\right)-\displaystyle\sum_{k=1}^M\overline{\mathbb{S}}^{2k}:\left(\dfrac{\partial\dot{\boldsymbol{\bar{\mu}}^k}}{\partial\dot{\boldsymbol{\bar{\varepsilon}}^1}}\right)  & \cdots & \cdots & -\overline{\mathbb{E}}^2:\left(\dfrac{\partial\boldsymbol{\dot{\varepsilon}}^o}{\partial\dot{\boldsymbol{\bar{\varepsilon}}^M}}\right)-\displaystyle\sum_{k=1}^M\overline{\mathbb{S}}^{2k}:\left(\dfrac{\partial\dot{\boldsymbol{\bar{\mu}}^k}}{\partial\dot{\boldsymbol{\bar{\varepsilon}}^M}}\right)  \\
-\overline{\mathbb{E}}^3:\left(\dfrac{\partial\boldsymbol{\dot{\varepsilon}}^o}{\partial\dot{\boldsymbol{\bar{\varepsilon}}^1}}\right)-\displaystyle\sum_{k=1}^M\overline{\mathbb{S}}^{3k}:\left(\dfrac{\partial\dot{\boldsymbol{\bar{\mu}}^k}}{\partial\dot{\boldsymbol{\bar{\varepsilon}}^1}}\right)  & \cdots & \cdots & \vdots\\
\cdots & \cdots & \cdots & \mathbb{I}-\overline{\mathbb{E}}^M:\left(\dfrac{\partial\boldsymbol{\dot{\varepsilon}}^o}{\partial\dot{\boldsymbol{\bar{\varepsilon}}^M}}\right)-\displaystyle\sum_{k=1}^M\overline{\mathbb{S}}^{Mk}:\left(\dfrac{\partial\dot{\boldsymbol{\bar{\mu}}^k}}{\partial\dot{\boldsymbol{\bar{\varepsilon}}^M}}\right)  
\end{bmatrix}_{M\times M} \label{eq:85}
\end{equation}
Furthermore, Eq. (\ref{eq:85}) can be simplified as 
\begin{equation}
\left[\dfrac{\partial\mathbf{R}}{\partial\mathbf{U}}\right]=
\renewcommand\arraystretch{2.0}
\begin{bmatrix}
\mathbb{I}-\overline{\mathbb{E}}^1:\left(\dfrac{\partial\boldsymbol{\dot{\varepsilon}}^o}{\partial\dot{\boldsymbol{\bar{\varepsilon}}}^1}\right)-\overline{\mathbb{S}}^{11}:\left(\dfrac{\partial\dot{\boldsymbol{\bar{\mu}}^1}}{\partial\dot{\boldsymbol{\bar{\varepsilon}}}^1}\right) & \cdots & \cdots & -\overline{\mathbb{E}}^1:\left(\dfrac{\partial\boldsymbol{\dot{\varepsilon}}^o}{\partial\dot{\boldsymbol{\bar{\varepsilon}}}^M}\right)-\overline{\mathbb{S}}^{1M}:\left(\dfrac{\partial\dot{\boldsymbol{\bar{\mu}}^M}}{\partial\dot{\boldsymbol{\bar{\varepsilon}}}^M}\right) \\
-\overline{\mathbb{E}}^2:\left(\dfrac{\partial\boldsymbol{\dot{\varepsilon}}^o}{\partial\dot{\boldsymbol{\bar{\varepsilon}}}^1}\right)-\overline{\mathbb{S}}^{21}:\left(\dfrac{\partial\dot{\boldsymbol{\bar{\mu}}^1}}{\partial\dot{\boldsymbol{\bar{\varepsilon}}}^1}\right)  & \cdots & \cdots & -\overline{\mathbb{E}}^2:\left(\dfrac{\partial\boldsymbol{\dot{\varepsilon}}^o}{\partial\dot{\boldsymbol{\bar{\varepsilon}}}^M}\right)-\overline{\mathbb{S}}^{2M}:\left(\dfrac{\partial\dot{\boldsymbol{\bar{\mu}}^M}}{\partial\dot{\boldsymbol{\bar{\varepsilon}}}^M}\right)  \\
-\overline{\mathbb{E}}^3:\left(\dfrac{\partial\boldsymbol{\dot{\varepsilon}}^o}{\partial\dot{\boldsymbol{\bar{\varepsilon}}}^1}\right)-\overline{\mathbb{S}}^{31}\left(\dfrac{\partial\dot{\boldsymbol{\bar{\mu}}^1}}{\partial\dot{\boldsymbol{\bar{\varepsilon}}}^1}\right)  & \cdots & \cdots & \vdots\\
\cdots & \cdots & \cdots & \mathbb{I}-\overline{\mathbb{E}}^M:\left(\dfrac{\partial\boldsymbol{\dot{\varepsilon}}^o}{\partial\dot{\boldsymbol{\bar{\varepsilon}}}^M}\right)-\overline{\mathbb{S}}^{MM}:\left(\dfrac{\partial\dot{\boldsymbol{\bar{\mu}}^M}}{\partial\dot{\boldsymbol{\bar{\varepsilon}}}^M}\right)  
\end{bmatrix}_{M\times M}  \label{eq:86}
\end{equation}
\begin{center}
\includegraphics[width=0.85\textwidth]{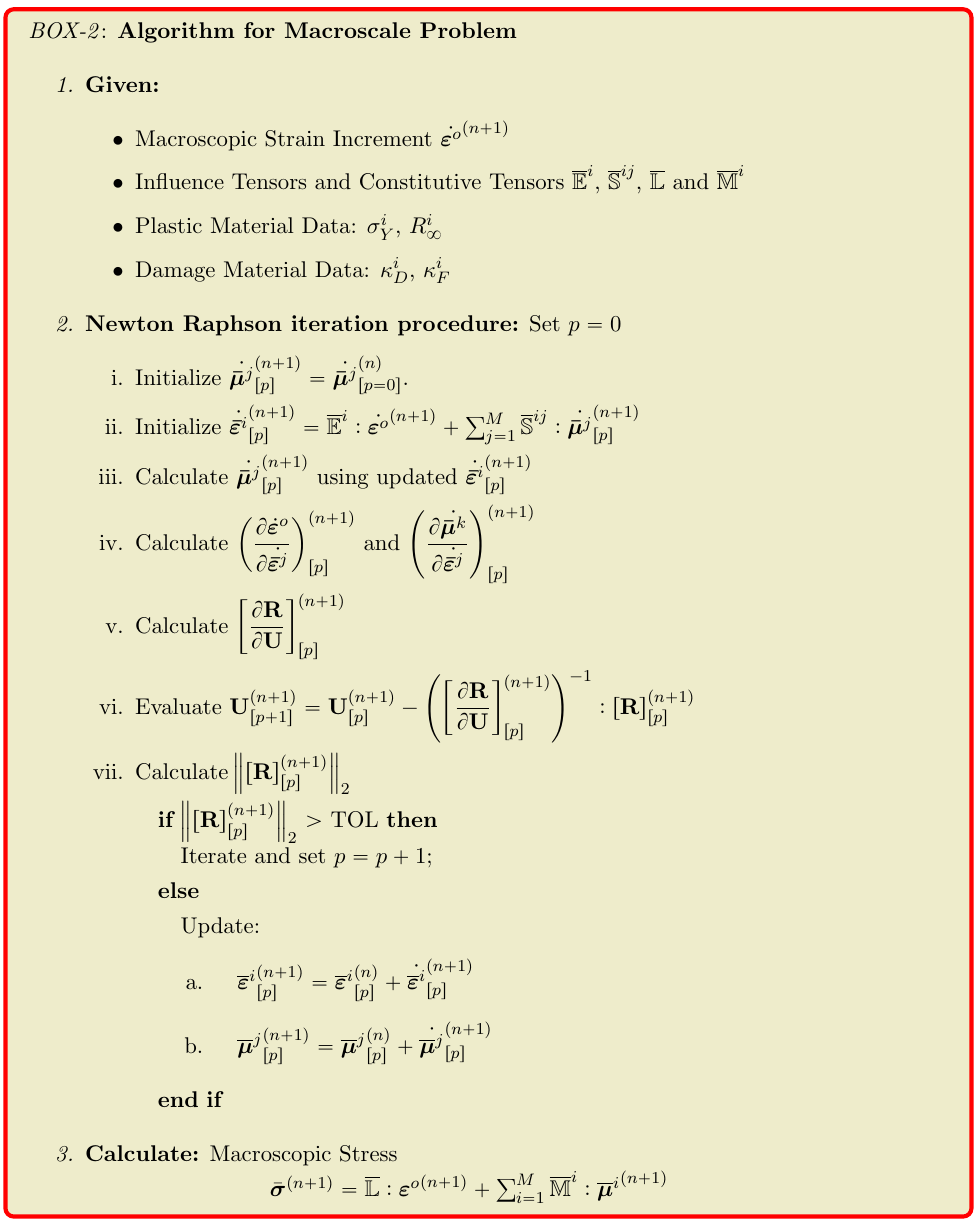}
\end{center}
\textit{BOX-2} shows the algorithm for updating the macroscale stress based on the abovementioned procedure. Eq. (\ref{eq:86}) indicates the need for calculating $\left(\dfrac{\partial\dot{\boldsymbol{\bar{\mu}}^M}}{\partial\dot{\boldsymbol{\bar{\varepsilon}}^M}}\right)$ and $\left(\dfrac{\partial\dot{\boldsymbol{\varepsilon}^o}}{\partial\dot{\boldsymbol{\bar{\varepsilon}}}^M}\right)$ which are obtained as explained in the \ref{appendix1} and \ref{appendix2}.\color{black}

\begin{figure}[H]
\centering
\includegraphics[width=0.85\textwidth]{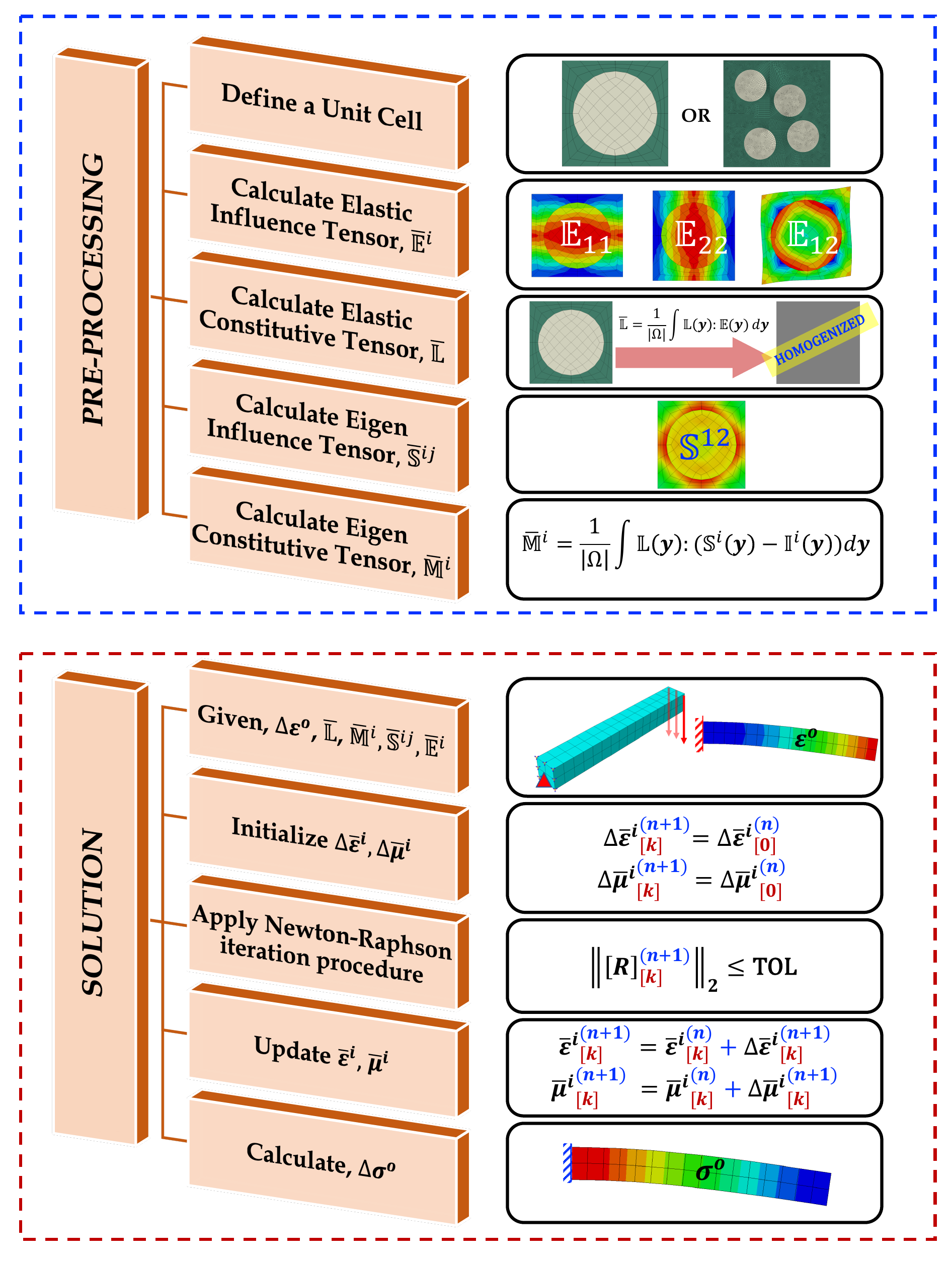}
\caption{Illustration for the implementation of $\mathtt{E}^2$-TFA based homogenisation procedure, which graphically represents two stages of the numerical execution: 1). Preprocessing stage and 2). Macroscale solution stage.}
\label{fig:4}
\end{figure} 
\section{Numerical Implementations}\label{sec:6}
Fig. \ref{fig:4} illustrates the sequential implementation of $\mathtt{E}^2$-TFA based homogenisation procedure, which represents two stages of the numerical execution: 1). Preprocessing and 2). Macroscale solution. The proposed methodology is implemented in the finite element method-based framework. A two-stage verification is performed, i.e., first, the RVE simulations are carried out under monotonic and cyclic loading conditions using $\mathtt{E}^2$-TFA. Later, multi-scale analyses are carried out for different composites under complex loading to validate the code. Macroscale analysis results are compared with experimental data available in the literature. \color{black}Table \ref{table:0} summarises the various numerical studies carried out in this work to demonstrate the ability of the proposed methodology.\color{black} For RVE scale simulations, periodic boundary conditions are imposed in directions other than the applied load direction.
\begin{table}[h]
\captionsetup{width=0.75\textwidth}
    \caption{\color{black}Numerical studies carried out using $\mathtt{E}^2$-TFA}
 \centering
\begin{tabular}{c p{3.75cm} c p{8cm}}
\toprule
\textbf{No.} & \textbf{Name of the study} & \textbf{Domain} & \textbf{Purpose}\\
\midrule
1 & RVE Simulations & 2D & To demonstrate the eigenstrain prediction capability and accurate post-damage response.\\
2 & Plate with a hole & 3D & To check the laminate response under tensile load along the fiber and transverse directions.\\
3 & Crack propagation in double notch plate & 3D & To capture the crack trajectory in a composite under monotonic and cyclic loading. \\
4 & Damage in 10$^\circ$ composite plate & 3D  & To predict the experimentally observed damage maps and stress/strain behaviour.\\
5 & Plate under impact load & 3D & To investigate the $\mathtt{E}^2$-TFA prediction capability under complex loading scenarios such as impact and comparison with classical TFA. \\
\bottomrule
\end{tabular} \label{table:0}
\end{table}
\color{black}
\subsection{Numerical Implementation-1: RVE Simulations}
For verification of the proposed $\mathtt{E}^2$-TFA methodology, RVE simulations are carried out to capture the damage and inelastic behaviour at the microscale. A 2D domain of unit size containing a single fiber is applied. The diameter of the fiber is considered to correspond to the volume fraction of {41\%}\color{black}. The details of matrix and fiber material data are shown in Table \ref{table:1}.      

\begin{table}[h]
\captionsetup{width=0.75\textwidth}
    \caption{Material properties of fiber and matrix used for RVE simulations. The data used is nearly the same as for glass-epoxy composite. Source: \cite{singh2017reduced}}
    
\centering
\begin{tabular}{lcc}
\toprule
\textbf{Material Property} & \textbf{Fiber} & \textbf{Matrix} \\
\midrule
Volume Fraction & 0.41 & 0.59\\
Elastic Modulus [MPa]& 80,000 & 2,670 \\
Poisson's Ratio & 0.3 & 0.3 \\
Yield Strength [MPa] & - & 26 \\
Hardening Modulus [MPa] & - & 500 \\
Damage Initiation Strain [\%] & - & 0.9 \\
Damage Failure Strain [\%] & - & 3.15 \\
\bottomrule
\end{tabular} \label{table:1}
\end{table}
This 2D domain is meshed with plane strain elements. For calculation of mechanical transformation tensor, $\mathbb{\overline{E}}^1$ and $\mathbb{\overline{E}}^2$, linear elastic finite element analyses of RVE subjected to displacement boundary conditions per each component of macroscale strain, $\boldsymbol{\varepsilon}^o$ are performed. The strain components in the RVE elements are evaluated for 3 cases. Each load case is equivalent to the applied unit macroscale strain component along with periodic boundary conditions defined from Eq. (\ref{eq:6-1}) to Eq. (\ref{eq:6-3}). 
\begin{equation}\label{eq:6-1}
\text{CASE}\hspace{2mm}1\qquad
\begin{cases} 
   \tilde{{u}}_1(\boldsymbol{y})\bigg\vert_{\boldsymbol{y}\in\partial\Omega^{bc}} &=  \tilde{{u}}_1(\boldsymbol{y})\bigg\vert_{\boldsymbol{y}\in\partial\Omega^{ad}}+ {\varepsilon}^o_{11}\times \left(y_1\big\vert_{\partial\Omega^{bc}}-y_1\big\vert_{\partial\Omega^{ad}}\right) \\
    \tilde{{u}}_2(\boldsymbol{y})\bigg\vert_{\boldsymbol{y}\in\partial\Omega^{bc}} &=  \tilde{{u}}_2(\boldsymbol{y})\bigg\vert_{\boldsymbol{y}\in\partial\Omega^{ad}} \\
    \tilde{\boldsymbol{u}}(\boldsymbol{y})\bigg\vert_{\boldsymbol{y}\in\partial\Omega^{ab}} &=  \tilde{\boldsymbol{u}}(\boldsymbol{y})\bigg\vert_{\boldsymbol{y}\in\partial\Omega^{dc}}
   \end{cases}
\end{equation}
\begin{equation}\label{eq:6-2}
\text{CASE}\hspace{2mm}2\qquad
\begin{cases} 
    \tilde{\boldsymbol{u}}(\boldsymbol{y})\bigg\vert_{\boldsymbol{y}\in\partial\Omega^{bc}} &=  \tilde{\boldsymbol{u}}(\boldsymbol{y})\bigg\vert_{\boldsymbol{y}\in\partial\Omega^{ad}}
    \vspace{2mm}\\
   \tilde{{u}}_1(\boldsymbol{y})\bigg\vert_{\boldsymbol{y}\in\partial\Omega^{dc}} &=  \tilde{{u}}_1(\boldsymbol{y})\bigg\vert_{\boldsymbol{y}\in\partial\Omega^{ab}} \\
    \tilde{{u}}_2(\boldsymbol{y})\bigg\vert_{\boldsymbol{y}\in\partial\Omega^{dc}} &=  \tilde{{u}}_2(\boldsymbol{y})\bigg\vert_{\boldsymbol{y}\in\partial\Omega^{ab}} + {\varepsilon}^o_{22}\times \left(y_2\big\vert_{\partial\Omega^{dc}}-y_2\big\vert_{\partial\Omega^{ab}}\right)
   \end{cases}
\end{equation}
\begin{equation}\label{eq:6-3}
\text{CASE}\hspace{2mm}3\qquad
\begin{cases} 
   \tilde{{u}}_1(\boldsymbol{y})\bigg\vert_{\boldsymbol{y}\in\partial\Omega^{bc}} &=  \tilde{{u}}_1(\boldsymbol{y})\bigg\vert_{\boldsymbol{y}\in\partial\Omega^{ad}}\\
    \tilde{{u}}_2(\boldsymbol{y})\bigg\vert_{\boldsymbol{y}\in\partial\Omega^{bc}} &=  \tilde{{u}}_2(\boldsymbol{y})\bigg\vert_{\boldsymbol{y}\in\partial\Omega^{ad}} + {\varepsilon}^o_{12}\times \left(y_1\big\vert_{\partial\Omega^{bc}}-y_1\big\vert_{\partial\Omega^{ad}}\right) \\
    \tilde{\boldsymbol{u}}(\boldsymbol{y})\bigg\vert_{\boldsymbol{y}\in\partial\Omega^{ab}} &=  \tilde{\boldsymbol{u}}(\boldsymbol{y})\bigg\vert_{\boldsymbol{y}\in\partial\Omega^{dc}}
   \end{cases}
\end{equation}\color{black}

The results obtained in terms of strain for each case correspond to the elastic influence tensor, $\mathbb{E}(\boldsymbol{y})$ for an element of RVE located at $\boldsymbol{y}$. For each partition (taking 2 partitions), elastic influence tensors, $\mathbb{\overline{E}}^1$ and $\mathbb{\overline{E}}^2$ are calculated using Eq. (\ref{eq:48}). The homogenized property tensor, $\mathbb{\overline{L}}$ is obtained from calculated $\mathbb{E}(\boldsymbol{y})$ and local material data $\mathbb{L}(\boldsymbol{y})$ as defined in Eq. (\ref{eq:65}) and  Eq. (\ref{eq:66}). Furthermore, the effective property tensors for eigenstrains as $\mathbb{\overline{M}}^1$ and $\mathbb{\overline{M}}^2$ are evaluated from Eq. (\ref{eq:72a}) by employing $v^1_f$, $v^2_f$ and $\mathbb{\overline{L}}$. Finally, Eq. (\ref{eq:76a}) gives four eigenstrain influence tensors as $\mathbb{\overline{S}}^{11}$, $\mathbb{\overline{S}}^{12}$, $\mathbb{\overline{S}}^{21}$ and $\mathbb{\overline{S}}^{22}$ using $\mathbb{\overline{E}}^1$ and $\mathbb{\overline{E}}^2$. Fig. \ref{fig:homo_tensor} shows all the calculated tensors using the material data mentioned in Table \ref{table:1}.   
\begin{figure}[H]
\centering
\includegraphics[width=0.75\textwidth]{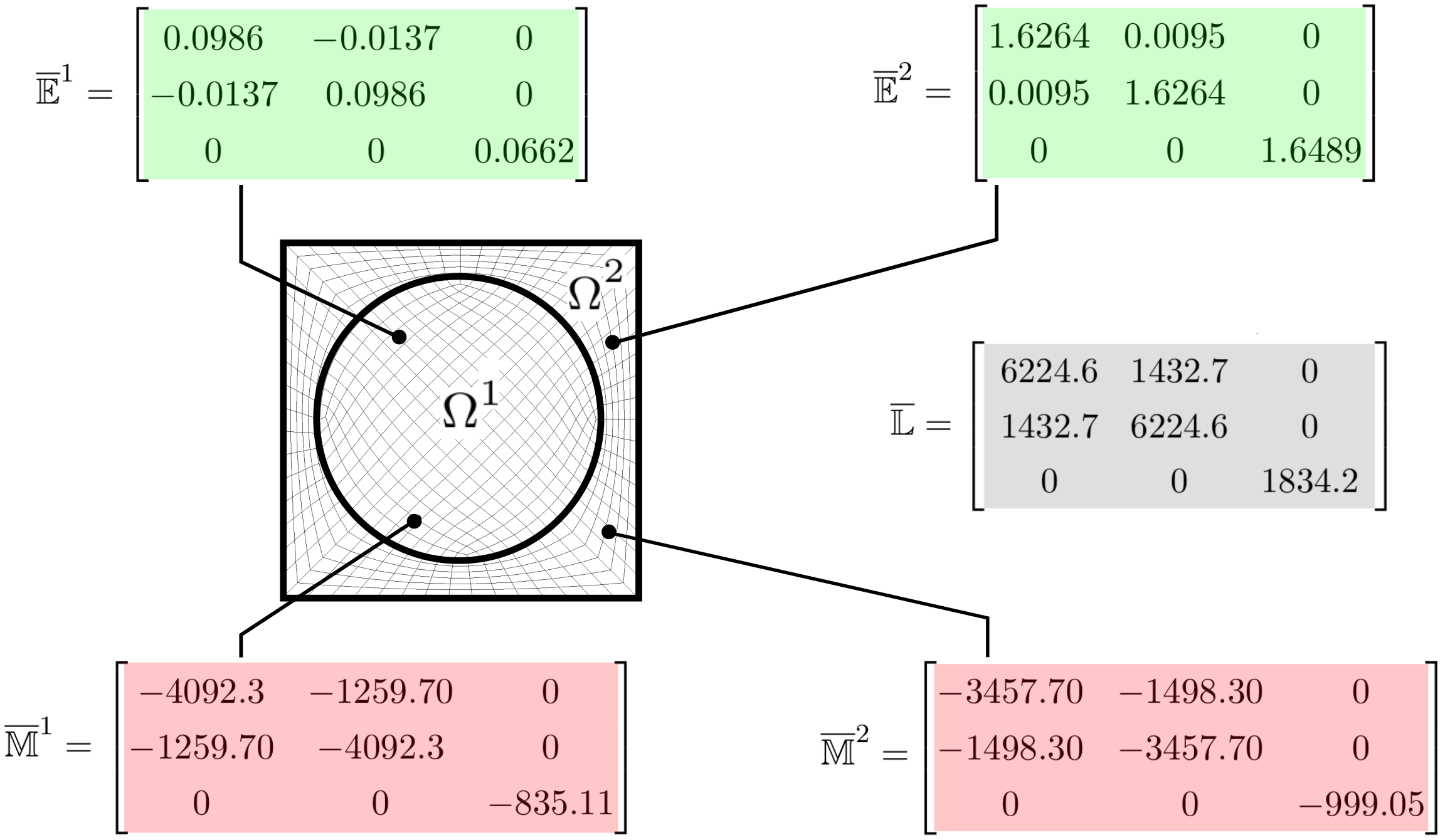}
\caption{Calculated elastic influence tensors ($\mathbb{\overline{E}}^1$ and $\mathbb{\overline{E}}^2$ shown with green background) and homogenized tensors ($\mathbb{\overline{L}}$ shown with a grey background, $\mathbb{\overline{M}}^1$ and $\mathbb{\overline{M}}^2$ shown with pink background) corresponding to two partitioned domain ($\Omega^1$ and $\Omega^2$) using the material data mentioned in Table \ref{table:1}.}
\label{fig:homo_tensor}
\end{figure}
\begin{table}[h]
\captionsetup{width=0.75\textwidth}
    \caption{Three types of models used for studying the homogenized response of RVE. The categorization is based on the constitutive response of the material. }
    \label{tab-b}
\centering
\begin{tabular}{lcc}
\toprule
\textbf{No.} & \textbf{Damage} & \textbf{Plasticity} \\
\midrule
Model-1 & \checkmark & $\bigtimes$\\
Model-2 & $\bigtimes$ & \checkmark\\
Model-3 & \checkmark & \checkmark \\
\bottomrule
\end{tabular}
\end{table}

After the preprocessing stage, a finite-size homogenized domain is checked when subjected to uniaxial tension, as shown in Fig. {\ref{fig:monotonic}A. This homogenous domain is investigated for monotonic (see Fig. {\ref{fig:monotonic}A) and reversed (see Fig.  {\ref{fig:cyclic}}A) loading conditions. The studies are carried out for three types of microconstituents constitutive responses named Model-1, Model-2 and Model-3, as shown in Table {\ref{tab-b}. Model-1 considers damage of microconstituents into account in the absence of inelastic strain. Fig. {\ref{fig:monotonic}B depicts the variation of macroscale stress, stress and damage in the matrix with applied macroscale strain and {\ref{fig:monotonic}C  shows the variation of eigenstrain in the matrix for Model-1. As soon as the damage in the material initiates, eigenstrain in the matrix starts increasing, and upon reaching the state of full damage where the stiffness becomes zero, it starts to rise like the elastic strain. The linear stress-strain law for softening of matrix material is applied using Eq. (\ref{eq:36}) and (\ref{eq:37}}) for the damage evolution formulation, as mentioned in Section \ref{sec:3}}. \color{black}The stress-strain response of RVE is analyzed with results calculated using direct numerical simulations (DNS) for a heterogeneous FE model when subjected to monotonic and reversed loading conditions. One partition per phase TFA results are compared with the FE model, where the matrix domain is meshed with four elements. The TFA was performed with a set of properties for the constituents as mentioned in Table \ref{table:1}, whereas, in FEA, $\kappa_D$ and $\kappa_F$ are taken to equalize the fracture energy (area under the stress-strain diagram) calculated from TFA \citep{singh2023representative}. \color{black}The comparison, as shown in Fig. {\ref{fig:monotonic}} and {\ref{fig:cyclic}}, demonstrates a good match between $\mathtt{E}^2$-TFA preprocessing results and direct numerical simulations.\color{black} Similarly, the variation of these quantities is examined considering the inelasticity of phases in the absence of damage in Model-2, as shown in Fig. {\ref{fig:monotonic}D and {\ref{fig:monotonic}E, which manifests the same variation of eigenstrain as plastic strain in the matrix subdomain. Linear isotropic hardening law is employed for yield surface evolution in the matrix (as per Eq. (\ref{eq:22}) and (\ref{eq:30})) using the parameters mentioned in Table \ref{table:1}. In Model-3, both damage and plasticity of the phases are taken into consideration. Fig. {\ref{fig:monotonic}F and {\ref{fig:monotonic}G demonstrate the capability of the proposed formulation not only to capture macroscale stress but also it gives damage, inelastic and eigenstrain for each subdomain at the RVE level.

The identical homogenized medium is analyzed under reversed uniaxial loading conditions for a single cycle of applied displacement, as shown in Fig. \ref{fig:cyclic}A. The purpose of performing this analysis is to investigate the ability of the proposed formulation to obtain the residual strain and stress in the material. The studies are executed for the same three models with different constitutive behaviour, as in Table \ref{table:1}. For Model-1, the material is loaded till the state of partial damage, and then the loading has been reversed as illustrated in Fig. {\ref{fig:cyclic}}B and {\ref{fig:cyclic}}C, which also demonstrates that, in the absence of inelasticity, the homogenized material and matrix show the full recovery of developed eigenstrain and elastic strain. On the contrary,  residual eigenstrain/plastic strain appears in the material at the end of the cycle for Model-2 (see Fig. {\ref{fig:cyclic}}D and {\ref{fig:cyclic}}E). In Model-3 analysis, eigenstrain includes both the damage-equivalent and plasticity components. As soon as the matrix yielding starts, the eigenstrain begins to grow. The growth of eigenstrain is enhanced with damage. After reaching a certain level of damage, the loading in the material is reversed, which stagnates the plastic strain, whereas the eigenstrain in the matrix decreases with the reversal of the load. 

Overall, these RVE studies exhibit that the suggested $\mathtt{E}^2$-TFA based multiscale framework has the capability to simulate the mechanical response of RVE as an equivalent homogenized material. Although no absolute validation of the results has been conducted, the presented results for RVE are found good qualitatively, and no post-damage stiffness effects are observed. 
\begin{figure}[H]
\centering
\includegraphics[width=0.85\textwidth]{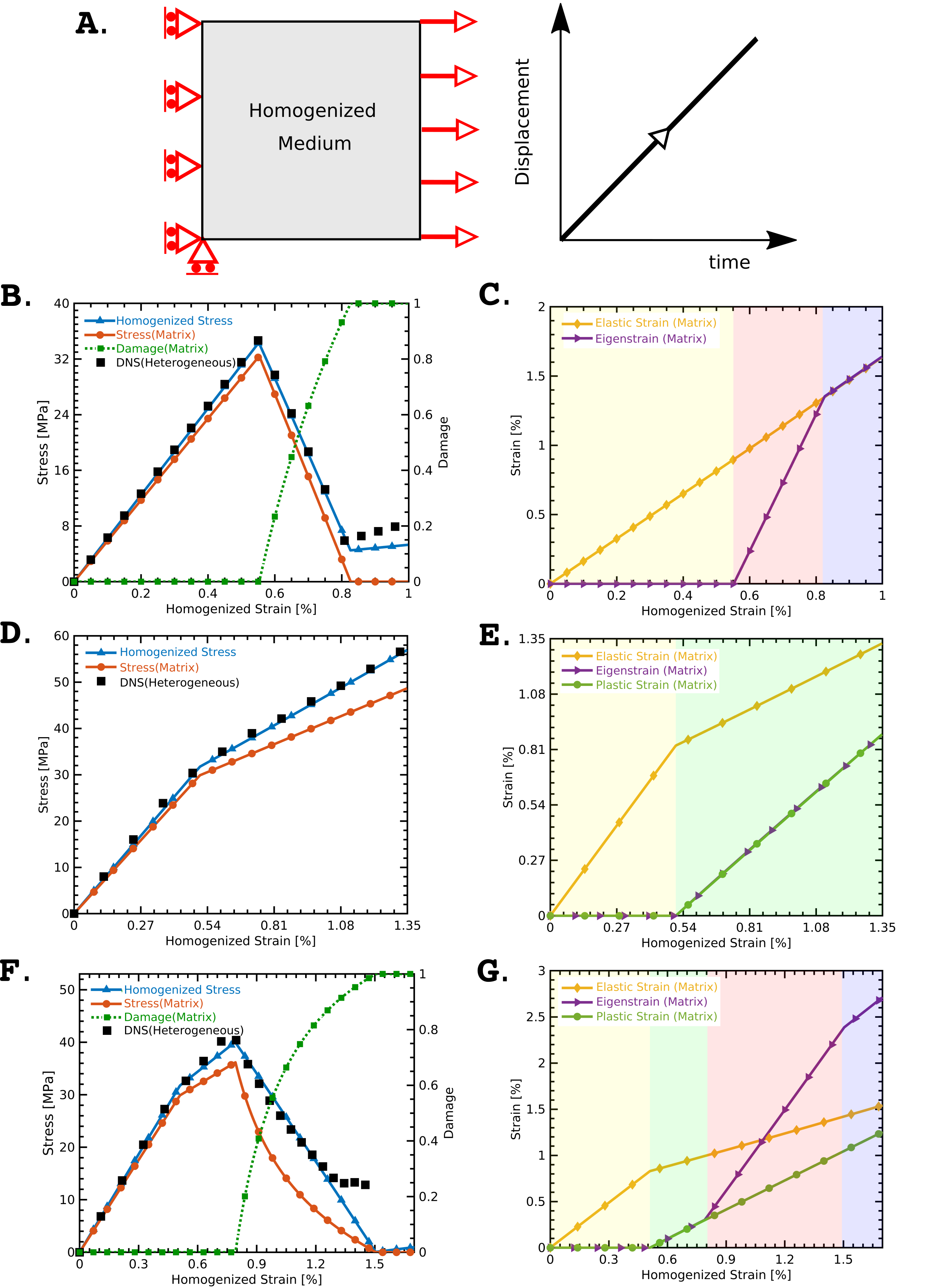}
\caption{Mechanical response of the homogenized medium subjected to monotonic loading whereas subfigures depict A. loading and boundary conditions applied, B. Stress-strain plot for homogenized material and matrix considering damage only, C. Variation of damage equivalent eigenstrain and elastic strain in the matrix (damage only case), D. Stress-strain plot for homogenized material and matrix considering plasticity only, E. Variation of various strains to applied homogenized strain (plasticity only case), F. Stress-strain plot for homogenized material and matrix considering damage and plasticity, G. Variation of various strains to applied homogenized strain (damage + plasticity case)}
\label{fig:monotonic}
\end{figure}
\begin{figure}[H]
\centering
\includegraphics[width=0.85\textwidth]{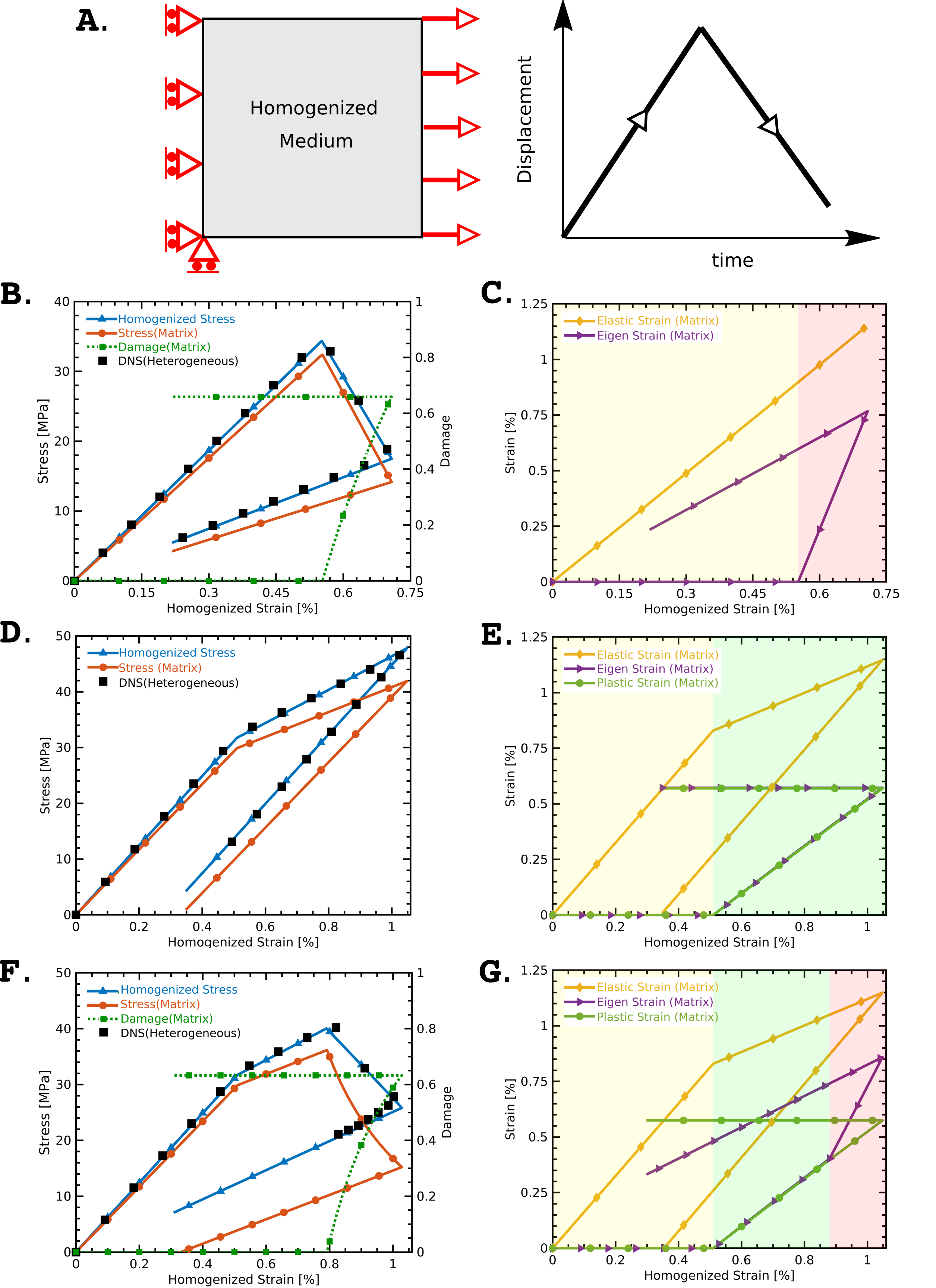}
\caption{Mechanical response of the homogenized medium subjected to reversed loading whereas subfigures depict A. loading and boundary conditions applied, B. Stress-strain plot for homogenized material and matrix considering damage only, C. Variation of damage equivalent eigenstrain and elastic strain in the matrix (damage only case), D. Stress-strain plot for homogenized material and matrix considering plasticity only, E. Variation of various strains to applied homogenized strain (plasticity only case), F. Stress-strain plot for homogenized material and matrix considering damage and plasticity, G. Variation of various strains to applied homogenized strain (damage + plasticity case)}
\label{fig:cyclic}
\end{figure}           
\subsection{Numerical Implementation-2: Plate with a hole problem}
In order to verify the proposed $\mathtt{E}^2$-TFA methodology, a 3-dimensional composite plate with a centre hole is analysed. The geometry and FE model is illustrated in Fig. \ref{fig:pwah1}A and \ref{fig:pwah1}B. The quarter portion of the geometry is modelled with symmetry boundary conditions applied on the faces along the cut-planes. This one-fourth domain is discretised with 4564 eight-noded quadrilateral elements. Unit thickness has been considered in an out-of-plane direction with single-ply configuration. The top boundary is subjected to displacement-controlled loading as shown in Fig. \ref{fig:pwah1}B. 
\begin{figure}[H]
\centering
\includegraphics[width=0.75\textwidth]{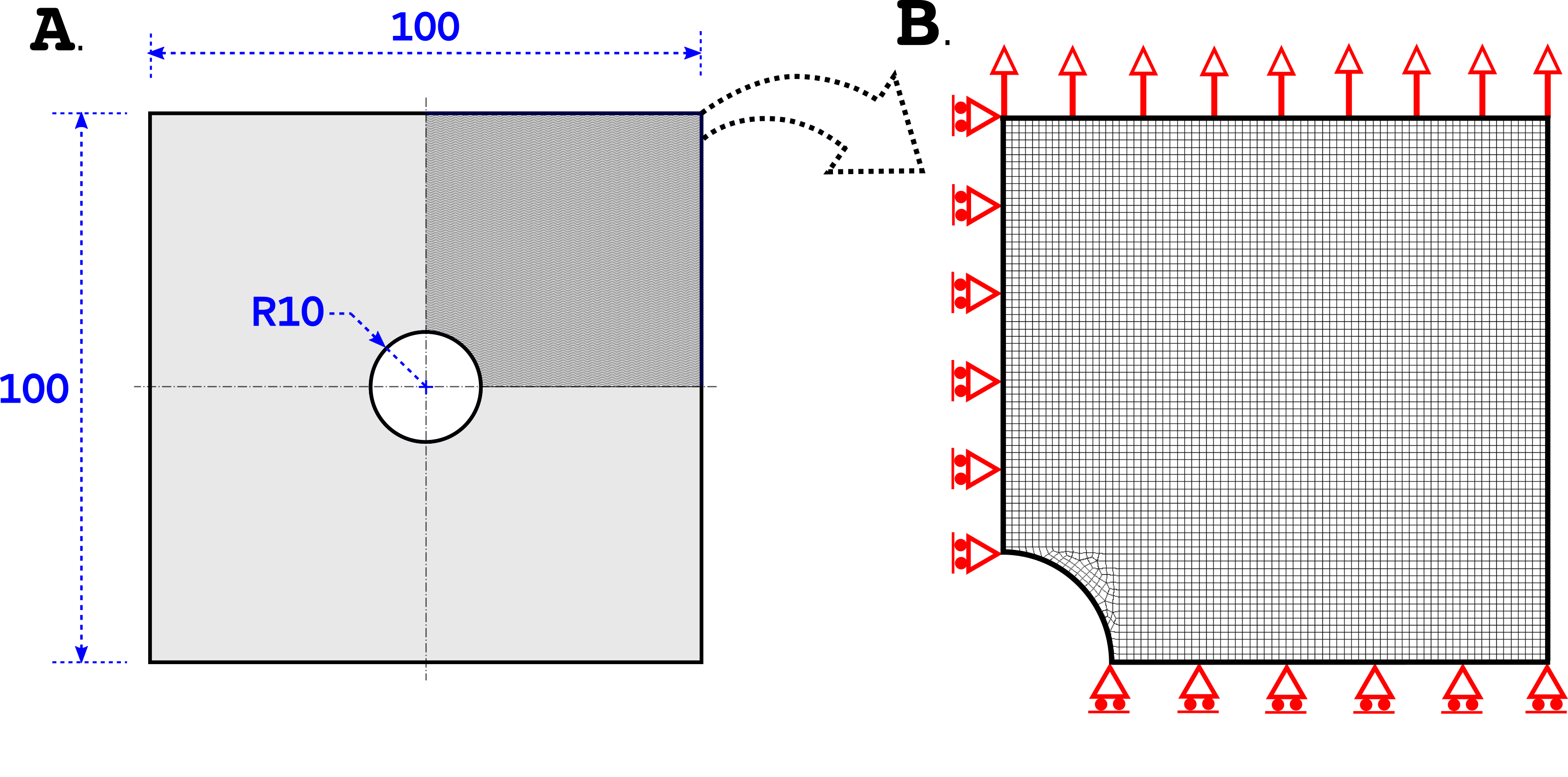}
\caption{Model illustration for open hole specimen simulation $\textbf{A.}$ Geometrical description of the plate of size 100 $\times$ 100 with a hole of radius 10 at the centre; shaded area shows the quarter model. $\textbf{B.}$ FE description of the model with boundary and loading conditions; symmetry conditions applied at the cut boundary with displacement applied at the top edge. Geometrical measurements are in $mm$.}
\label{fig:pwah1}
\end{figure}
The plate is analysed for two different fiber orientations:
\begin{enumerate}
	\item CASE-1; the load is applied along the fibers.
	\item CASE-2; the load is applied transverse to fibers.
\end{enumerate} 
For CASE-2, the local coordinate system is rotated by $90^\circ$ about the $x_1$-axis. No fiber failure is expected when the load acts in the transverse direction. The fiber volume fraction is considered as 49\%, and a single-fiber, two-partitioned microscale model is utilised for obtaining the influence tensors during the pre-processing stage.
\begin{table}[h]
\captionsetup{width=0.75\textwidth}
    \caption{Material properties of fiber and matrix used for RVE simulations. The data used is nearly the same as for glass-epoxy composite. Source: \cite{singh2017reduced}}
    \label{tab-a}
\centering
\begin{tabular}{lcc}
\toprule
\textbf{Material Property} & \textbf{Fiber} & \textbf{Matrix} \\
\midrule
Volume Fraction & 0.49 & 0.51\\
Elastic Modulus [MPa]& 80,000 & 2,670 \\
Poisson's Ratio & 0.3 & 0.3 \\
Yield Strength [MPa] & - & 30 \\
Hardening Modulus [MPa] & - & 200 \\
Damage Initiation Strain [\%] & 0.00315 & 0.01515 \\
Damage Failure Strain [\%] & 0.1575 & 0.7575 \\
\bottomrule
\end{tabular} \label{table:3}
\end{table}
 The elastic modulus for fiber and matrix is considered the same as glass-epoxy composite as in Table \ref{table:3}. Before carrying out the solution phase for the open hole specimen, the mechanical and eigenstrain influence tensors are computed using the same procedure adopted for "\textit{Validation-1: RVE simulations}" or by following the algorithm mentioned as "\textit{BOX-1: Algorithm for Microscale Problem}" in Section 5.1.        
 
 The results of macroscale analysis are shown in Fig. \ref{fig:pwah4} for CASE-1, where the tensile load is acting along the fiber direction. Fig. \ref{fig:pwah4}A shows the force-displacement variation, and in Fig. \ref{fig:pwah4}B, the matrix and fiber damage plots are illustrated. These damage maps are obtained essentially at three displacements marked as $\textbf{1}$, $\textbf{2}$ and $\textbf{3}$ in force-displacement plot (see Fig. \ref{fig:pwah4}). 
\begin{figure}[H]
\centering
\includegraphics[width=0.835\textwidth]{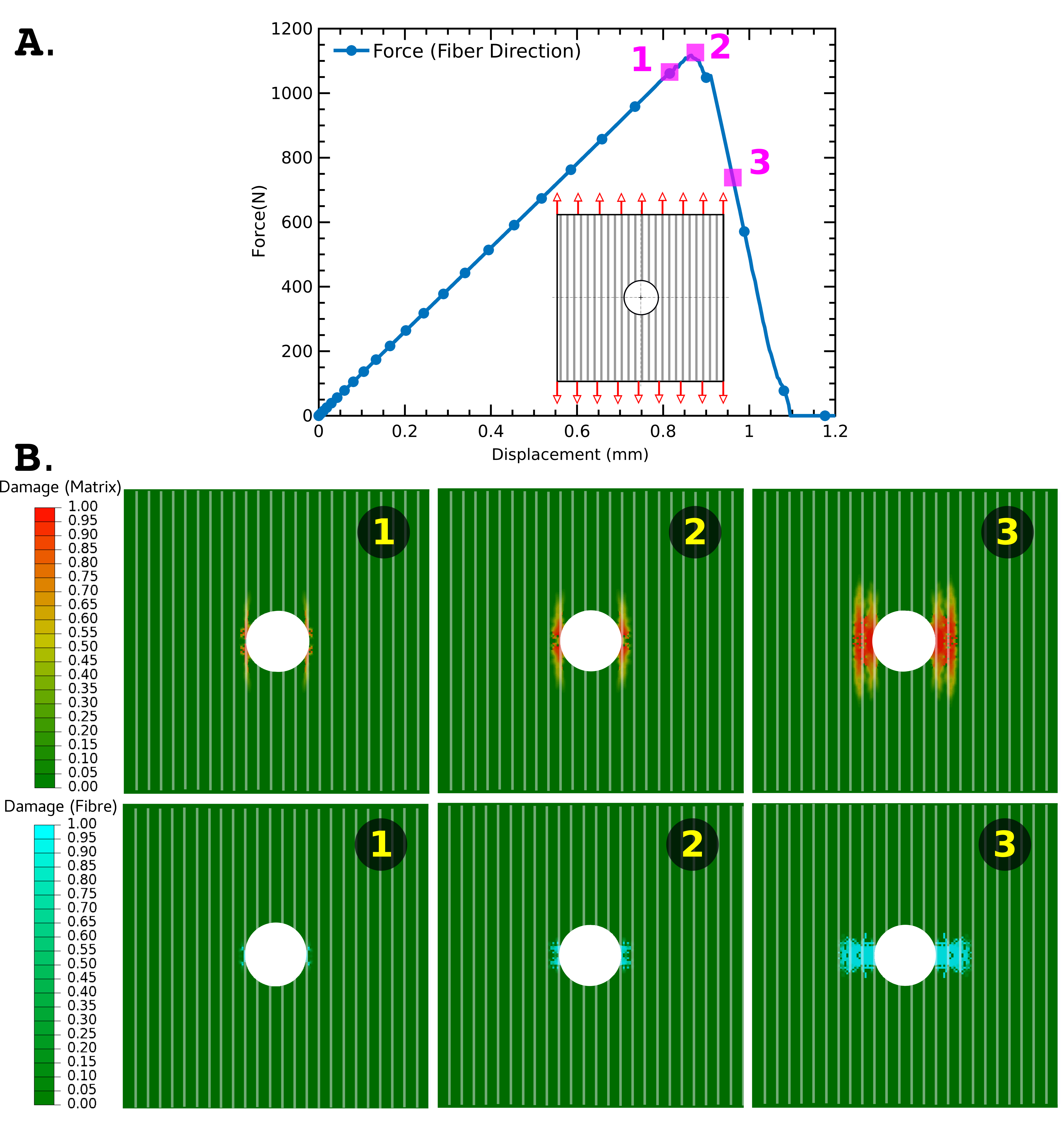}
\caption{Results for CASE-1; $\textbf{A.}$ Force-displacement response of the composite when the load is applied along the fiber $\textbf{B.}$ Damage maps for three points marked on the force-displacement plot as $\textbf{1}$, $\textbf{2}$ and $\textbf{3}$ where the red region corresponds to the matrix damage. The Cyan region corresponds to the fiber damage in the composite.}
\label{fig:pwah4}
\end{figure}

At stage 1, predominantly, the matrix damage starts with little evidence of fiber failure. As soon as the damage grows in the stress-concentrated regions (stage-2), the fibers begin to break. This point corresponds to the maximum force capacity of the plate, and beyond this point, the material starts to soften. \color{black}It can be observed from the direction of matrix damage growth, especially at stage 1 and stage 2, that the matrix damage starts to propagate parallel to the fibers and fiber damage grows across the fibers as expected. Further increase of load causes growth of matrix and fiber damage in the normal direction to the applied load which ultimately results in the complete fracture of open hole specimen in the $90^\circ$ direction. \color{black}The vertical lines and dots, shown in Fig. \ref{fig:pwah4} and Fig. \ref{fig:pwah5} respectively, are essentially the indicators for the fiber direction, and these are not part of the modelled equivalent homogeneous domain. Similar behaviour is observed for fiber damage growth where the direction of propagation is largely across the fibers. In CASE-2, three identical levels of the displacements, marked as 1, 2, and 3 (see Fig. \ref{fig:pwah5}) are selected for investigating the damage. It is observable that damage occurs primarily in the matrix region, and growth of the failure field happens perpendicular to the direction of loading. However, as anticipated, no breakage of fibers is noticed under transverse loading. 

After simulating these two cases, it can be concluded that the proposed $\mathtt{E}^2$-TFA provides realistic multiscale simulation results for complex morphological domains and has the capability of capturing the fiber and matrix damage growth.
\begin{figure}[H]
\centering
\includegraphics[width=0.9\textwidth]{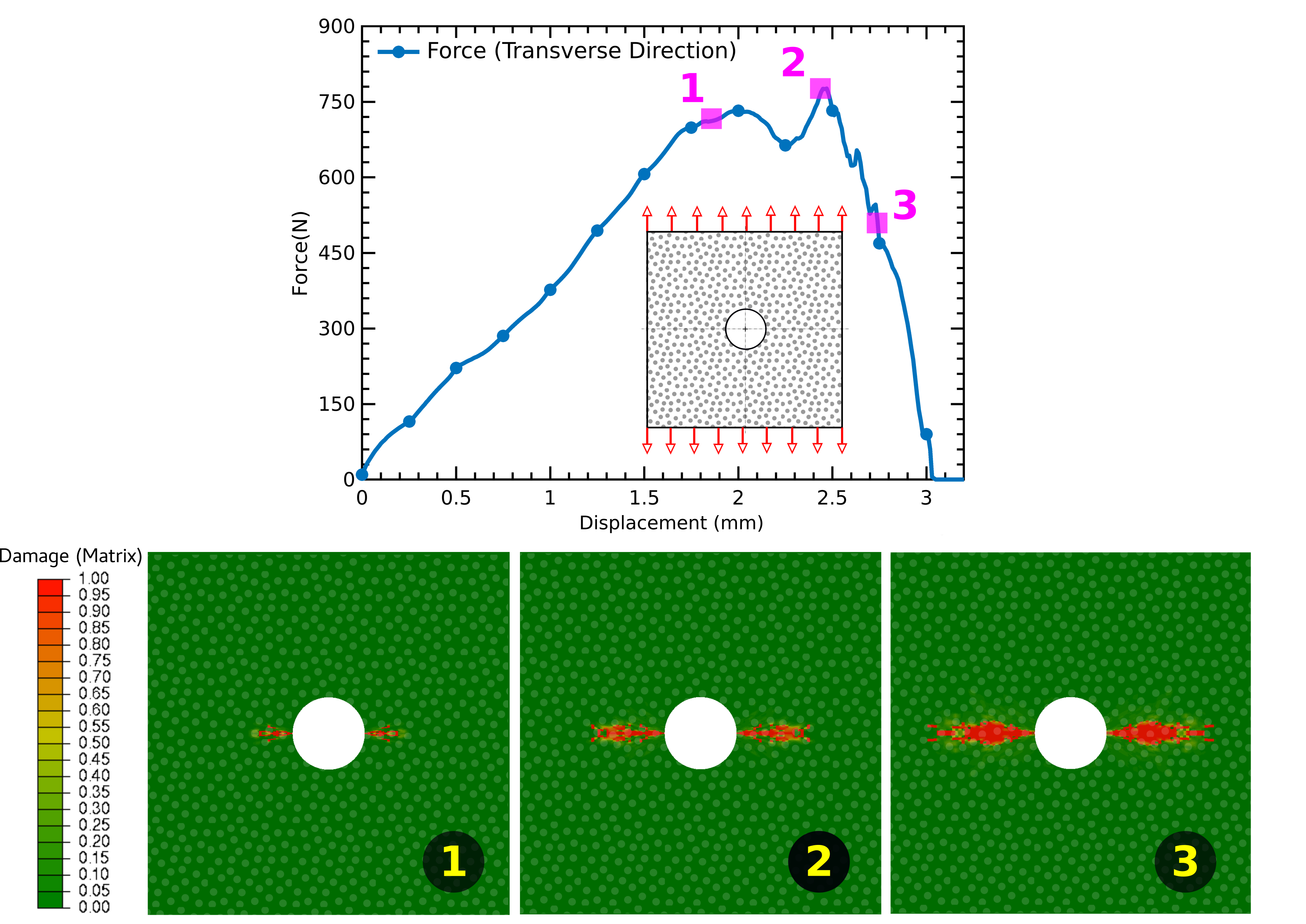}
\caption{Results for CASE-2; $\textbf{A.}$ Force-displacement response of the composite when the load is applied along the fiber $\textbf{B.}$ Damage maps for three points marked on the force-displacement plot as $\textbf{1}$, $\textbf{2}$ and $\textbf{3}$ where the red region corresponds to the matrix damage in the composite.}
\label{fig:pwah5}
\end{figure}

\subsection{Numerical Implementation-3: Crack propagation in double notch plate}
In another test case, a double-notch specimen of the composite layer is considered. The overall size of the specimen is 30 mm $\times$ 50 mm $\times$ 1 mm with two notches of depth 8 mm from the edge and 1 mm in width. A vertical offset of 5 mm between the notches is introduced as depicted in Fig. \ref{fig:double_notch_1}A. The reason for introducing the offset is to simulate the curved trajectory of the fracture path developing from the tip of one notch to another notch. Obviously, it also depends on the nature and direction of the load it is subjected to. Therefore, the bottom edge of the model is constrained to move out of the plane, whereas the displacement-controlled load is applied on the top edge in the $x_1$-direction to simulate the shear mode of deformation. Fig. \ref{fig:double_notch_1}B shows the FE mesh where the domain has 9875 eight-noded quadrilateral elements. 

The microscopic properties of the composite are used the same as mentioned in Table \ref{table:3}. After performing the pre-processing analyses using microscale material data for obtaining the influence and homogenised tensors, the model is assessed further for two loading conditions:
\begin{enumerate}
	\item Monotonic
	\item Cyclic
\end{enumerate}
\subsubsection{Monotonic shear loading}
Under monotonic loading conditions, an incrementally increasing displacement is applied at the top face of the domain as shown in Fig. \ref{fig:double_notch_1}B. Fig. \ref{fig:double_notch_2} exhibits the overall response of the double-notched plate in terms of (1). Damage contour in the material (2). Force-displacement variation. Since the fiber orientation is considered in the $x_3$-direction (out-of-plane), the damage plot governs mainly the matrix failure. It is observable that the damage starts from both notches simultaneously and merges into each other finally, which leads to a complete loss of stiffness of the structure. Force-displacement plot manifests the softening behaviour of the material and a sudden decrease in the load-carrying capability of the material after reaching a peak level of force.

\begin{figure}[H]
\centering
\includegraphics[width=0.65\textwidth]{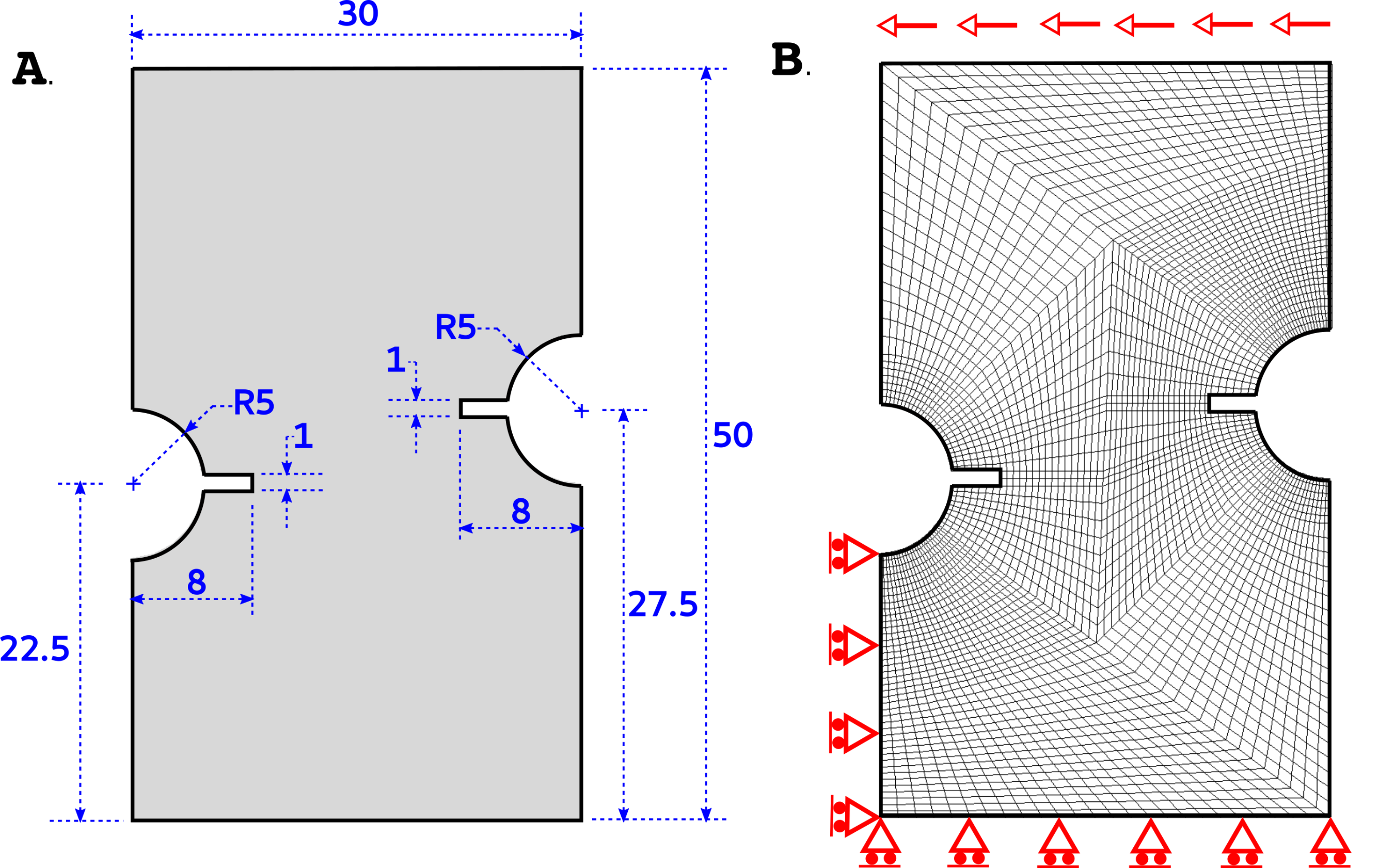}
\caption{Model illustration for double notch specimen simulation $\textbf{A.}$ Geometrical description of the plate of size 30 mm $\times$ 50 mm $\times$ 1 mm with two notches of depth 8 mm and width 1 mm at an offset of 5 mm. $\textbf{B.}$ FE description of the model with boundary and loading conditions; symmetry conditions applied at the bottom face, and displacement applied at the top edge simulating the shear mode of deformation. Geometrical measurements are in $mm$.}
\label{fig:double_notch_1}
\end{figure}

\begin{figure}[H]
\centering
\includegraphics[width=0.75\textwidth]{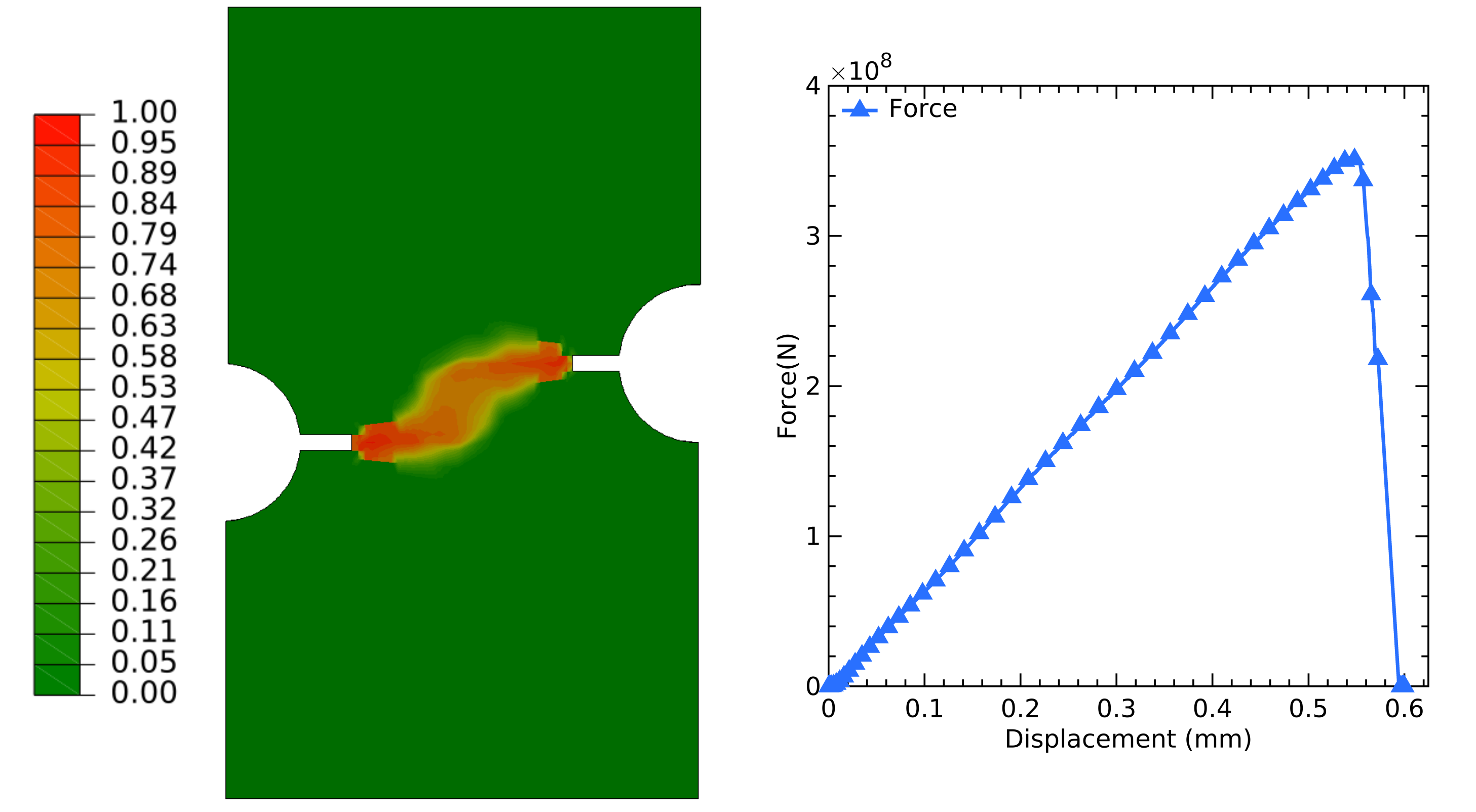}
\caption{Damage contour plot (on the left) for double notch specimen where the damage state variable varies from 0 (Zero), which corresponds to "\textit{No damage state}" to 1 (One), which depicts "\textit{Fully damaged state}". Force-displacement variation (On the right) for double notch specimen, which illustrates the maximum load-bearing capability of the composite and complete loss of stiffness as crack fronts meet from both the notches.}
\label{fig:double_notch_2}
\end{figure}

Furthermore, the damage variation at the three locations, as shown in Fig. \ref{fig:double_notch_3}, in the domain is investigated to check its variation:  
\begin{enumerate}
	\item $\textbf{Location 1}$: Tip of the left notch
	\item $\textbf{Location 2}$: Geometrical centre of the domain
	\item $\textbf{Location 3}$: Tip of the right notch
\end{enumerate}
At these three locations, damage variation with applied displacement on the top face, as shown in Fig. \ref{fig:double_notch_3}, confirms that the damage initiates at both the notch-tips (location 1 and 3) simultaneously, and damage growth rates at these locations are almost the same. As soon as the damage variable at location 2 becomes '1', the specimen completely loses its load-carrying capacity. This confirms that the complex fracture trajectories, as Fig. \ref{fig:double_notch_2} depicts, can be captured realistically using the proposed $\mathtt{E}^2$-TFA methodology.  
 
\begin{figure}[H]
\centering
\includegraphics[width=0.75\textwidth]{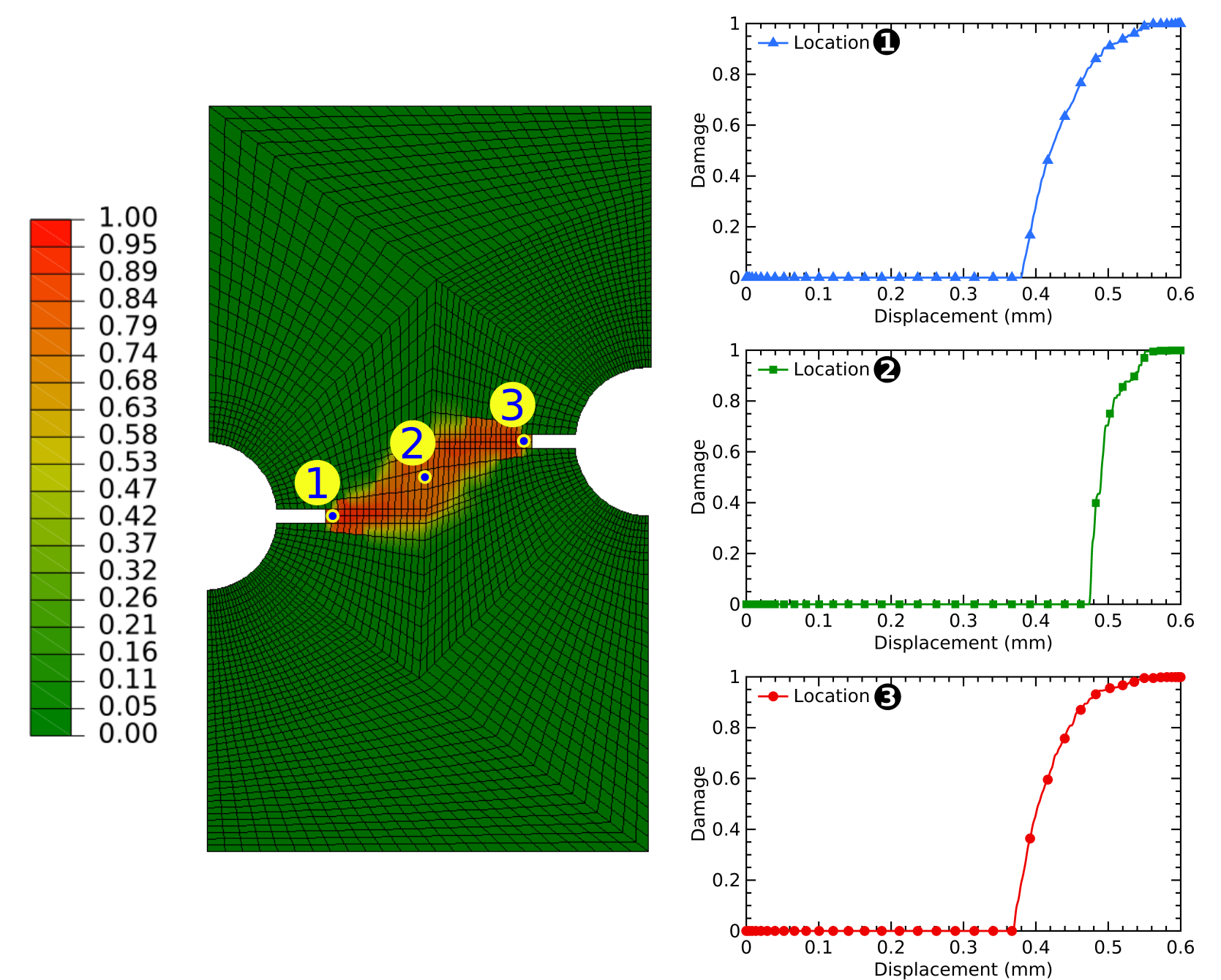}
\caption{The three locations marked \textbf{1} (at the tip of left side notch), \textbf{2} (at the middle of the specimen) and \textbf{3} (at the tip of right side notch) of the double notch specimen shown on the left side, are investigated for obtaining the damage variation with applied displacement corresponding to the shear mode of deformation. Damage-displacement plots are shown on the right side.}
\label{fig:double_notch_3}
\end{figure}

\subsubsection{Cyclic shear loading}
In cyclic loading, applied displacement is characterised as loading, unloading, and reverse loading paths as shown in Fig. \ref{fig:double_notch_4}, where it varies linearly with respect to time for three cycles under unequal maximum displacement amplitude. Again, the growth of the failure zone is monitored, especially at maximum and zero displacement points.   
\begin{figure}[H]
\centering
\includegraphics[width=0.65\textwidth]{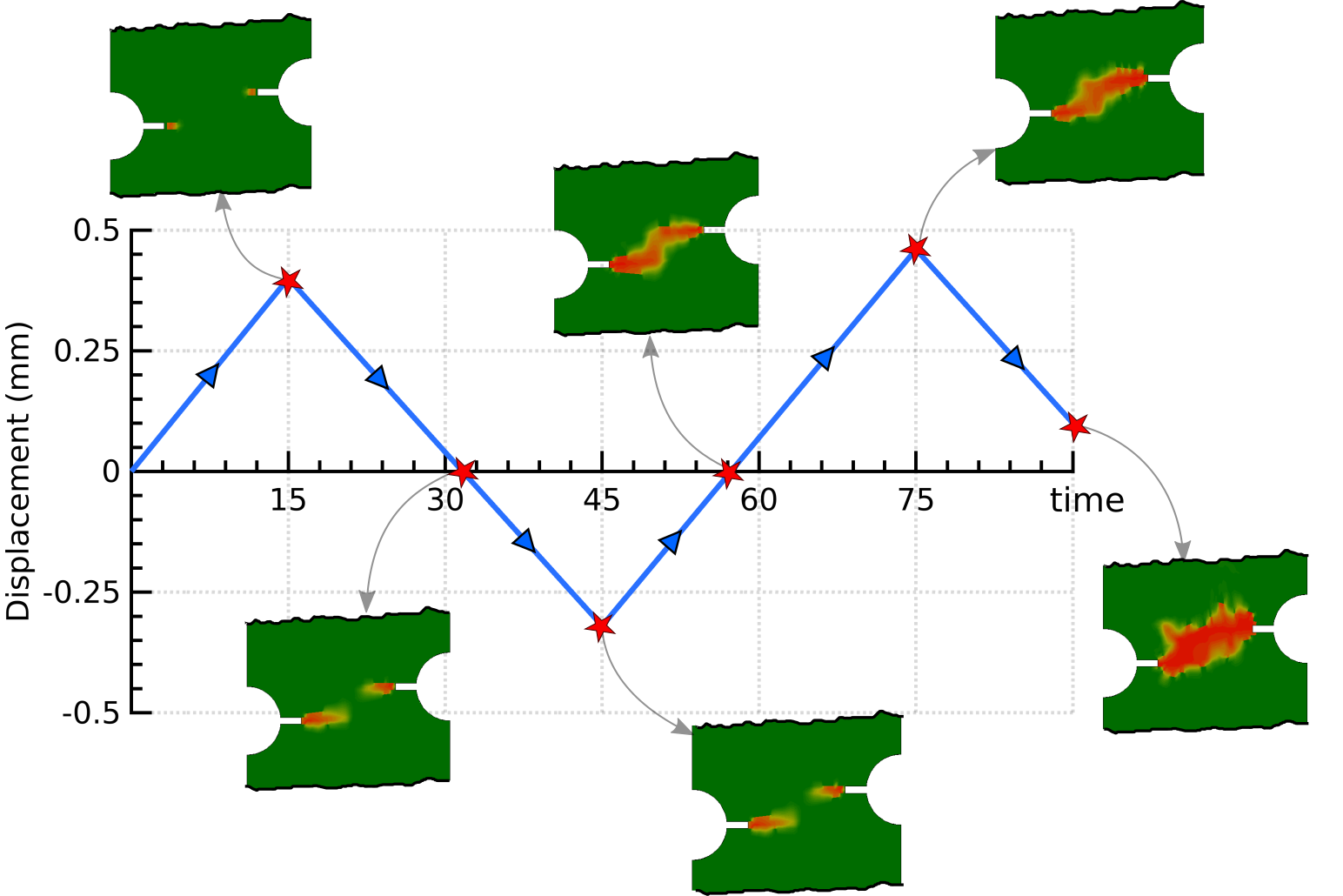}
\caption{Applied displacement variation with respect to time and the damage plots near the notched region at various stages (marked with red stars) of the applied displacement, which portrays the growth of damage from the undamaged state (at the initial stage) to the fully damaged state (at the end of the cycle). Here the green region belongs to undamaged material, and the red region shows fully damaged material. The time along $x$-axis is in "$msec$"}
\label{fig:double_notch_4}
\end{figure}
Fig. \ref{fig:double_notch_4} portrays that the damage path progresses gradually with time. Overall, the purpose of simulating this double-notch specimen subjected to the condition of the transient load is two-fold; First, to demonstrate the competence of recommended methodology for transient loading, and second, to witness the inertial effects which are clearly evident from the propagation of fracture during the time of decreasing displacement.
\subsection{Numerical Implementation and Experimental Validation-4: Damage in 10$^\circ$ composite plate}
Nonlinear shear stress-strain response of glass fiber composites using experimental tests, reported by \cite{van2006modelling1}, is used for the assessment of the proposed method to capture the non-linear stress-strain data \citep{van2006modelling2}. The sketch of the model with indicators of fiber direction 10$^\circ$ is represented in Fig. \ref{fig:10_degree_1}. The model reproduces the off-axis test of $[10^\circ]_8$ glass epoxy composite. The specimen size is 100 mm $\times$ 20 mm without modeling the gripping region and with both ends cut at $54^\circ$ to horizontal (see Fig. \ref{fig:10_degree_1}A). The experimentally prescribed $x_1$-displacements at both ends are mimicked in simulation by fixing one end of the specimen and applying the displacement on the other as depicted in Fig. \ref{fig:10_degree_1}B. \color{black} Here, the whole domain is meshed with 3D full integration linear elements. \color{black} Fiber direction is considered in the model by rotating the elemental coordinate system by $10^\circ$ in the clockwise direction.  
\begin{figure}[H]
\centering
\includegraphics[width=0.75\textwidth]{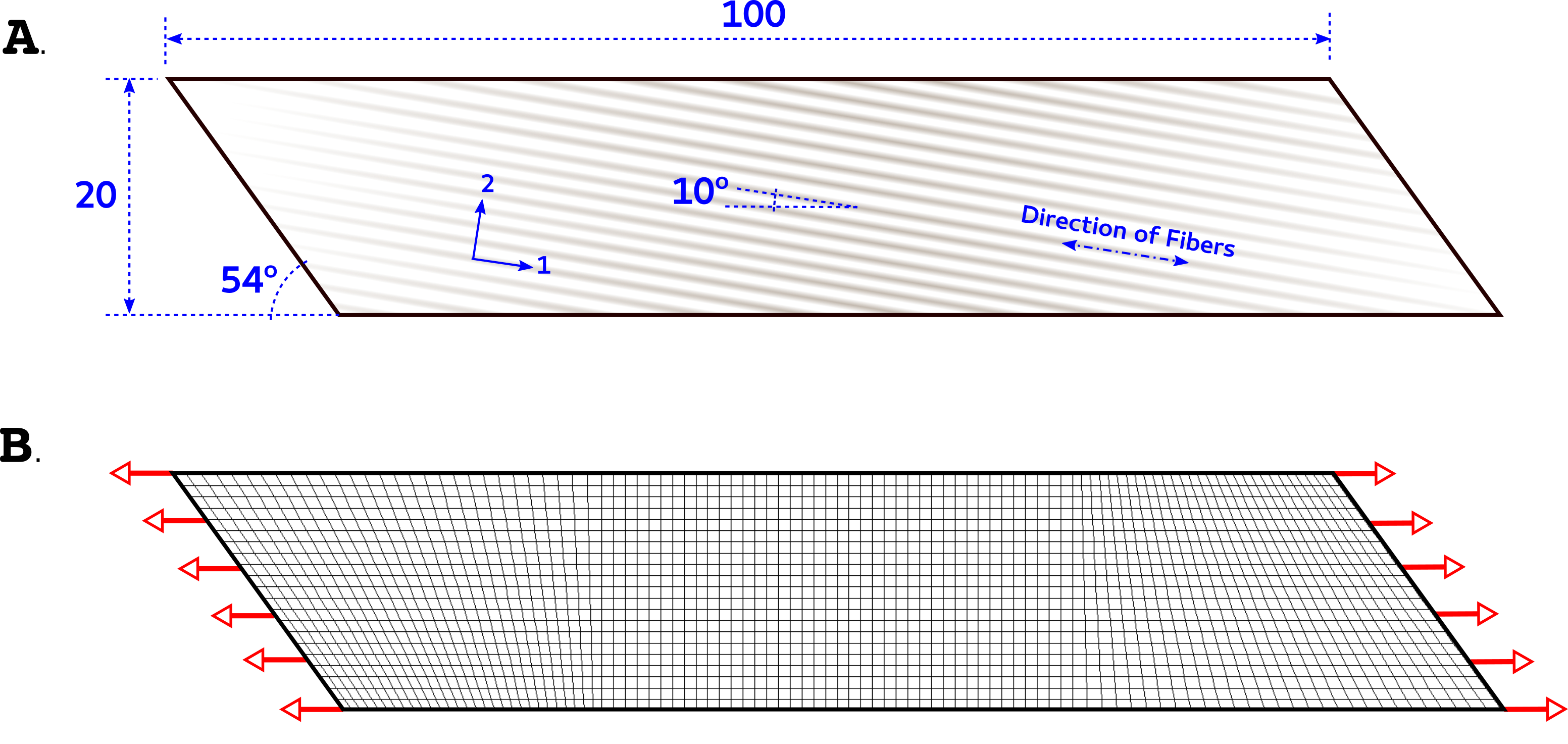}
\caption{Model illustration for $10^\circ$ composite plate simulation $\textbf{A.}$ Geometrical description of the plate of size 100 $\times$ 20 with the side faces inclined at $54^\circ$. fibers are oriented along a direction of $10^\circ$ to the horizontal shown schematically. $\textbf{B.}$ FE description of the model with boundary and loading conditions; displacement boundary condition is applied in the horizontal direction to both sides of the domain. Geometrical measurements are in $mm$.}
\label{fig:10_degree_1}
\end{figure}
The microscopic material properties for the constituents, as summarised in Table \ref{table:3}, are used except 50\% fiber volume fraction is considered. Microscale studies are performed for the calculation of influence tensors using 2-partitioned single fiber RVE. The directional moduli and other elastic material properties are determined from $\mathbb{L}$ tensor and compared with experimental data furnished in \citet{van2006modelling1}. Table \ref{table:5} shows the comparison of experimental data \citep{van2006modelling1} with the preprocessing results, which exhibits a good match.
\begin{table}[h]
\captionsetup{width=0.85\textwidth}
    \caption{Calculated material data and its comparison with experimentally reported values by \cite{van2006modelling1}}
    \label{tab-a}
\centering
\begin{tabular}{lcc}
\toprule
\textbf{Material Property} & \textbf{Experimental} & \textbf{Numerical} \\
 In-plane  &  \citep{van2006modelling1} & $\mathtt{E}^2$-TFA Preprocessing\\
\midrule
$E_{11}$ [GPa]& 42.4 & 41.4 \\
$E_{22}$ [GPa] & 14.2 & 14.7 \\
$\nu_{12}$ & 0.245 & 0.210 \\
$G_{12}$ [GPa] & 5.13 & 5.98 \\
\bottomrule
\end{tabular} \label{table:5}
\end{table}

\begin{figure}[H]
\centering
\includegraphics[width=0.75\textwidth]{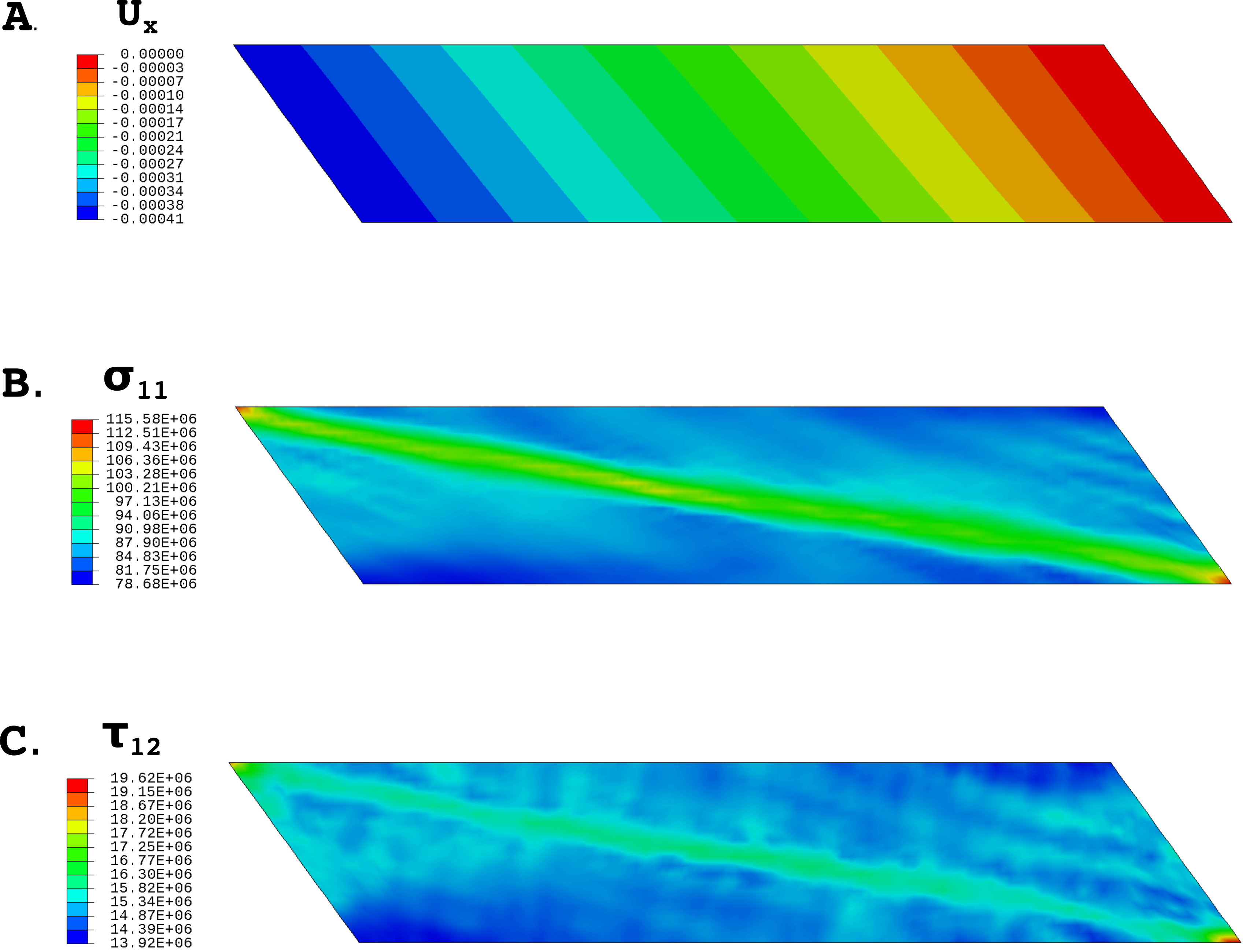}
\caption{Contour plot of $\textbf{A.}$ displacement in the horizontal direction, U$_x$; $\textbf{B.}$ stress in the fiber-direction, $\sigma_{11}$; $\textbf{C.}$ in-plane shear stress, $\tau_{12}$.  Here 1 is the material direction along the fiber, and 2 is the in-plane material direction transverse to the fiber. Unit of displacement is '$m$' and unit of stress is '$Pa$'.}
\label{fig:10_degree_2}
\end{figure}
Macroscale analysis results are presented in Fig. \ref{fig:10_degree_2} in terms of displacement and stress contours, plotted in material directions (1-direction along fibers and 2-direction is transverse to fibers) at an intermediate time step. Fig. \ref{fig:10_degree_2}A reveals the applied displacement in $x_1$-direction which basically endorses the applied loading. High tensile and shear stresses are established along the fiber directions along the diagonal (see Fig. \ref{fig:10_degree_2}B and C), which matches with the data reported in \citep{van2006modelling1, van2006modelling2} except the vicinity of ends where some localized high-stress regions are found. 
\begin{figure}[H]
\centering
\includegraphics[width=1\textwidth]{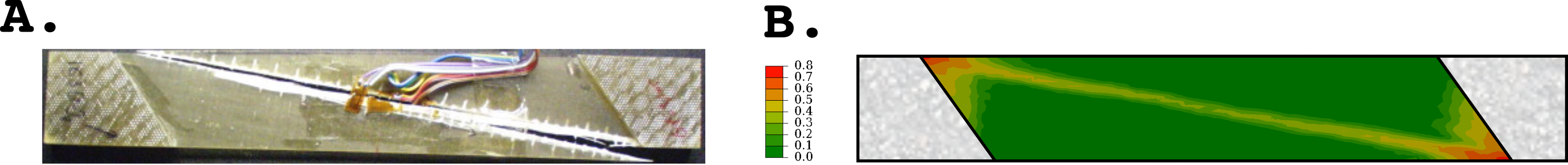}
\caption{Comparison of $\textbf{A.}$ the fractured specimen (Source: \cite{van2006modelling1}) with $\textbf{B.}$ the damage plot predicted from the multiscale calculations, which demonstrates the excellent match of the numerically calculated contour with the experimentally obtained fracture map.}
\label{fig:10_degree_3}
\end{figure}
Fig. {\ref{fig:10_degree_3}B shows the damage contour from the simulations, which reveals the good agreement with the experimentally obtained failure map (see Fig. \ref{fig:10_degree_3}A).    
	
Lastly, the strain time histories are plotted for $\epsilon_{11}$, $\epsilon_{22}$, and $\gamma_{12}$ (see Fig. \ref{fig:10_degree_4}) and compared with experimental data \citep{van2006modelling1}. \color{black}Fig. \ref{fig:10_degree_4} depicts that the shear response of the composites shows the nonlinear stress-strain variation in experiments, whereas the linear behaviour has been demonstrated for in-plane fiber and transverse directions. It has been typically observed that long-fiber polymer composites exhibit nonlinearity under shear loading only. In the proposed formulation, this macroscale nonlinearity can be simulated by considering the inelastic constitutive model for matrix material at the microscale, where the fiber domain is modelled with linear elastic material. This microscopic inelasticity leads to a nonlinear macroscopic response in shear and transverse directions caused by the matrix-dominated deformation modes. Therefore, capturing the nonlinear response solely for shear loading will always be challenging with multiscale formulations. In the present study, the inelastic properties of the matrix are calibrated, and the $\mathtt{E}^2$-TFA computed macroscopic material data is compared with experiments as shown in Table \ref{table:5}. Using those properties, it has been found that the proposed formulation can capture the almost linear variation of $\varepsilon_{11}$ and $\varepsilon_{22}$. Although the results demonstrate that $\gamma_{12}$ variation is almost linear in simulations, notable overlap with the experiments has been found, which endorses the proficiency of the proposed multiscale $\mathtt{E}^2$-TFA methodology. \color{black}      
\begin{figure}[H]
\centering
\includegraphics[width=0.6\textwidth]{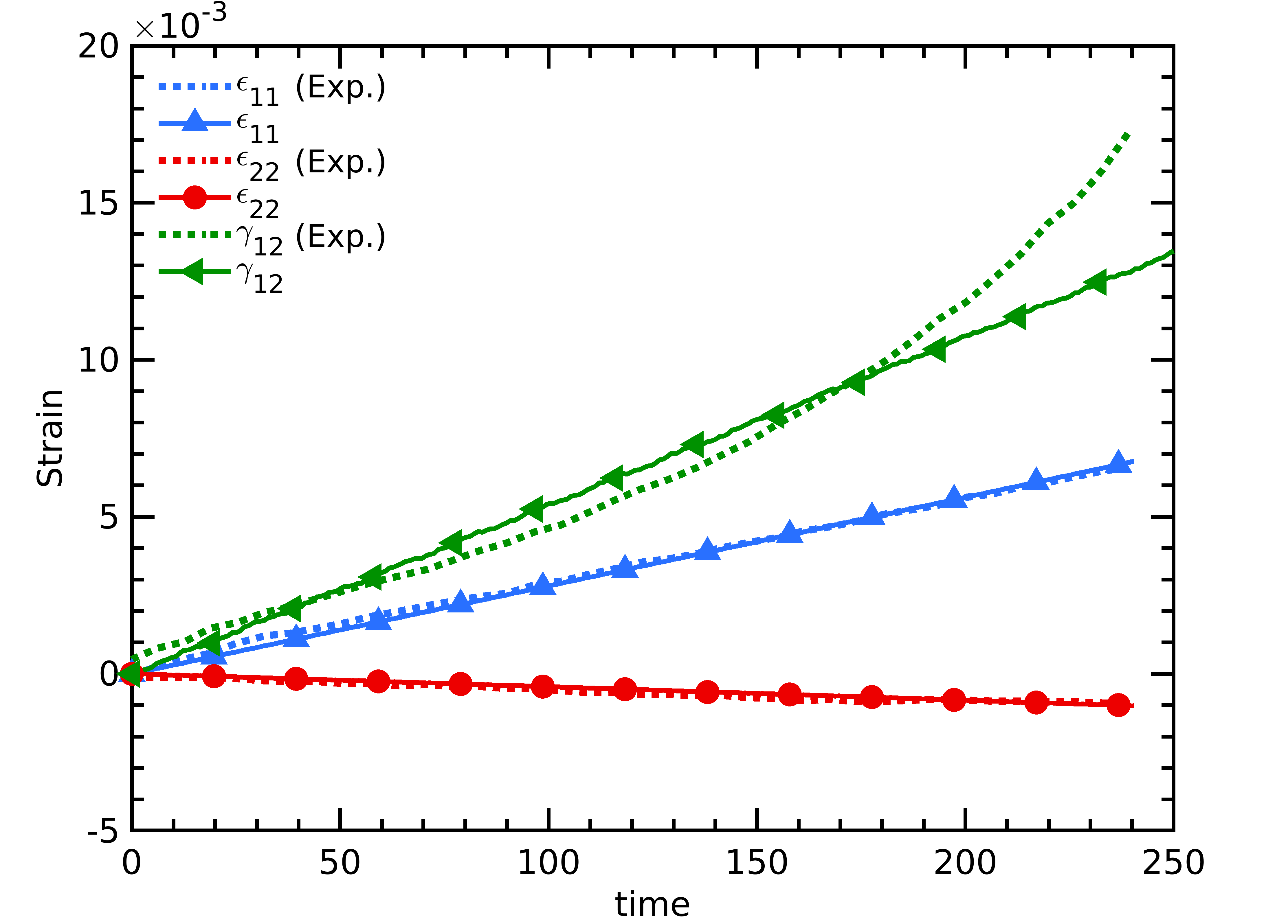}
\caption{Comparison of the variation of numerically calculated strain components $\varepsilon_{11}$, $\varepsilon_{22}$ and $\gamma_{12}$ with the experimentally reported variation (Source: \cite{van2006modelling1, van2006modelling2}). Here 1 is the material direction along the fiber, and 2 is the in-plane material direction transverse to the fiber.}
\label{fig:10_degree_4}
\end{figure}

\subsection{Numerical Implementation and Experimental Validation-5: Composite plate under impact}
The behaviour of E-glass/epoxy composite plate subjected to impact load is simulated and compared with experimental data \citep{singh2017reduced}. A laminate with 3.85 mm thickness and 8 plies of layup sequence $[0^\circ/90^\circ/0^\circ/90^\circ]_s$ is modelled. The plate is positioned between the top and bottom steel frames, creating an open effective area measuring 125 mm by 125 mm, as shown in Fig. \ref{fig:impact_model}. The impactor is modelled as a rigid hemisphere with 13 kg mass and $\Phi$ 12.7 mm diameter. In simulation, quarter symmetry is utilized to shorten the computation time by considering the geometric configuration and loading conditions. The symmetry conditions are imposed at the cut faces of the laminate and frame. The simulations are performed at 14 J and 20 J impact energy levels corresponding to 1.458 m/s and 1.780 m/s impact velocities. A pressure is applied at the top face to represent the preload that is exerted on the frame to tighten the plate whereas the bottom face of the frame is fixed in all directions. A traction-displacement-based cohesive interaction is applied between two adjacent plies to simulate the interlaminar damage. These cohesive surfaces have been assigned Mode-I, II, and III fracture energies, which serve as regulating parameters for interlaminar damage propagation. A maximum stress-based criterion is defined at the mating surfaces of the adjacent plies. Mode-I, II and III fracture energies are used to simulate the interlaminar damage propagation. A damage variable that can be described as a function of fracture energy emanates when the damage starts.

Fig. \ref{fig:impact_force} compares the contact force-time histories obtained from the experiments and the simulations. $\mathtt{E}^2$-TFA evaluated force is also compared with numerical results reported by \cite{singh2017reduced} which are calculated using classical TFA based framework. Stiffer behaviour of the laminate was observed in the simulation performed by \cite{singh2017reduced} during the rebound phase which could be attributed to the spurious post-damage microscopic response of classical TFA-based methods.    It is evident from the plots that $\mathtt{E}^2$-TFA estimates the rebound force-time history better than the previous approach, and overall, the prediction matches with the experiments for 14 J and 20 J impact energies. The rebound phase is the post-peak period when the impactor moves upward.

The damage maps from the proposed formulation, yielded at 14 J and 20 J are contrasted with experiments and previously reported results by \cite{singh2017reduced} as shown in Fig. \ref{fig:impact_damage}. Interlaminar damage/delamination in the laminate and intralaminar matrix cracking majorly around the point of impact is assessed. No fiber damage is found at both impact energies. Therefore, the damage maps correspond to the interlaminar or intralaminar matrix damage. In the simulations shown in Fig. \ref{fig:impact_damage}, the blue contour refers to the zero damage zone whereas the red region depicts the fully interlaminar/intralaminar damaged region. Again, the $\mathtt{E}^2$-TFA predictions are found to be better than the previously reported simulations and match with the experiments in terms of size and shape.
\begin{figure}[H]
\centering
\includegraphics[width=0.85\textwidth]{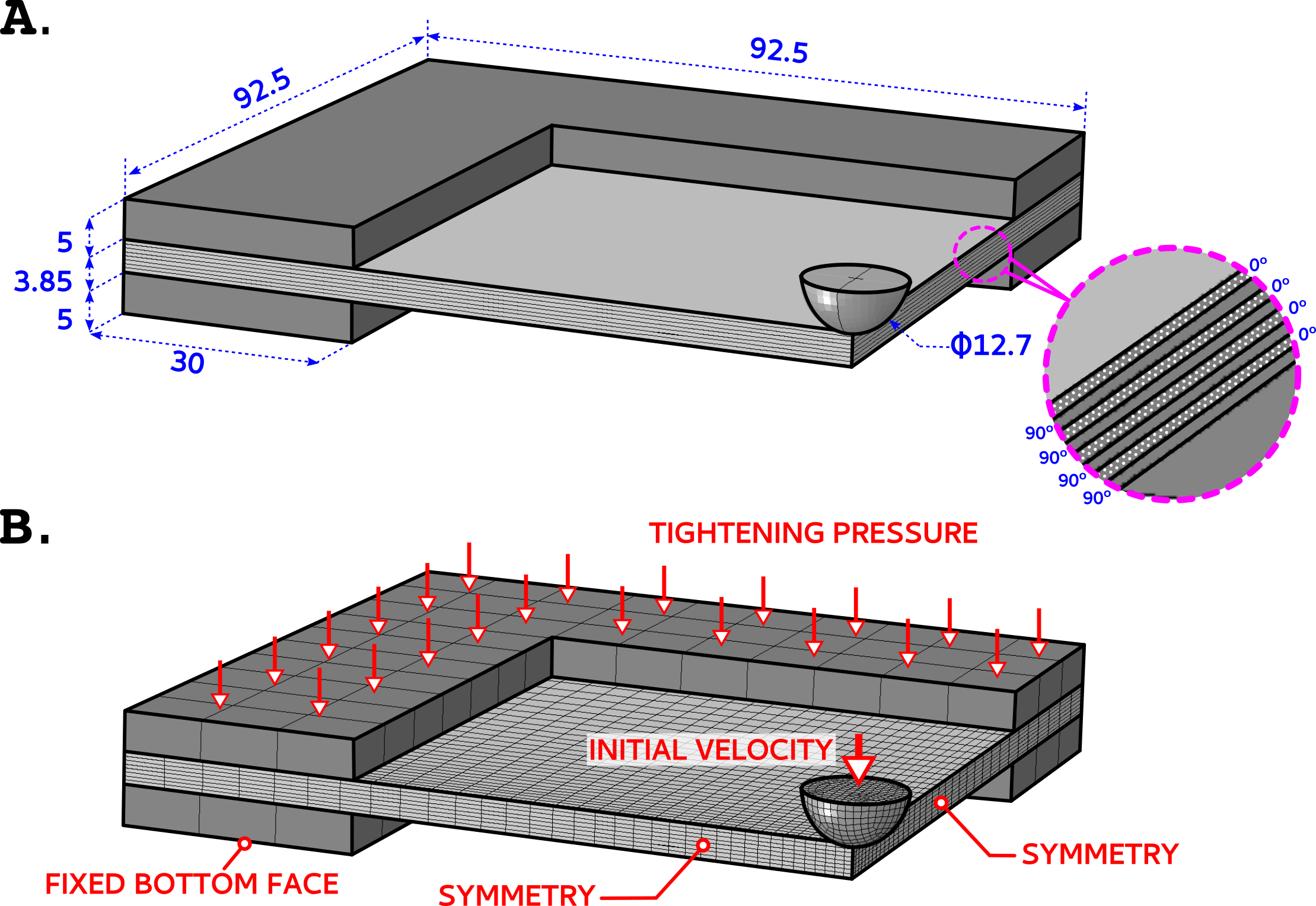}
\caption{\color{black}Model illustration for impact simulation \textbf{A.} Geometrical description of plate with ungripped size 125 mm $\times$ 125 mm $\times$ 3.85 mm modeled one-fourth and impactor of size $\Phi$ 12.7 mm. The layup for the layers is shown in the inset. \textbf{B.} FE description of the model with loading and boundary conditions; symmetry conditions are applied on the cut faces, tightening pressure is applied top face of the gripper, and the bottom face is fixed. Initial velocity is given to the impactor. Geometrical measurements are in $mm$}
\label{fig:impact_model}
\end{figure}
   
\begin{figure}[H]
\centering
\includegraphics[width=1\textwidth]{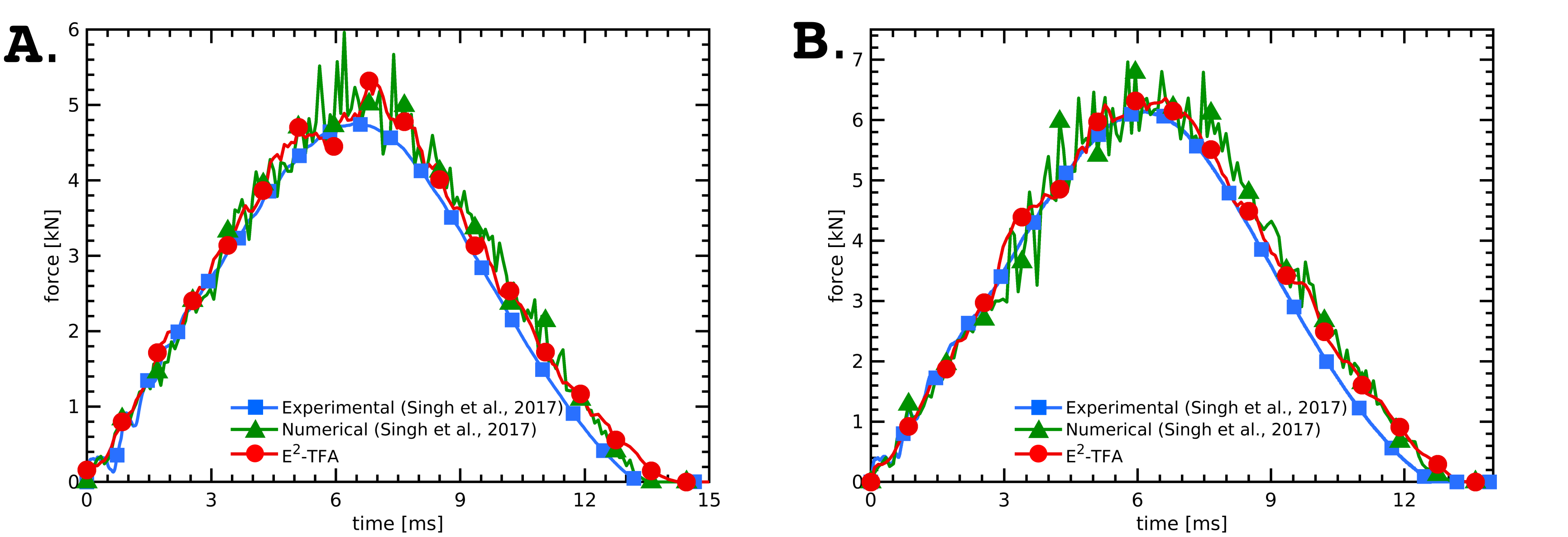}
\caption{\color{black}Comparison of $\mathtt{E}^2$-TFA computed force-time histories (shown in red) at \textbf{A.} 14 J, and \textbf{B.} 20 J impact energies with experimental data \citep{singh2017reduced} (shown in blue). Classical TFA-based numerical results obtained by \cite{singh2017reduced} (shown in green) are compared with $\mathtt{E}^2$-TFA.}
\label{fig:impact_force}
\end{figure}

\begin{figure}[H]
\centering
\includegraphics[width=0.85\textwidth]{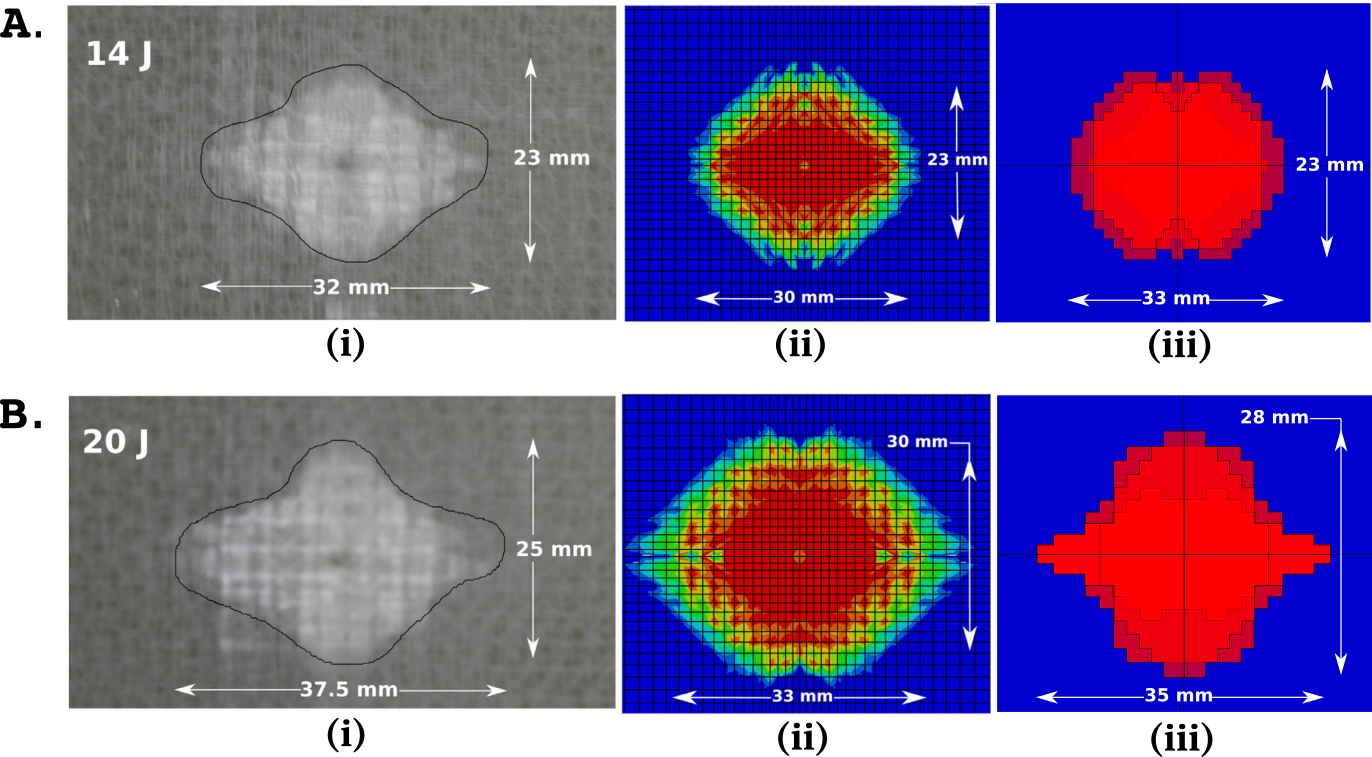}
\caption{\color{black}Comparison of \textbf{(i)} experimentally determined damage area with \textbf{(ii)} and \textbf{(iii)} the simulation results. \textbf{A.} depicts the results for 14 J impact energy and \textbf{B.} shows the results for 20 J impact energy. Here, \textbf{(ii)} the damage areas calculated using TFA-based formulation by \cite{singh2017reduced} are also compared with \textbf{(iii)} $\mathtt{E}^2$-TFA computed damage maps for 14 J and 20 J impact energies.}
\label{fig:impact_damage}
\end{figure}\color{black}
\section{Conclusions}\label{sec:7}
A homogenisation method, $\mathtt{E}^2$-TFA, for obtaining the mechanical response of elastoplastic composite material in the presence of damage is proposed in this work. This technique appraises the microscopic eigenstrain field, which accounts for intra-phase damage and inelastic strains using a continuum damage mechanics (CDM) based framework. The rate equations of internal variables are instituted for the evolution of damage and plastic strain by introducing a dissipation potential. The method characterises a simple and numerically effective approach for calculating preprocessing data, 1). elastic and eigen influence tensors ($\mathbb{\overline{E}}$ and $\mathbb{\overline{S}}$); 2). homogenised constitutive tensors ($\mathbb{\overline{L}}$ and $\mathbb{\overline{M}}$) and assuming a piecewise constant eigenstrain field in the microscale domain for reduction of order. The proposed procedure leads to the mitigation of post-damage stiffness issues often noticed in TFA-based techniques when applied to softening materials. The required information for solving the macroscale problem is acquired by adopting a direct approach, as mentioned in Section 5. The number of variables for defining the eigenstrain field induced by plasticity and damage of the microconstituents is optimised.   

The $\mathtt{E}^2$-TFA has been implemented systematically in two stages a). Preprocessing stage, and b). Solution stage. First, this two-stage approach was applied to carry out the RVE studies. The equivalent homogenised medium was checked under monotonic and reversed loads, and special constitutive responses of the phase materials were considered for separately investigating 1). plasticity 2). elastic damage, and 3). elastoplastic damage equivalent eigenstrain variation. These three-step studies, thereby, demonstrated not only the predictive capability of the formulation but also the accuracy of the results by alleviating the problem of post-damage pseudo stiffness results. Moreover, the process of calculating the preprocessing data has been simplified, and, especially, the direct approach of obtaining the influence tensors and homogenised properties is noteworthy.   

Next, the macroscale study of open-hole composite specimens was performed for two configurations. In the first configuration, the fibers are oriented along the axis of the hole, whereas for the second configuration fiber direction was aligned perpendicular to the hole axis. These give rise to dissimilar loading scenarios and two different modes of failure. fiber and matrix damage patterns have been captured distinctly, which proves the excellent qualitative predictivity of $\mathtt{E}^2$-TFA.

In another numerical application, the complex damage trajectories have been studied using $\mathtt{E}^2$-TFA for a double-notch composite specimen when subjected to both monotonic and cyclic loading histories. The model was tested under a shear dominant mode of deformation. The $\mathtt{E}^2$-TFA predictions were good in terms of crack path direction. Consequently, it has been observed that the damage initiation and growth at stress-concentrated locations under cyclic load were satisfactorily captured with visible effects of interia on fracture evolution.

In addition, the validation of $\mathtt{E}^2$-TFA was performed by simulating the response of $10^\circ$ glass epoxy composite under tensile load and results are compared with experiments done by \cite{van2006modelling1}. \color{black}Again, the fractured plane was found matching with the predicted damage surface, and strain histories were reproduced to a large extent. 

Eventually, the proposed formulation was used to predict the impact response of composite laminate and the results were compared with numerical evaluations by \cite{singh2017reduced}. Simulation demonstrates the improvement in the prediction capability with $\mathtt{E}^2$-TFA compared to classical TFA-based techniques. A significant match between experiments and simulations manifests the $\mathtt{E}^2$-TFA as a valuable homogenisation technique in multiscale research.\color{black}    

The author wants to emphasise that even though all of the numerical studies were carried out using two-phase single-fiber RVEs, the extension to multi-phase materials with randomly distributed microscale architecture is anticipated to be a simple extension of the work with no implementation difficulties. \color{black}The computational cost of random composites analysis may necessitate the use of an effective partitioning approach, which could be the subject of future research. \color{black}The isotropic nature of the damage used in the proposed formulation disregards the specific shapes of the micro-cracks. Several anisotropic damage models are available in literature \citep{levasseur2011two, fritzen2011nonuniform, morin2017micromechanics, pei2022anisotropic, ren2023micro, liu2023anisotropic} and incorporating anisotropic damage in $\mathtt{E}^2$-TFA approach will be explored in future.\color{black} The rate-dependent effects will be another upcoming improvement to the formulation. fiber-reinforced composites show viscoelastic/viscoplastic behaviour when subjected to different strain rates, which are important to capture for dynamic applications. For phase materials, rate-dependent elastoplastic constitutive framework can be used instead of rate-independent plasticity models, as stated in Section \ref{sec:3}. Furthermore, the proposed formulation regards inter-phase damage successfully; however, subsequent explorations are needed in order to capture the intra-phase damage mode, and it is the subject of the author's current study. 

Nonetheless, $\mathtt{E}^2$-TFA can be effectively utilized as a multiscale scheme for efficiently and accurately estimating the response of the composite material.
   
\appendix
\section{Calculation of $\left(\dfrac{\partial\dot{\boldsymbol{\bar{\mu}}^M}}{\partial\dot{\boldsymbol{\bar{\varepsilon}}^M}}\right)$}\label{appendix1} 
Constitutive equation for each partition yields: 
\begin{equation}
	\boldsymbol{\dot{\bar{\sigma}}}={\mathbb{L}}^{\alpha}:(\boldsymbol{\dot{\bar{\varepsilon}}}-\dot{\boldsymbol{\bar{\mu}}}) \label{eq:87}
\end{equation} 
which can be written for eigenstrain as 
\begin{equation}
	\dot{\boldsymbol{\bar{\mu}}}=\boldsymbol{\dot{\bar{\varepsilon}}}-({{\mathbb{L}}^{\alpha}})^{-1}:\boldsymbol{\dot{\bar{\sigma}}} \label{eq:88}
\end{equation}
Derivative of eigenstrain increment with respect to total strain increment is expressed as 
\begin{equation}
	\left(\dfrac{\partial\dot{\boldsymbol{\bar{\mu}}}}{\partial\dot{\boldsymbol{\bar{\varepsilon}}}}\right)=\mathbb{I}-({{\mathbb{L}}^{\alpha}})^{-1}:\left(\dfrac{\partial\boldsymbol{\dot{\bar{\sigma}}}}{\partial\dot{\boldsymbol{\bar{\varepsilon}}}}\right) \label{eq:89}
\end{equation}
which further demands calculating tangent stiffness, $\left(\dfrac{\partial\dot{\boldsymbol{\bar{\sigma}}}}{\partial\dot{\boldsymbol{\bar{\varepsilon}}}}\right)$ (see \ref{appendix2}).
\section{Calculation of $\left(\dfrac{\partial\dot{\boldsymbol{\bar{\sigma}}}}{\partial\dot{\boldsymbol{\bar{\varepsilon}}}}\right)$}\label{appendix2}
For a phase material, effective stress can be expressed in terms of total strain as:
\begin{equation}
	\boldsymbol{\dot{\tilde{\sigma}}}=\mathbb{L}^{ep}:\boldsymbol{\dot{\varepsilon}}  \label{eq:90}
\end{equation}
where $\mathbb{L}^{ep}$ is an elastoplastic constitutive tensor. Using Eq. (\ref{constitutive}), (\ref{eq:10}) and (\ref{eq:11}), the derivative of Eq. (\ref{eq:90}) can be expressed as:
\begin{equation}
	\od{\boldsymbol{\dot{\sigma}}}{\boldsymbol{\dot{\varepsilon}}}=\left(\mathbb{L}(\mathbb{D})+\frac{\partial{\mathbb{L}(\mathbb{D})}}{\partial{\boldsymbol{\varepsilon}^e}}:\boldsymbol{\dot{\varepsilon}}^e\right):\mathbb{L}^{-1}:\mathbb{L}^{ep}  \label{eq:91}
\end{equation}
where $\dfrac{\partial{\mathbb{L}(\mathbb{D})}}{\partial{\boldsymbol{\varepsilon}^e}}$ is obtained from following:
\begin{equation}
	\frac{\partial{\mathbb{L}(\mathbb{D})}}{\partial{\boldsymbol{\varepsilon}^e}}=\left(\dfrac{\partial\mathbb{L}(\mathbb{D})}{\partial\mathbb{D}}\right):\left(\dfrac{\partial\mathbb{D}}{\partial\omega}\right)\times\left(\dfrac{\partial\omega}{\partial{\kappa}}\right):\left(\dfrac{\partial\kappa}{\partial{\boldsymbol{\varepsilon}^e}}\right)  \label{eq:92}
\end{equation}
and $\mathbb{L}^{ep}$ is derived as: 
\begin{equation}
	\mathbb{L}^{ep}=\frac{\mathrm{d}{\boldsymbol{\dot{\tilde{\sigma}}}}}{\mathrm{d}{\boldsymbol{\varepsilon}}}=\mathbb{L}-\dfrac{\left(\mathbb{L}:\dfrac{\partial \mathfrak{F}^P}{\partial \boldsymbol{\sigma}}\right)\otimes\left(\dfrac{\partial \mathfrak{F}^P}{\partial \boldsymbol{\sigma}}:\mathbb{L}\right)}{\left(\dfrac{\partial \mathfrak{F}^P}{\partial \boldsymbol{\sigma}}:\mathbb{L}:\dfrac{\partial \mathfrak{F}^P}{\partial \boldsymbol{\sigma}}\right)-\left(\dfrac{\partial \mathfrak{F}^P}{\partial {r}}\right):\left(\dfrac{2}{3}\dfrac{\partial \mathfrak{F}^P}{\partial \boldsymbol{\sigma}}:\dfrac{\partial \mathfrak{F}^P}{\partial \boldsymbol{\sigma}}\right)}  \label{eq:93}
\end{equation}
\section{Calculation of $\left(\dfrac{\partial\dot{\boldsymbol{\varepsilon}^o}}{\partial\dot{\boldsymbol{\bar{\varepsilon}}}^M}\right)$}\label{appendix3}
For the calculation of $\left(\dfrac{\partial\dot{\boldsymbol{\varepsilon}^o}}{\partial\dot{\boldsymbol{\bar{\varepsilon}}}^M}\right)$, Eq. (\ref{eq:55}) is rewritten as: 
\begin{equation}
	\dot{\boldsymbol{\bar{\varepsilon}}^i}-\overline{\mathbb{E}}^i:\dot{\boldsymbol{{\varepsilon}}}^o- \sum_{j=1}^M\overline{\mathbb{S}}^{ij}:\dot{\boldsymbol{\bar{\mu}}^{j}}=0   \label{eq:94}
\end{equation}
Derivative of Eq. (\ref{eq:94}) with respect to $\dot{\boldsymbol{\varepsilon}^o}$ gives:
\begin{equation}
	\left(\dfrac{\partial\dot{\boldsymbol{\bar{\varepsilon}}}^i}{\partial\dot{\boldsymbol{\varepsilon}^o}}\right)=\overline{\mathbb{E}}^i- \sum_{j=1}^M\overline{\mathbb{S}}^{ij}:\left(\dfrac{\partial\dot{\boldsymbol{\bar{\mu}}^j}}{\partial\dot{\boldsymbol{\varepsilon}^o}}\right)   \label{eq:95}
\end{equation}
Eq. (\ref{eq:95}) can be expressed as
\begin{equation}
	\left(\dfrac{\partial\dot{\boldsymbol{\bar{\varepsilon}}}^i}{\partial\dot{\boldsymbol{\varepsilon}^o}}\right)=\overline{\mathbb{E}}^i- \sum_{j=1}^M\overline{\mathbb{S}}^{ij}:\left(\dfrac{\partial\dot{\boldsymbol{\bar{\mu}}^j}}{\partial\dot{\boldsymbol{\bar{\varepsilon}}}^j}\right):\left(\dfrac{\partial\dot{\boldsymbol{\bar{\varepsilon}}}^j}{\partial\dot{\boldsymbol{\varepsilon}^o}}\right)     \label{eq:96}
\end{equation}
and simplified as 
\begin{equation}
	\left[\mathbb{I}\delta_{ij}+\sum_{j=1}^M\overline{\mathbb{S}}^{ij}:\left(\dfrac{\partial\dot{\boldsymbol{\bar{\mu}}^j}}{\partial\dot{\boldsymbol{\bar{\varepsilon}}}^j}\right)\right]:\left(\dfrac{\partial\dot{\boldsymbol{\bar{\varepsilon}}}^j}{\partial\dot{\boldsymbol{\varepsilon}^o}}\right)=\overline{\mathbb{E}}^i \label{eq:97}
\end{equation}
Finally, we have
\begin{equation}
	\left(\dfrac{\partial\dot{\boldsymbol{\varepsilon}^o}}{\partial\dot{\boldsymbol{\bar{\varepsilon}}}^j}\right)={(\overline{\mathbb{E}}^i)}^{-1}:\left[\mathbb{I}\delta_{ij}+\sum_{j=1}^M\overline{\mathbb{S}}^{ij}:\left(\dfrac{\partial\dot{\boldsymbol{\bar{\mu}}^j}}{\partial\dot{\boldsymbol{\bar{\varepsilon}}}^j}\right)\right] \label{eq:98}
\end{equation}
\color{black}

\section*{Acknowledgement}
This work was supported by a research grant (ARDB/01/1051983/M/I, Project No. 1983) of the Aeronautical Research \& Development Board, India. The content is solely the responsibility of the authors and does not necessarily represent the official views of the funding organisation.
\newpage 
\bibliography{bibfile}
\end{document}